\documentclass[aps,a4,floatfix,nofootinbib]{revtex4}  

\usepackage[polutonikogreek,english]{babel}

\usepackage{fmtcount} 
\usepackage{graphicx}
\usepackage{float}
\usepackage{psfrag}
\usepackage{fixmath}
\usepackage{amsmath, amsfonts, amssymb, amsthm, amscd} 
\usepackage{marvosym} 
\usepackage{yfonts}  
\DeclareGraphicsExtensions{ps,eps,ps.gz,frh.eps,ml.eps}
\usepackage{amstext,amssymb,amsbsy}
\usepackage{braket}
\usepackage{color}
\usepackage{relsize}
\usepackage{mathrsfs} 
\usepackage{mathcomp} 
\usepackage{wasysym} 
\usepackage[makeroom]{cancel}
\usepackage{upgreek}
\usepackage[all]{xy}
\usepackage{ dsfont }
\usepackage{ stmaryrd }
\usepackage[nottoc,numbib]{tocbibind}
\usepackage{array}

\usepackage{url}
\usepackage{hyperref}

\def\XXint#1#2#3{{\setbox0=\hbox{$#1{#2#3}{\int}$ }
\vcenter{\hbox{$#2#3$ }}\kern-.6\wd0}}

\makeatletter

\newcommand{\Rmnum}[1]{\expandafter\@slowromancap\romannumeral #1@}
\makeatother

\makeatletter

\def\env@blscases{
  \let\@ifnextchar\new@ifnextchar
  \left.
  \def\arraystretch{1.2}
  \array{@{}l@{\quad}l@{}}
}
\makeatother

\makeatletter

\def\env@rcases{
  \let\@ifnextchar\new@ifnextchar
  \left.
  \def\arraystretch{1.2}
  \array{@{}l@{\quad}l@{}}
}
\makeatother
\DeclareMathOperator{\tr}{\operatorname{tr}}

\begin {document}
\vspace*{-30mm}
\title{Thermodynamics of isospin-asymmetric nuclear matter \\ from chiral effective field theory}
\author{Corbinian\ Wellenhofer$^{1}$}
\email[E-mail:~]{corbinian.wellenhofer@tum.de}
\author{Jeremy\ W.\ Holt$^{2}$}
\email[E-mail:~]{jwholt.phys@gmail.com}
\author{Norbert\ Kaiser$^{1}$}
\email[E-mail:~]{n.kaiser@ph.tum.de}
\affiliation{\vspace*{0.5mm} $^1$Physik Department, Technische Universit\"{a}t M\"{u}nchen, D-85747 Garching, Germany \\ 
$^2$Department of Physics, University of Washington, Seattle, WA 98195, USA }  
\date{July 2, 2015}

\begin{abstract}
The density and temperature dependence of the nuclear symmetry free energy is 
investigated using microscopic two- and three-body nuclear potentials constructed 
from chiral effective field theory. The nuclear force models and many-body 
methods are benchmarked to properties of isospin-symmetric nuclear matter 
in the vicinity of the saturation density as well as the virial expansion of
the neutron matter equation of state at low fugacities. The free energy per particle 
of isospin-asymmetric nuclear matter is calculated assuming a 
quadratic dependence of the interaction contributions on the isospin asymmetry. The 
spinodal instability at subnuclear densities is examined in detail. 
\end{abstract}
\maketitle

\vspace*{-11mm}
%----------------------------------------------------
\section{Introduction}\label{sec0}
%----------------------------------------------------
\vspace*{-2mm}

Determining the thermodynamic equation of state (EoS) of nuclear matter is a central objective in modern nuclear theory. 
The isospin-asymmetry dependence of the EoS is essential for many phenomena in nuclear physics and astrophysics \cite{Steiner:2004fi}. Next-generation radioactive beam 
facilities \cite{Balantekin:2014opa,Li2008113} studying the reactions and structure of 
exotic neutron-rich isotopes in particular provide motivation to improve our microscopic description of highly isospin-asymmetric nuclear matter. To a certain extent, the isospin-asymmetry dependence of the EoS is described by the so-called symmetry free energy. In this work, starting from microscopic calculations of the EoS of isospin-symmetric nuclear matter and pure neutron matter using two- and three-body chiral nuclear interactions, we examine in detail the density and temperature dependence of the symmetry free energy. 
Furthermore, we construct the EoS of isospin-asymmetric nuclear matter, and study the behavior of the nuclear liquid-gas instability as the proton fraction is decreased. The present work is a first step toward the development 
of a chiral effective field theory thermodynamic equation of state across the temperatures, 
densities and isospin asymmetries relevant for describing astrophysical phenomena and 
the matter produced experimentally in heavy-ion collisions at moderate energies.

Chiral effective field theory ($\chi$EFT) provides the basis for the
study of strongly-interacting matter at the energy scales characteristic of normal nuclei 
\cite{Machleidt:2011zz,Holt:2013fwa,Epelbaum:2008ga}. In $\chi$EFT, microscopic nuclear interactions are organized in a 
systematic expansion, with many-nucleon forces naturally included. The low-energy constants parametrizing the 
interactions are generally fixed by high-precision fits to nucleon-nucleon scattering phase shifts and properties of 
light nuclei. Employing chiral interactions in calculations of nuclear many-body systems then gives pure predictions
without additional fine tuning, and theoretical uncertainties can be estimated 
\cite{Epelbaum:2008ga,Bogner:2009bt,Tews:2012fj,Furnstahl:2014xsa,Sammarruca:2014zia} by varying the resolution scale, 
the fitting procedures applied to fix the low-energy constants, and the chiral order of the nuclear potentials. In the case of isospin-symmetric nuclear matter,
semiempirical constraints from the zero-temperature saturation energy, density and incompressibility as well as the critical point of the nuclear liquid-gas 
phase transition have been reproduced with low-momentum chiral nuclear forces in many-body perturbation theory 
\cite{Coraggio:2014nvaa,paper1}. This motivates a study of isospin-asymmetric nuclear matter, which is by comparison
much less constrained by experimental data.

The symmetry free energy 
$\bar F_{\rm sym}(T,\rho)$ is defined as the difference between the free energy per particle in homogeneous isospin-symmetric matter (SNM) and pure neutron matter (PNM):
\begin{align} \label{Fsym00}
\bar{F}_{\text{sym}}(T,\rho)=\bar{F}(T,\rho,\delta=1)-\bar{F}(T,\rho,\delta=0),
\end{align}
where $T$ is the temperature, $\rho=\rho_\text{n}+\rho_\text{p}$ is the total nucleon density, and 
$\delta=(\rho_\text{n}-\rho_\text{p})/(\rho_\text{n}+\rho_\text{p})$ is the isospin-asymmetry parameter (with $\rho_{\text{n}/\text{p}}$ the neutron/proton density).
The free energy per particle of homogeneous
nuclear matter with proton fraction $Y_{\text{p}}=(1-\delta)/2$ can be written as
\begin{align} \label{Fsym0}
\bar{F}(T,\rho,\delta)= \bar{F}(T,\rho,\delta=0)+ \bar{F}_{\text{sym}}(T,\rho)\:\upbeta(T,\rho,\delta)\: \delta^2,
\end{align}
where $\upbeta(T,\rho,\delta=1)=1$.
If isospin-symmetry breaking effects are neglected $\upbeta(T,\rho,\delta)$  is an even function of $\delta$.
It has been validated in various microscopic
many-body calculations (see e.g., Refs.\ \cite{Bombaci:1991zz,PhysRevC.89.025806}) that 
the isospin-asymmetry dependence of $\bar{F}(T,\rho,\delta)$ at zero 
temperature is approximately quadratic to high accuracy over the entire range $0<\delta \leq1$, i.e., $\upbeta(T=0,\rho,\delta)\simeq 1$.
At finite temperatures however
the free Fermi gas contribution to $\bar{F}(T,\rho,\delta)$ contains large terms with quartic and higher
powers of $\delta$, which we quantify explicitly in this work. By comparison, in initial calculations we have found that at the densities and temperatures relevant for the nuclear liquid-gas phase transition the interaction 
contributions give rise to weaker nonquadratic terms and will therefore be assumed to have a quadratic dependence on $\delta$ in this work.
Future research will address the accuracy of this approximation in greater detail.

The liquid-gas phase transition in isospin-asymmetric nuclear matter (ANM) involves \textit{isospin distillation}: in the transition region the system separates into two phases whose proton concentrations deviate from the global value $Y_\text{p}$, with $0\leq Y_\text{p}^{\text{gas}}< Y_\text{p}$ and $Y_\text{p}< Y_\text{p}^{\text{liquid}} < 0.5$ for the case $Y_\text{p}<0.5$. These distillation effects are a generic property of first-order phase transitions in binary thermodynamic systems.
If isospin-symmetry breaking effects are neglected, neutrons and protons are thermodynamically indistinguishable in SNM. Hence, thermodynamically SNM is a pure substance, and there is no isospin distillation for $\delta=0$.
The new features of the phase transition in ANM were discussed in detail in Refs. \cite{Ducoin:2005aa,Hempel:2013tfa,Muller:1995ji,Chomaz:2003dz,PhysRevC.67.041602} using different phenomenological models of the nuclear force.
The transition region is comprised of regions of metastable and unstable single-phase equilibrium, corresponding to different dynamical phase separation mechanisms: nucleation and spinodal decomposition \cite{0034-4885-50-7-001,Abraham197993}.
In this work we focus mostly on the spinodal which delineates the inner region of thermodynamic instability where no metastable state can exist. The evolution of the unstable spinodal region with increasing isospin asymmetry is analyzed in terms of the trajectory of the critical temperature $T_c(\delta)$.
Moreover, we determine the neutron drip point in cold nuclear matter and the fragmentation temperature above which no self-bound drop of liquid nuclear matter can exist.
It should be emphasized that in the present paper we discuss the liquid-gas instability of \textit{infinite nuclear matter}, i.e., bulk nucleonic matter without Coulomb interactions. The inclusion of surface energies and the Coulomb repulsion of protons is required for an accurate description of the matter produced in intermediate-energy heavy-ion collisions.
In neutron stars and core-collapse supernovae the realization of the nuclear liquid-gas instability is strongly affected by the presence of a (highly incompressible) charge neutralizing background of electrons (and myons) \cite{PhysRevC.75.065805,PhysRevLett.98.131102}. 
In particular, the competition between nuclear and Coulomb interactions (frustration) entails the fomration of mesoscopic inhomogeneities with nontrivial spatial structures. These so-called \textit{pasta phases} have been studied extensively in the literature \cite{Ravenhall:1983uh,Hashimoto,Pethick:1998qv,PhysRevC.69.045804,Maruyama:2005vb,
Watanabe:2004tr,Avancini:2012bj,Schneider:2013dwa}. 
Including these effects as well as the presence of few-nucleon bound-states \cite{Horowitz:2005nd,PhysRevC.82.045802,PhysRevC.81.015803} at very low densities represents a future challenge.

The paper is organized as follows. In Sec.\ \ref{sec1} we recall the main results for the EoS of SNM obtained in Ref.\ \cite{paper1}, and show results for additional derived thermodynamic quantities, i.e., the entropy per nucleon and the internal energy per nucleon.
In Sec.\ \ref{sec2} we extend the calculations to pure neutron
matter. The zero-temperature results are compared to those from recent quantum Monte Carlo simulations while 
the finite-temperature EoS at low densities is compared to the virial expansion.
In Sec.\ \ref{sec3} we investigate the temperature and density dependence of  
the symmetry free energy, entropy, and internal energy.
The thermodynamics of isospin-asymmetric nuclear matter is studied in Sec.\ \ref{sec4}. In particular, we examine in detail the dependence on isospin asymmetry of the EoS of a free nucleon gas. Finally, Sec.\ \ref{sec5} provides a short summary.

\vspace*{-0.35cm}
%----------------------------------------------------
\section{Isospin-symmetric nuclear matter} \label{sec1}
%----------------------------------------------------
\vspace*{-2mm}

In Ref.\ \cite{paper1} we calculated the free energy per nucleon in infinite homogeneous SNM using the Kohn-Luttinger-Ward \cite{Kohn:1960zz,Luttinger:1960ua} many-body perturbation series including contributions up to second order, i.e.,
\begin{align} \label{KLWseries}
\bar{F}(T,\rho,\delta=0)=\bar{F}_{0}(T,\mu_0)+\bar{F}_{\text{rel}}(T,\mu_0)+\lambda \bar{F}_1(T,\mu_0)+\lambda^2 \bar{F}_2(T,\mu_0)+\mathcal{O}(\lambda^3).
\end{align}
Here, $\lambda$ counts the number of interaction insertions, $\bar{F}_{0}(T,\mu_0)$ corresponds to a nonrelativistic free nucleon gas, and $\bar{F}_{\text{rel}}(T,\mu_0)$ is a correction term which together with $\bar{F}_{0}$ reproduces the properties of a relativistic free nucleon gas over a wide range of densities and temperatures \cite{Fritsch:2002hp}. The first- and second-order terms $\bar{F}_1(T,\mu_0)$ and $\bar{F}_2(T,\mu_0)$ receive contributions from both the two-body and the three-body nuclear force. The second-order term $\bar{F}_2(T,\mu_0)$ includes
(temperature and density dependent) self-energy corrections, and has been evaluated by approximating the three-nucleon interaction with a temperature and density dependent effective two-body potential
(for details see Refs.\ \cite{paper1,Holt:2009ty,PhysRevC.82.014314,PhysRevC.83.031301,PhysRevC.79.054331}. 
Explicit formulas for the different contributions in Eq. (\ref{KLWseries}) are given in Ref.\ \cite{paper1}.
The effective one-body chemical potential $\mu_0$ is in one-to-one correspondence with the nucleon density via
\begin{align} \label{rhoNUC}
\rho(T,\mu_0) =
\frac{1}{\pi^2} \sum_{\tau} \int \limits_0^{\infty}\! \mathrm{d}k \, k^2 \bigg[1+
\exp{k^2/2M-\mu_0\over T}\bigg]^{-1},
\end{align}
where $\tau\in \{-1/2,1/2\}$ is the isospin projection quantum number and $M\simeq 938.9\,\text{MeV}$ is the average nucleon mass.
For Eq. (\ref{KLWseries}) to be sufficiently converged at second order in $\lambda$, 
low-momentum interactions have to be used, i.e., interactions with restricted resolution in coordinate space (corresponding to an ultraviolet cutoff in momentum space). 
The various sets of N3LO (i.e., fourth order in the chiral expansion) two-body and N2LO three-body chiral low-momentum interactions used in Ref.\ \cite{paper1} 
correspond to different regularization methods, resolution scales $\Lambda$, and 
low-energy constants. 
For interactions constructed at resolution scales $\Lambda\leq 450\,\text{MeV}$ appropriate perturbative behavior was found.
The SNM equation of state obtained from the sets of two- and three-body potentials denoted by n3lo414 ($\Lambda=414\,\text{MeV}$) and n3lo450 ($\Lambda=450\,\text{MeV}$), respectively, (see Refs.\ \cite{Coraggio:2007mc,Coraggio:2014nvaa,Coraggio13} for details) agree with empirical constraints from the zero-temperature saturation energy, density and incompressibility \cite{Brockmann:1990cn,Blaizot,Khoa:1997zz,Youngblood:1999zza}, and with estimates for the critical point of the nuclear liquid-gas phase transition obtained through the analysis of data from multifragmentation, fission and compound nuclear decay experiments \cite{Karnaukhov:2008be,PhysRevLett.89.212701,Ogul:2002ka,Elliott:2013pna}. The values of these quantities obtained from n3lo414 and n3lo450 in Ref.\ \cite{paper1} are displayed in Table \ref{table:observables1}. \footnote{Note that the value of the so-called critical compressibility factor is $Z_c= P_c/(T_c\,\rho_c)\simeq 0.29$ for both n3lo414 and n3lo450; this is very similar to the values of $Z_c$ of various atomic or molecular fluids \cite{bookZc}, but differs from the value $Z_c=0.375$ corresponding to equations of state of the van der Waals--Berthelot type \cite{0143-0807-9-1-010}.}

\vspace*{-0.1cm}

\begin{figure}[H] 
\centering
\vspace*{-3.6cm}
\hspace*{-2.2cm}
\includegraphics[width=1.15\textwidth]{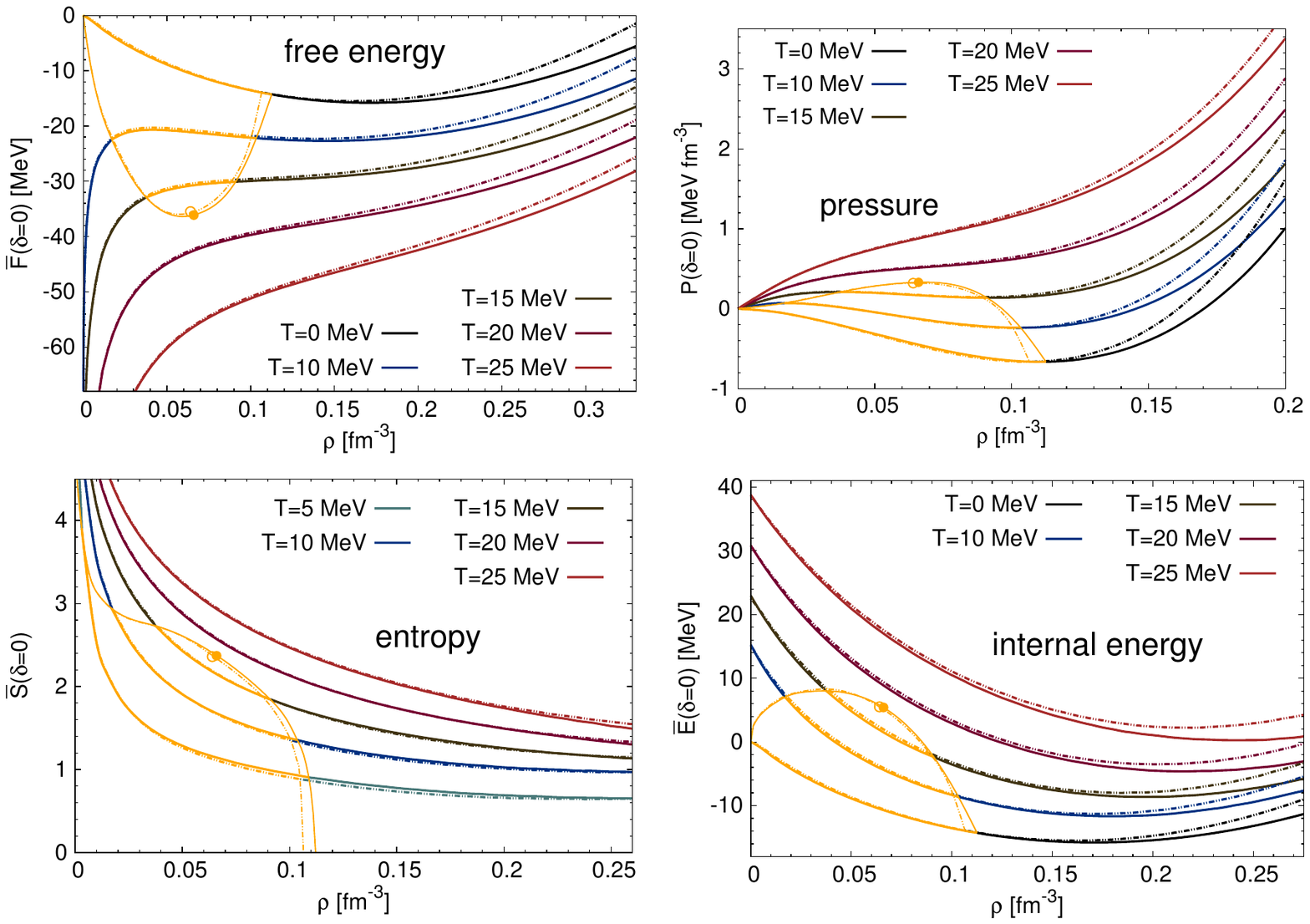} 
\vspace*{-11.8cm}
\caption{(Color online) Results for the free energy per nucleon $\bar F(T,\rho,\delta=0)$, the pressure $P(\rho,T,\delta=0)$, the entropy per nucleon $\bar S(T,\rho,\delta=0)$ and the internal energy per nucleon $\bar E(\rho,T,\delta=0)$ in isospin-symmetric nuclear matter. The uncertainty bands correspond to calculations using two different sets of chiral low-momentum two- and three-body interactions, n3lo414 (solid lines) and n3lo450 (dash-dot lines). 
The unstable spinodal region is marked out explicitly. The critical point is shown as a circle (full circle for n3lo414, open circle for n3lo450). The zero-temperature endpoint of the low-density part of the spinodal is located at $\rho\simeq 2\cdot 10^{-4}\,\text{fm}^{-3}$.}
\label{plots_sec0}
\end{figure}

\vspace*{-0.3cm}
From the free energy per nucleon the pressure and the entropy per nucleon follow via standard thermodynamic relations:
\begin{align} \label{KLW}
P(T,\rho,\delta=0)=\rho^2 \frac{\partial \bar{F}(T,\rho,\delta=0)}{\partial \rho},
\;\;\;\;\;\;\;\;\;\;
\bar{S}(T,\rho,\delta=0)=-\frac{\partial \bar{F}(T,\rho,\delta=0)}{\partial T}.
\end{align}
The internal energy per nucleon is given by $\bar{E}=\bar{F}+T\bar{S}$.
The results for these quantities are shown in Fig.\ \ref{plots_sec0} for temperatures in the range $T=0-25 \,\text{MeV}$. 
The spinodal region\footnote{In SNM the unstable spinodal region corresponds to $(\partial P/\partial \rho)_T \leq 0$, with $(\partial P/\partial \rho)_T =0$ on the spinodal, cf.\ Sec.\ \ref{sec41}.}  where the homogeneous (i.e., single-phase constrained) system is unstable with respect to infinitesimal density fluctuations is shown explicitly.
Note that for low temperatures the region of negative pressure extends into the metastable region (cf.\ also Fig.\ \ref{plots_sec4_5}), which is a generic property of liquids that are self-bound at low temperatures and a well-known feature of superheated molecular liquids \cite{bookML}.

{\vspace{-1mm}}
\begin{table}[H]
\begin{center}
\begin{minipage}{14cm}
\setlength{\extrarowheight}{1.5pt}
\hspace*{-10mm}
\begin{tabular}{c  ccc c c c}
\hline  \hline
&\;\;\;\;\;\;\;\; \;\;\;
$\bar E_{\text{sat}}\,(\text{MeV})$ &\;\;\;\; \;\;\;$\rho_{\text{sat}}\,(\text{fm}^{-3})$ &\;\;\;\;\;\;\; $K\,(\text{MeV})$ &\;\;\;\;\;\;\;
$T_c\,(\text{MeV})$&\;\;\;\;\;\;\;$\rho_c\,(\text{fm}^{-3})$&\;\;\;\;\;\;\;$P_c
\,(\text{MeV fm}^{-3})$
   \\ \hline 
n3lo414   &\;\;\;\;\;\;\;\;\;\;\;
-15.79   &\;\;\;\;\;\;\; 0.171  &\;\;\;\;\;\;\; 223   &\;\;\;\;  \;\;\;
17.4 &\;\;\;\; \;\;\; 0.066 &\;\;\;\;\;\;\; 0.33  \\ 
n3lo450   &\;\;\;\;\;\;\;\;\;\;\;
-15.50   &\;\;\;\;\;\;\;  0.161  &\;\;\;\;\;\;\; 244   &\;\;\;\; \;\;\; 
17.2     &\;\;\;\; \;\;\; 0.064  &\;\;\;\;\;\;\; 0.32 \\ 
 \hline \hline 
\end{tabular}
\end{minipage}
{\vspace{0mm}}
\begin{minipage}{14cm}
\caption
{Zero-temperature saturation energy $\bar E_{\text{sat}}$, density $\rho_{\text{sat}}$ and 
incompressibility $K$, as well as the critical temperature $T_c$, density $\rho_c$ and pressure $P_c$ of the 
liquid-gas phase transition in isospin-symmetric nuclear matter from the sets of chiral two- and three-body 
nuclear interactions n3lo414 and n3lo450.}
\label{table:observables1}
\end{minipage}
\end{center}
\end{table}

\vspace{-14mm}

\section{Pure neutron matter}\label{sec2}
\vspace*{-2mm}

\begin{figure}[H] 
\centering
\vspace*{-3.5cm}
\hspace*{-2.2cm}
\includegraphics[width=1.15\textwidth]{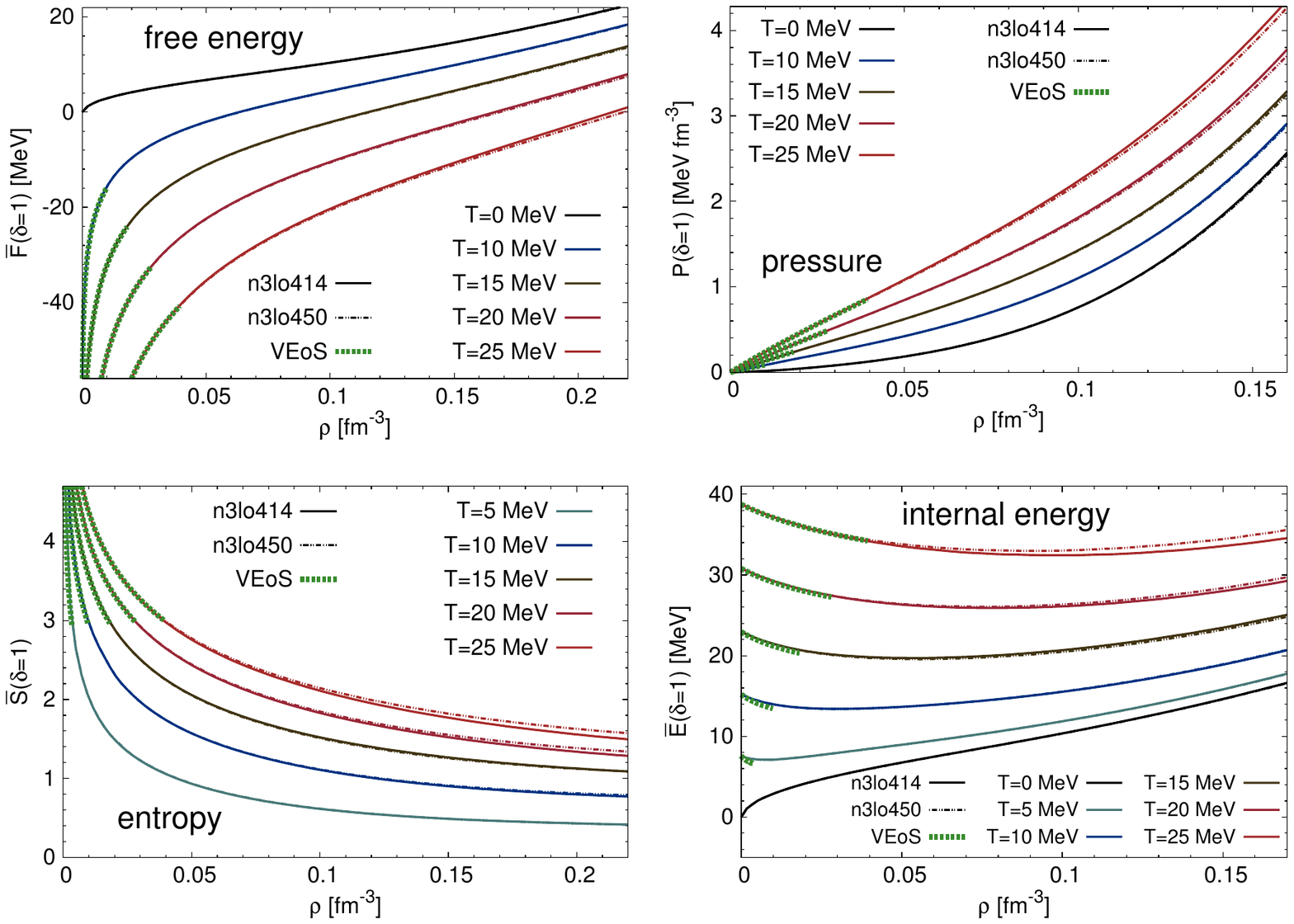} 
\vspace*{-11.6cm}
\caption{(Color online) Results for the free energy per particle $\bar F(T,\rho,\delta=1)$, the pressure $P(\rho,T,\delta=1)$, the entropy per particle $\bar S(T,\rho,\delta=1)$ and the internal energy per particle $\bar E(\rho,T,\delta=1)$ in pure neutron matter. The solid lines show the results from n3lo414, the dash-dot lines the n3lo450 results. 
The thick dashed lines correspond to the
model-independent virial equation of state (VEoS) determined from neutron-neutron scattering phase shifts.
The VEoS lines end where the fugacity is $z=0.5$. }
\label{plots_sec2_1}
\end{figure}

The free energy per particle in PNM, $\bar{F}(T,\rho,\delta=1)$, is obtained by restricting the isospin sum(s) in Eq. (\ref{rhoNUC}) and in the different contributions in Eq. (\ref{KLWseries}). Moreover, in PNM the three-body contributions proportional to the low-energy constants $c_{E,D}$ and $c_4$ are absent \cite{Holt:2009ty}.
The results for the free energy per particle, the pressure, the entropy per particle and the internal energy per particle in PNM are shown in Fig.\ \ref{plots_sec2_1}.
Note that the uncertainty bars obtained by varying the resolution scale (n3lo414 vs. n3lo450) increase with temperature and are significantly reduced as compared to the SNM results in Ref.\ \cite{paper1}, which is due to the decreased magnitude of nuclear interactions in PNM. 
At very low temperatures the internal energy per particle increases monotonically with increasing density (as required by the absence of a liquid-gas instability in PNM), but otherwise there is a local minimum at finite density.

%------------------------------------------------FIGURE sec2-1
%--------------------------------------------END_FIGURE sec2-1

At very low energies, where higher partial waves are unimportant, the interaction between neutrons is characterized by the
neutron-neutron scattering length $a_s$. In the regime where $a_s\simeq -19\,\text{fm}$ is large compared to the interparticle separation, $1 \ll |k_F a_s|$ (with $k_F$ the Fermi momentum), a perturbative approach to neutron matter is not reliable. 
The model-independent virial equation of state (VEoS) computed by Horowitz and Schwenk in Ref.\ \cite{Horowitz:2005zv} from neutron-neutron scattering phase-shifts provides a benchmark for perturbative calculations of low-density neutron matter at nonzero temperature.
In the virial expansion, the grand canonical expressions for the pressure and the density are expanded in powers of the fugacity $z=\exp(\mu/T)$, leading to
\begin{align}
P(T,z,\delta=1)=\frac{2T}{\uplambda^3}  \Big(z+z^2 b_{2}(T)+\mathcal{O}(z^3) \Big),
\;\;\;\;\;\;\;\;\;\;
\rho(T,z,\delta=1)=\frac{2}{\uplambda^3}
\Big(z+2 z^2 b_{2}(T)+\mathcal{O}(z^3) \Big),
\end{align}
where $\mu$ is the chemical potential, and $\uplambda=\sqrt{(2 \pi)/(M T)}$ is the neutron thermal wavelength.
The second
virial coefficient is given by
\begin{align}
b_2(T)=\frac{1}{2^{1/2} \pi T} \int_{0}^{\infty} \!\!\!\! dE \,\exp[-E/(2T)]\, \delta_{\text{tot}}(E)-2^{-5/2},
\end{align}
where $\delta_{\text{tot}}(E)$ is the sum of the isospin-triplet elastic scattering phase shifts
at laboratory energy $E$.
From the pressure and density as functions of the fugacity the
free energy per particle $\bar F$, entropy per particle $\bar S$, and internal energy per particle $\bar E$ follow again from standard thermodynamic relations (see Ref.\ \cite{Horowitz:2005zv} for details). The results for these quantities are shown as green dashed lines in Fig.\ \ref{plots_sec2_1}.\footnote{Note that we have added the relativistic correction term to the VEoS lines. In particular, the zero-density limit of the internal energy per particle is
given by 
$\bar E(T,\rho,\delta)\xrightarrow{\rho\rightarrow 0} 
\bar E_{0}(T,\rho,\delta)\,|_{\rho\rightarrow 0}+\bar E_{\text{rel}}(T,\rho,\delta)\,|_{\rho\rightarrow 0}=3T/2+15T^2/(8M)$. This agrees with the expansion in powers of $T$ of the internal energy per particle of a relativistic classical ideal gas, $\bar E_{\text{classical}}=3T+M K_1(M/T)/K_2(M/T)$ [where $K_{1,2}(M/T)$ are modified Bessel functions].
} 
One sees that in the case of $\bar F$, $P$ and $\bar S$ there are almost no visible deviations between the VEoS and the perturbative results. This seemingly perfect agreement is however misleading, because the discrepancies corresponding to the different treatment of 
the interactions in the virial and the perturbative approach
are overpowered by the large size of the (nonrelativistic) free Fermi gas contribution. The deviations are more 
transparent in the case of the internal energy per particle due to cancellations of the free Fermi gas
terms in the free energy and entropy. The virial and perturbative results are closer at larger temperatures,
since the EoS is less sensitive to the physics of large scattering lengths at higher momentum scales.

%-----------------------------------------------------------------------------
%------------------------------------------------FIGURE sec2-2
\begin{figure}[H] 
\centering
\vspace*{-4.0cm}
\hspace*{0.7cm}
\includegraphics[width=1.30\textwidth]{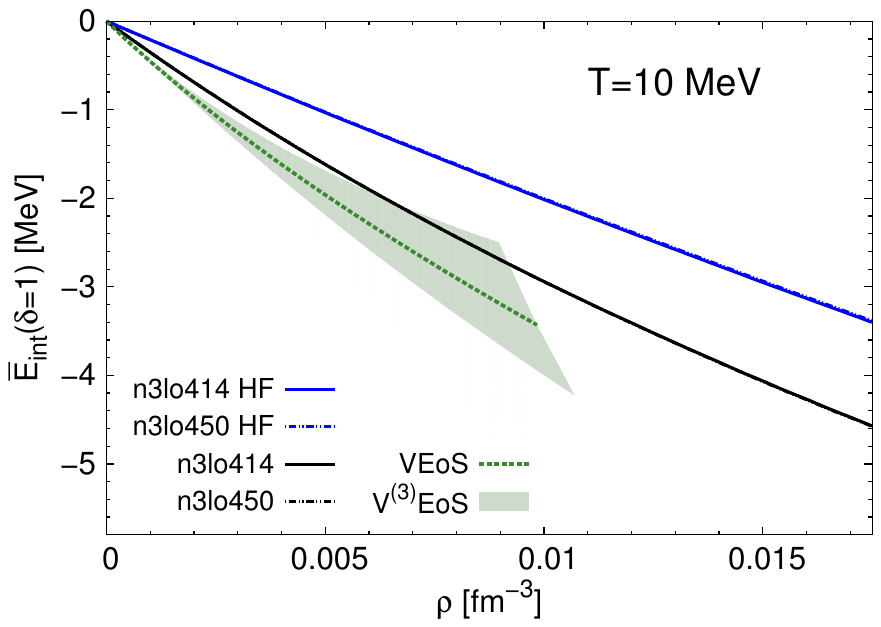} 
\vspace*{-19.8cm}
\caption{(Color online) Interaction contribution to the internal energy per particle, $\bar E_{\text{int}}
(\rho,T,\delta=1)$, in pure neutron matter at $T=10\,\text{MeV}$ and low densities. The different lines 
correspond to the results from microscopic chiral nuclear interactions at first order (labeled ``HF'') and 
second order in Eq.\ (\ref{KLWseries}) as well as the virial expansion truncated at second order (VEoS) 
and with uncertainty bands obtained by estimating the third-order term. The n3lo414 and n3lo450 results 
are almost identical.}
\label{plots_sec2_3}
\end{figure}
%--------------------------------------------END_FIGURE sec2-2
The differences between the virial and the perturbative results for the internal energy per particle 
are examined more closely in Fig.\ \ref{plots_sec2_3} for $T=10 \,\text{MeV}$. To depict the deviations 
more clearly we have subtracted the noninteracting contributions, i.e., the quantity shown is $\bar E_{\text{int}}
= \bar E-\bar E_{0}-\bar E_{\text{rel}}$. The virial results include uncertainty bands obtained from estimating 
the neglected third virial coefficient as $|b_3(T)|\leq |b_2(T)|/2$. We also show the perturbative results at the 
Hartree-Fock level [first order in Eq.\ (\ref{KLWseries})]. 
One sees that compared to the Hartree-Fock results the inclusion of second-order 
contributions leads to much closer agreement with the virial expansion. The second-order calculation still slightly underpredicts the attractive interaction
contributions, in contrast to the pseudopotential approach based on nucleon-nucleon scattering phase 
shift data that was explored in Ref.\ \cite{Rrapaj:2014yba}.
We conclude that while the perturbative approach cannot fully capture the 
large scattering length physics of low-density neutron matter, the resulting errors are reasonably small when 
second-order contributions are included.

\begin{figure}[H] 
%-----------------------------------------------------------------------------
\centering
\vspace*{-4.0cm}
\hspace*{0.7cm}
\includegraphics[width=1.3\textwidth]{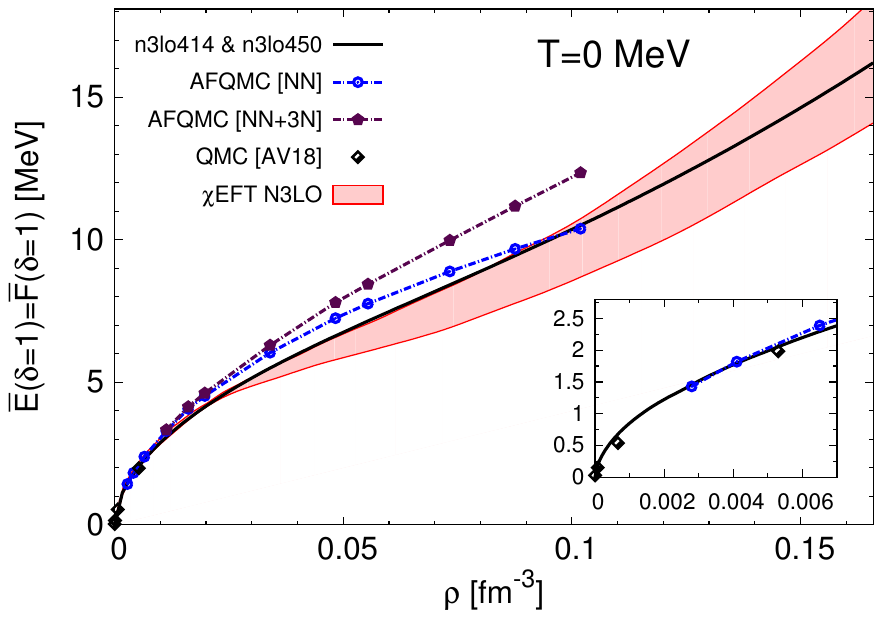} 
\vspace*{-19.8cm}
\caption{(Color online) Energy per particle in pure neutron matter at zero temperature, 
$\bar E(T=0,\rho,\delta=1)$, obtained from various many-body methods (see text for details). 
The inset magnifies the behavior at very low densities where quantum Monte Carlo simulations,
labeled ``QMC [AV18]'', are expected to be most accurate. }
\label{plots_sec2_2}
\end{figure}

In recent years, the zero-temperature EoS of PNM from chiral nuclear interactions has been studied by numerous authors 
within various many-body frameworks
\cite{Tews:2012fj,Kruger:2013kua,Wlazlowski2014,Hebeler:2014ema,PhysRevLett.112.221103,Gezerlis13,
Hagen2014,PhysRevC.90.054322,Coraggio13,PhysRevC.88.054312,PhysRevC.88.064005,PhysRevC.82.014314,Gezerlis:2014zia}. 
We compare our results to results obtained from perturbative calculations with various chiral interactions by the Darmstadt group (red band in Fig. 8 in Ref.\ \cite{Kruger:2013kua}) in Fig.\ \ref{plots_sec2_2}.
In addition to the N2LO chiral three-neutron forces, their calculations also include all
N3LO three- and four-neutron interactions. 
The uncertainty bands in their results were obtained by allowing large variations of the low-energy constants 
parameterizing the many-neutron forces. One sees that the (almost overlapping) results from n3lo414 and n3lo450 lie within these bands. 
In Fig.\ \ref{plots_sec2_2} we also show results
obtained from auxiliary-field quantum Monte Carlo simulations with chiral N3LO two-body (AFQMC [NN]) and N3LO two-body plus N2LO three-body forces (AFQMC [NN+3N]) by Wlaz{\l}owski \textit{et al.} \cite{Wlazlowski2014}.
The perturbative and the AFQMC results are very similar at densities $\rho\lesssim 0.006\,\text{fm}^{-3}$, where both are in close agreement with the (fixed-node) quantum Monte Carlo calculations (based on the AV18 potential) of Gezerlis and Carlson \cite{Gezerlis:2007fs}.
However, at higher densities the EoS predicted by the AFQMC calculations (with three-body forces included) is significantly more repulsive. 
This discrepancy may be (partly) related to systematic errors in the AFQMC treatment (cf.\ also Ref.\ \cite{Gezerlis:2014zia}). 

%------------------------------------------------FIGURE sec2-3
%--------------------------------------------END_FIGURE sec2-3
%
%
%-----------------------------------------------------------------------------
%
%%%%%%%%%%%%%%%%%%%%%%%%%%%%%%%%%%%%%%%%%%%%%%%%%%%%%%%%%%%%%%%%%%%%%%%%%%%%%%%
\section{Symmetry free energy, entropy and internal energy} \label{sec3}
\begin{figure}[H] 
\centering
\vspace*{-4.9cm}
\hspace*{-2.2cm}
\includegraphics[width=1.15\textwidth]{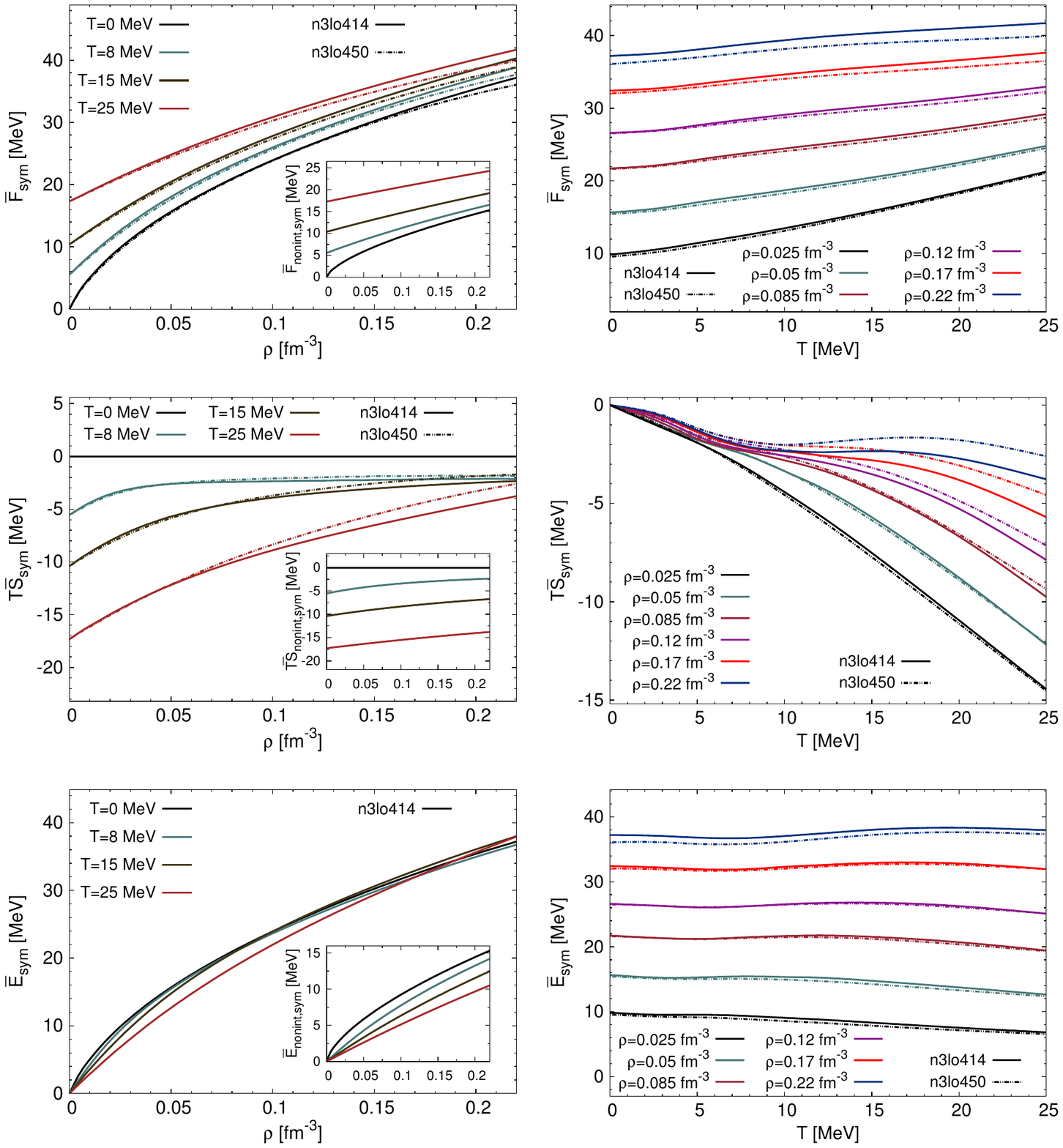} 
\vspace*{-3.9cm}
\caption{(Color online) Left column: results for the symmetry free energy $\bar F_{\text{sym}}(T,\rho)$, the symmetry entropy times temperature $T\bar S_{\text{sym}}(T,\rho)$, and the symmetry internal energy $\bar{E}_{\text{sym}}(T,\rho)$, plotted as functions of density.
The insets show the noninteracting (free Fermi gas) contribution to the different symmetry quantities.
Right column: $\bar F_{\text{sym}}(T,\rho)$,  $T\bar S_{\text{sym}}(T,\rho)$, and $\bar{E}_{\text{sym}}(T,\rho)$ as functions of temperature at different densities.
The lines are interpolated, with calculated data points at $T/\text{MeV}=0,3,5,8,10,12,15,20,25$. }
\label{plots_sec3_1}
\end{figure}

From the results for the free energy per particle in homogeneous SNM and PNM 
the symmetry free energy $\bar{F}_{\text{sym}}(T,\rho)$ is obtained via Eq.\ (\ref{Fsym00}).
The symmetry entropy and internal energy are related to the symmetry free energy via $\bar{S}_{\text{sym}}=-\partial \bar{F}_{\text{sym}}/\partial T$ and $\bar{E}_{\text{sym}}=\bar{F}_{\text{sym}}+T\bar{S}_{\text{sym}}$. 
The results for $\bar{F}_{\text{sym}}$, $T\bar{S}_{\text{sym}}$ and $\bar{E}_{\text{sym}}$ are shown as functions of density at different temperatures in the left column of Fig.\ \ref{plots_sec3_1}. 
In the insets we show the 
noninteracting contribution to these quantities, i.e., 
\begin{align} \label{Fsym-0}
\bar{F}_{\text{nonint},\text{sym}}(T,\rho)=\bar{F}_{0}(T,\rho,1)-\bar{F}_{0}(T,\rho,0)
+\bar{F}_{\text{rel}}(T,\rho,1)-\bar{F}_{\text{rel}}(T,\rho,0),
\end{align}
in the case of the symmetry free energy. 
In the right column of Fig.\ \ref{plots_sec3_1} we show $\bar F_{\text{sym}}(T,\rho)$,  $T\bar S_{\text{sym}}(T,\rho)$, and $\bar{E}_{\text{sym}}(T,\rho)$ as functions of temperature at different densities.

\newpage

\vspace*{0.1cm}
%_______________________________________________________
%------------------------------------------------FIGURE sec3_1
%--------------------------------------------END_FIGURE sec3_1
\newpage

%-----------------------------------------------------------------------------

One sees that in the considered range of densities and temperatures, $\bar{F}_{\text{sym}}$ is a 
monotonic increasing function of density and temperature. The density and temperature dependence 
of $T\bar{S}_{\text{sym}}$ is more involved. At low densities $T\bar{S}_{\text{sym}}$ decreases 
monotonically with $T$, but for densities $\rho\gtrsim 0.2\,\text{fm}^{-3}$ a local minimum is found 
at $T\sim 10\,\text{MeV}$. \footnote{We note that the temperature dependence of $\bar{F}_{\text{sym}}$ and $T\bar{S}_{\text{sym}}$ approaches linear behavior in the limit of vanishing density, $\bar{F}_{\text{sym}}(T,\rho \!\rightarrow \! 0)=-T\bar{S}_{\text{sym}}(T,\rho\! \rightarrow \! 0)=T\ln 2$.}
The results from n3lo414 and n3lo450 are very similar for densities well 
below nuclear saturation density, but at higher densities the dependence on the resolution scale
becomes significant. In particular, the decrease in the slope of $\bar{F}_{\text{sym}}$ with increasing 
density is more pronounced in the n3lo450 results.
The $T$ dependence of $T\bar{S}_{\text{sym}}(T,\rho)$ approximately balances that of 
$\bar{F}_{\text{sym}}(T,\rho)$, and as a result their sum, the symmetry internal energy 
$\bar{E}_{\text{sym}}$, increases with density but varies only very little with temperature.
At densities near nuclear saturation density the deviations of 
$\bar{E}_{\text{sym}}(T,\rho\simeq \rho_{\text{sat}})$ from its value at zero temperature are 
below $0.5\,\text{MeV}$.

\begin{figure}[H] 
\centering
\vspace*{-3.0cm}
\hspace*{-2.2cm}
\includegraphics[width=1.15\textwidth]{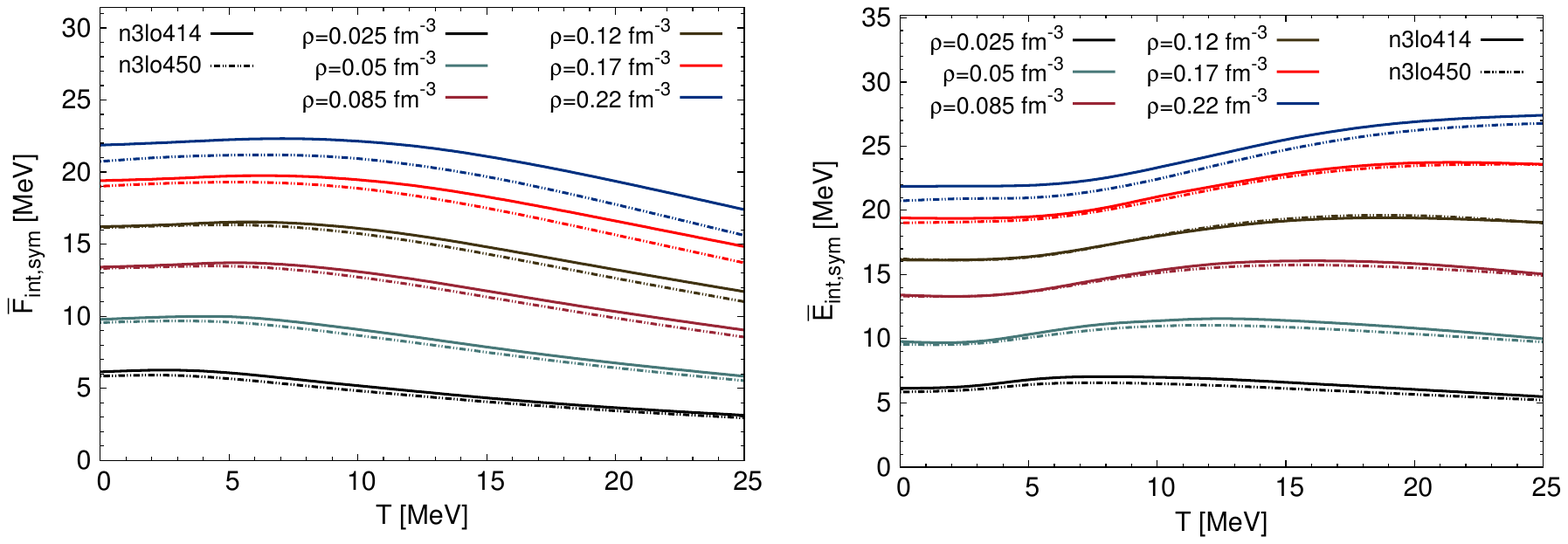} 
\vspace*{-18.2cm}
\caption{(Color online) Temperature dependence of the interaction contributions to the symmetry free energy, $\bar{F}_{\text{sym,int}}(T,\rho)$, and the symmetry internal energy, $\bar{E}_{\text{sym,int}}(T,\rho)$  at different densities. The lines are interpolated, with calculated data points at $T/\text{MeV}=0,3,5,8,10,12,15,20,25$. }
\label{plots_sec3_2}
\end{figure}

In Fig.\ \ref{plots_sec3_2}
we show the symmetry quantities with the noninteracting contributions subtracted, i.e.,  
\begin{align} \label{Fsym-int}
\bar{F}_{\text{int,sym}}(T,\rho)=\bar{F}_{\text{sym}}(T,\rho)-\bar{F}_{\text{nonint},\text{sym}}(T,\rho),
\end{align}
as functions of temperature at different densities.
In both cases the interaction contributions tend to counteract the temperature dependence of noninteracting contributions, cf.\ the insets in Fig.\ \ref{plots_sec3_1}. In the case of $\bar{F}_{\text{sym}}$ (and also $T\bar{S}_{\text{sym}}$) the noninteracing contributions dominate, but in the case of $\bar{E}_{\text{sym}}$ the size of the noninteracting contributions and the ones from chiral nuclear interactions is more balanced, and the $T$ dependence of both contributions approximately cancels each other, leading to the observed approximate temperature independence at densities near nuclear saturation density.

\vspace*{0.7cm}
%------------------------------------------------FIGURE sec3-2
%--------------------------------------------END_FIGURE sec3-2
%
%-----------------------------------------------------------------------------
\vspace*{0.7cm}

\begin{figure}[H] 
\centering
\vspace*{-3.4cm}
\hspace*{0.7cm}
\includegraphics[width=1.3\textwidth]{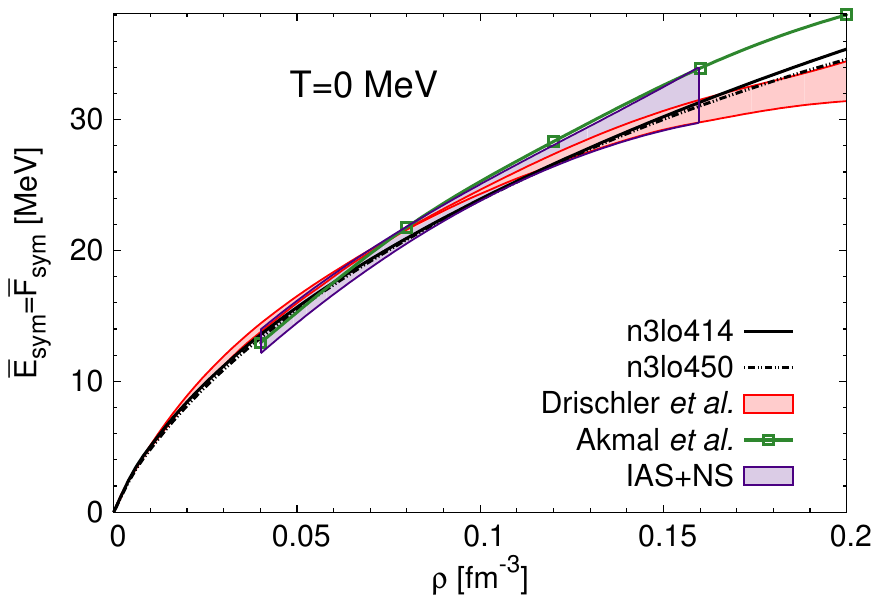} 
\vspace*{-20.7cm}
\caption{(Color online) Symmetry (free) energy as a function of density at zero temperature, $\bar{F}_{\text{sym}}(T=0,\rho)$. The results from the chiral nuclear interactions n3lo414 and n3lo450 are compared to those of Drischler \textit{et al.} \cite{PhysRevC.89.025806} and Akmal \textit{et al.} \cite{Akmal:1998cf}. Also shown are
empirical constraints from the analysis of isobaric analog states and neutron skins (IAS+NS).
}
\label{plots_sec3_3}
\end{figure}

In Fig.\ \ref{plots_sec3_3} we compare our results for the symmetry (free) energy at zero temperature
to the results obtained by Drischler \textit{et al.} \cite{PhysRevC.89.025806} from calculations of the 
EoS of neutron-rich matter using several renormalization group--evolved chiral nuclear interactions. 
For comparison we also show the results from microscopic calculations within a variational approach 
by Akmal \textit{et al.} \cite{Akmal:1998cf} based on the AV18 two-body and the Urbana UIX three-body potential. \footnote{The results by Akmal \textit{et al.} include relativistic boost corrections as well as an artificial correction term added to reproduce the empirical saturation point of SNM (``corrected'' in Table VI. and ``A18$ + \delta \! v + $UIX*\,'' in Table VII. in Ref.\ \cite{Akmal:1998cf}).} While the results of Drischler \textit{et al.} are 
compatible with our results, the calculations by Akmal \textit{et al.} predict a symmetry energy that deviates visibly from the n3lo414 and n3lo450 results. In Fig.\ \ref{plots_sec3_3} we also show 
recent empirical constraints obtained from the analysis of isobaric analog states and 
neutron skins (IAS+NS) \cite{Danielewicz20141}. One sees that the n3lo414 and n3lo450 results lie in 
the IAS+NS bands in the entire constrained density region $0.04\lesssim \rho/\text{fm}^{-3}\lesssim 0.16$.

%-----------------------------------------------------------------------------
%------------------------------------------------FIGURE sec3-3
%--------------------------------------------END_FIGURE sec3-3
%
%-----------------------------------------------------------------------------

For densities close to nuclear saturation density the symmetry (free) energy at zero temperature is usually expanded around $J=\bar F_{\text{sym}}(T=0,\rho_{\text{sat}})$ in terms of $x=(\rho/\rho_{\text{sat}}-1)/3$:
\begin{align} \label{JLK}
\bar F_{\text{sym}}(T=0,\rho) = J+Lx+\frac{1}{2} K_{\text{sym}} x^2
+\mathcal{O}(x^3),
\end{align}
where $L=3\rho_\text{sat}\partial \bar F_{\text{sym}}(T=0,\rho)/\partial \rho \,|_{\rho=\rho_{\text{sat}}}$ is called the slope parameter, and $K_{\text{sym}}=9\rho_\text{sat}^2\partial^2 \bar F_{\text{sym}}(T=0,\rho)/\partial \rho^2 \,|_{\rho=\rho_{\text{sat}}}$ the symmetry incompressibility.
The density where the ground state energy per particle in isospin-asymmetric nuclear matter has a local minimum is related to the parameters in the above expansion via $\rho_{\text{sat}}(\delta)\simeq \rho_{\text{sat}} [1-3L\,\delta^2/K]$ (cf.\ Ref.\ \cite{PhysRevC.80.014322}).\footnote{The densities $\rho_{\text{sat}}(\delta)$ correspond to stable self-bound states only for isospin asymmetries up to the neutron drip point, $\delta \leq \delta_{\text{ND}}$, cf.\ Secs.\ \ref{sec42} and \ref{sec43}.} 
The corresponding incompressibility $K(\delta)$ obeys the approximate relation
\begin{align} \label{isobaric}
K(\delta)\simeq K+K_{\tau}\, \delta^2, \;\;\;\;\;\;\;\;\;\; K_{\tau}=K_{\text{sym}}-6L,
\end{align}
where $K_{\tau}$ is usually called the isobaric incompressiblity.
In recent years, much effort has been invested in determining the parameters in Eqs. (\ref{JLK}). The empirical values of $J=29.0-32.7 \,\text{MeV}$ and to a lesser degree also $L=40.5-61.9  \,\text{MeV}$ are relatively well constrained (values from \cite{Lattimer:2012xj}, see also
\cite{Newton:2012ry,Newton:2014tga,Lattimer:2014sga,
Hebeler:2014ema,Tews:2012fj}),
whereas experimental determinations of $K_{\tau}$ suffer from large uncertainties. For instance, from measurements of neutron skin thicknesses \cite{PhysRevLett.102.122502} the value $K_{\tau}=-500_{-100}^{+125} \,\text{MeV}$ was obtained, which is compatible with the giant monopole resonance measured in Sn isotopes \cite{PhysRevLett.99.162503} giving $K_{\tau}=-550\pm 100 \,\text{MeV}$. 
Theoretical studies using a selection of Skyrme interactions however led to an estimate of
$K_{\tau}=-370\pm 120 \,\text{MeV}$ \cite{PhysRevC.80.014322}.\footnote{However, in each case a slightly different definition of $K_{\tau}$ is used, with the differences corresponding to higher-order terms in Eq. (\ref{isobaric}) and finite-size effects \cite{PhysRevC.80.014322}.} 
Our results for $J$, $L$ and $K_{\tau}$ are given in Table \ref{table:observables}. They are in agreement with the mentioned constraints.

\vspace*{0.7cm}
%
%--------------------------TABLE
\begin{table}[H]
\begin{center}
\begin{minipage}{14cm}
\setlength{\extrarowheight}{1.5pt}
\hspace*{15mm}
\begin{tabular}{c ccc}
\hline \hline 
        \;\;\;\;\;\;\; \;\;\;\;\;\;\;\;\;\;
&$J\,(\text{MeV})$ \;\;\;\; \;\;\;\;\;\;\;\;\;\;
& $L\,(\text{MeV})$ \;\;\;\; \;\;\;\;\;\;\;\;\;\;
& $K_{\tau}\,(\text{MeV})$   
\\   \hline  
n3lo414 \;\;\;\;\;\;\; \;\;\;\;\;\;\;\;\;\;
& $32.51$\;\;\;\; \;\;\;\;\;\;\;\;\;\;& $53.8$ \;\;\;\; \;\;\;\;\;\;\;\;\;\;& $-424$   
\\  
n3lo450 \;\;\;\;\;\;\; \;\;\;\;\;\;\;\;\;\;
& $31.20$\;\;\;\; \;\;\;\;\;\;\;\;\;\;& $48.2$ \;\;\;\; \;\;\;\;\;\;\;\;\;\;& $-434$   
\\  \hline \hline
\end{tabular}
\end{minipage}
{\vspace{0mm}}
\begin{minipage}{14cm}
\caption
{Value of the (zero-temperature) symmetry energy at saturation density $J$, the slope parameter $L$, and the isobaric incompressibility $K_{\tau}$, extracted from the results obtained from the sets of chiral nuclear two- and three-body interactions n3lo414 and n3lo450. }
\label{table:observables}
\end{minipage}
\end{center}
\end{table}
%--------------------------TABLE
%
%
%
%%%%%%%%%%%%%%%%%%%%%%%%%%%%%%%%%%%%%%%%%%%%%%%%%%%%%%%%%%%%%%%%%%%%%%%%%%%%%%%%%%
%%%%%%	%%%%%%%	%%%%%%	%%%%%%	%%%%%%%	%%%%%%
%%%%%%%%%%%%%%%%%%%%%%%%%%%%%%%%%%%%%%%%%%%%%%%%%%%%%%%%%%%%%%%%%%%%%%%%%%%%%%%%%%
%%%%%%	%%%%%%%	%%%%%%	%%%%%%	%%%%%%%	%%%%%%
%%%%%%%%%%%%%%%%%%%%%%%%%%%%%%%%%%%%%%%%%%%%%%%%%%%%%%%%%%%%%%%%%%%%%%%%%%%%%%%%%%
%
\newpage

\section{Thermodynamics of isospin-asymmetric nuclear matter}\label{sec4}
%
%%%%%%%%%%%%%%%%%%%%%%%%%%%%%%%%%%%%%%%%%%%%%%%%%%%%%%%%%%%%%%%%%%%%%%%%%%%%%%%%%%
In this section we examine the thermodynamic equation of state of isospin-asymmetric nuclear matter (ANM). 
The free energy per particle in homogeneous ANM is calculated as follows.
The dependence of the nonrelativistic free Fermi gas contributions on the isospin asymmetry $\delta$ is treated exactly, while the relativistic correction term and the interaction contributions are assumed to have a quadratic dependence on the isospin asymmetry:
\begin{align} \label{ANM-EoS}
\bar{F}(T,\rho,\delta)&\simeq
\bar{F}_{\text{0}}(T,\rho,\delta)
+\bar{F}_{\text{rel}}(T,\rho,0)
+\bar{F}_{\text{sym,rel}}(T,\rho)\, \delta^2
+\bar{F}_{\text{int}}(T,\rho,0)
+\bar{F}_{\text{sym,int}}(T,\rho)\, \delta^2.
\end{align}
This approach is motivated in Sec.\ \ref{sec40}, where we investigate in detail the dependence on isospin asymmetry of the noninteracting contributions $\bar{F}_{\text{0}}(T,\rho,\delta)$ and $\bar{F}_{\text{rel}}(T,\rho,\delta)$. 

In Sec.\ \ref{sec41} we then discuss the construction of the spinodal in ANM and present our results for the trajectory of the critical temperature. The dependence of the neutron and proton chemical potentials on isospin asymmetry is examined in Sec.\ \ref{sec42}, and we determine the neutron drip point in cold nuclear matter. 
Finally, in Sec.\ \ref{sec43} we show results for the free energy per particle and pressure in isospin-asymmetric nuclear matter and determine the values of $T$ and $\delta$ where an isolated drop of liquid nuclear matter becomes unstable.

\vspace*{0.6cm}
\subsection{Isospin dependence of free nucleon gas} \label{sec40}
Here we examine the $\delta$ dependence of the noninteracting contributions to the free energy per particle in homogeneous ANM.\footnote{As noted in Sec. \ref{sec0}, in initial calculations we have found that at the densities and temperatures relevant for the liquid-gas phase transition the interaction contributions give rise to comparatively weaker terms with quartic and higher powers in $\delta$.
The contributions from two-nucleon interactions $\bar F_{1,\text{NN}}$ and $\bar F_{2,\text{NN}}$ (which give the dominant contribution to $\bar F_{\text{int}}$ at subnuclear densities) were found to be quadratic in $\delta$ to high accuracy. In the case of $\bar F_{1,\text{3N}}$ we have found that higher-order terms in $\delta$ are more sizeable, but still small compared to those that emerge from $\bar{F}_{\text{0}}$ (at finite $T$). Future research will quantify the isospin-asymmetry dependence of the interaction contributions in more detail.} We compute the four leading terms in an expansion of $\bar{F}_{\text{0}}(T,\rho,\delta)$ and $\bar{F}_{\text{rel}}(T,\rho,\delta)$ in powers of $\delta^2$. The results show that the accuracy of the quadratic approximation for $\bar{F}_{\text{0}}(T,\rho,\delta)$ decreases significantly with increasing temperature, which 
necessitates the exact calculation of this contribution. The relativistic correction term on the other hand can be safely approximated via
$\bar{F}_{\text{rel}}(T,\rho,\delta)\simeq \bar{F}_{\text{rel}}(T,\rho,0)
+\bar{F}_{\text{sym,rel}}(T,\rho)\, \delta^2$.

The noninteracting contributions to the free energy density, $F_{0}=F^{\text{n}}_{0}+F^{\text{p}}_{0}$ and $F_{\text{rel}}=F^{\text{n}}_{\text{rel}}+F^{\text{p}}_{\text{rel}}$, can be expressed in terms of polylogarithms $\text{Li}_\nu(x)=\sum^{\infty}_{k=1} k^{-\nu} x^k$, i.e.,
\begin{align} \label{F0_Li}
F^{\text{n/p}}_{0}(T,\mu_0^{\text{n/p}})
&=- \alpha T^{5/2} \Big( \ln(-x_{\text{n/p}}) \;
\text{Li}_{3/2}(x_{\text{n/p}})
-  \text{Li}_{5/2}(x_{\text{n/p}}) \Big), \\
F^{\text{n/p}}_{\text{rel}}(T,\mu_0^{\text{n/p}})
&= \frac{15\alpha T^{7/2}}{8M}  \text{Li}_{7/2}(x_{\text{n/p}})
\end{align} 
where $\mu_0^{\text{n/p}}$ are the
neutron and proton effective one-body chemical potentials, $x_{\text{n/p}}=-\exp(\mu_0^{\text{n/p}}/T)$ and $\alpha=2^{-1/2}(M/\pi)^{3/2} $. For given values of $T$, $\rho$ and $\delta$ the effective one-body chemical potentials $\mu_0^{\text{n/p}}$ are uniquely determined by
$\rho_{\text{n/p}}=-\alpha T^{3/2} \text{Li}_{3/2}(x_{\text{n/p}})$.

The noninteracting contribution to the free energy per particle, $\bar F_{\text{nonint}}=(F_{0}+F_{\text{rel}})/(\rho_{\text{n}}+\rho_{\text{p}})$, as a function of temperature $T$, nucleon density $\rho$ and isospin asymmetry $\delta$ can be expanded\footnote{Odd-order terms in $\delta$ in the expansion of $\bar F_{\text{nonint}}$ arise only from the neutron-proton mass difference $\Delta M\simeq 1.4 \cdot 10^{-3} M$, which we neglect in this work.} in powers of $\delta^2$ around its value in SNM ($\delta=0$):
\begin{align} \label{Fsym1}
\bar{F}_{\text{nonint}}(T,\rho,\delta)=
\bar{F}_{\text{nonint}}(T,\rho,0)+ \sum_{n=1}^{\infty} \bar{B}_{\text{nonint},2n}(T,\rho)\, \delta^{2n}.
\end{align}
The various expansion coefficients  $\bar{B}_{\text{nonint},2n}$ are given by
\begin{align} \label{Fsym4}
\bar{B}_{\text{nonint},2n}(T,\rho)=\frac{1}{(2n)!}\frac{\partial^{2n} \bar{F}_{\text{nonint}}(T,\rho,\delta)}{\partial \delta^{2n}}\bigg|_{\delta=0}.
\end{align}
Setting $\delta=1$ in Eq. (\ref{Fsym1}) we obtain the noninteracting symmetry free energy as the sum
of 
the above coefficients, i.e., 
\begin{align} \label{Fsym5}
\bar{F}_{\text{nonint},\text{sym}}(T,\rho)=\sum_{n=1}^{\infty} \bar{B}_{\text{nonint},2n}(T,\rho)
=
\bar{F}_{\text{\text{nonint},sym}}(T,\rho) \sum_{n=1}^{\infty} \upbeta_{\text{nonint},2n}(T,\rho).
\end{align}
Here, we have introduced the weight factors $\upbeta_{\text{nonint},2n}=\bar{B}_{\text{nonint},2n}/\bar{F}_{\text{nonint},\text{sym}}$ as a means to specify the relative size of the different expansion coefficients.

The rules of multivariable calculus lead to the following expression for the $\delta$ derivative of order $n$ of $F^{\text{n/p}}_{i}$, $i\in\{0,\text{rel}\}$, at fixed density and temperature:
\begin{align} \label{deltaderiv}
\left(\frac{\partial^n F^{\text{n/p}}_{i}}{\partial \delta^n}\right)_{\!\!T,\rho}=-
\frac{(\pm 1)^{n}\,\alpha T^{5/2} (n-1)!}{(1\pm\delta)^{n}} \; \mathscr{Y}_{i}^{(n)}(x_{\text{n/p}}).
\end{align}
where the functions $\mathscr{Y}_{i}^{(n)}$ are defined recursively as 
\begin{align} \label{Xirecurs}
\mathscr{Y}_{i}^{(n)}(x_{\text{n/p}})=\frac{x_{\text{n/p}}}{\text{max}(n-1,1)}\,\frac{\text{Li}_{3/2}(x_{\text{n/p}})}{\text{Li}_{1/2}(x_{\text{n/p}})}\,
\frac{\partial}{\partial x_{\text{n/p}}}\mathscr{Y}_{i}^{(n-1)}(x_{\text{n/p}})
-(1-\updelta_{n,1})\;\mathscr{Y}_{i}^{(n-1)}(x_{\text{n/p}}),
\;\;\;\;\;\;\; n\geq 1,
\end{align}
with $\updelta_{k,l}$ the Kronecker delta. The expressions to start the recursion are
\begin{align}
\mathscr{Y}_{0}^{(0)}(x_{\text{n/p}})= \ln(-x_{\text{n/p}}) \;
\text{Li}_{3/2}(x_{\text{n/p}})
-  \text{Li}_{5/2}(x_{\text{n/p}}),
\;\;\;\;\;\;\;
\mathscr{Y}_{\text{rel}}^{(0)}(x_{\text{n/p}})=-\frac{15T}{8M}  \text{Li}_{7/2}(x_{\text{n/p}}).
\end{align}
One then obtains for $\bar B_{\text{nonint},2n}$ the expression
\begin{align}
\bar B_{\text{nonint},2n}(T,x)=\frac{T}{2n\, \text{Li}_{3/2}(x)} \;  \left(\mathscr{Y}_{0}^{(2n)}(x)+\mathscr{Y}_{\text{rel}}^{(2n)}(x)\right),
\end{align}
where $x=-\exp(\mu_0/T)$, with $\mu_0$ the nucleon effective one-body chemical potential, which is uniquely determined by $\rho=-2\alpha T^{3/2} \text{Li}_{3/2}(x)$.

\begin{figure}[H] 
\centering
\vspace*{-3.2cm}
\hspace*{-2.2cm}
\includegraphics[width=1.15\textwidth]{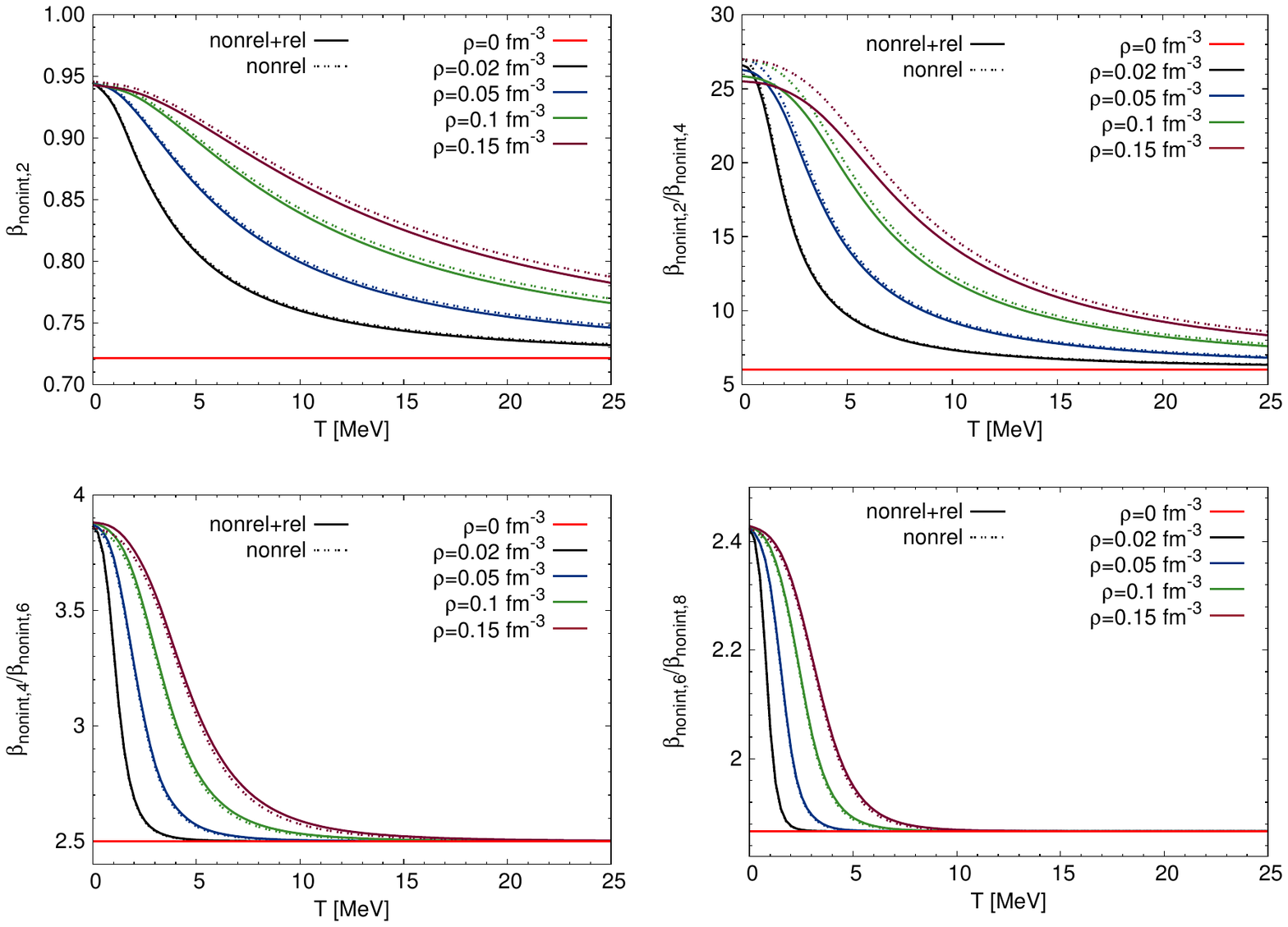} 
\vspace*{-11.6cm}
\caption{(Color online) Temperature dependence of the first weight factor $\upbeta_{\text{nonint},2}$ and the ratios
$\upbeta_{\text{nonint},2n}/\upbeta_{\text{nonint},2(n+1)}$ for $n=1,2,3$ at different densities, corresponding to the expansion of $\bar F_{\text{nonint}}(T,\rho,\delta)$ in powers of $\delta^2$. 
The full lines show the results with the relativistic correction term included, the dotted lines the nonrelativistic results. Note that the deviations between the relativistically improved and the nonrelativistic results decrease with increasing values of $n$, indicating the opposite convergence behavior of the expansion in powers of $\delta^2$ of $\bar F_0$ and $\bar F_{\text{rel}}$, respectively.}  
\label{plots_sec4_0}
\end{figure}

The results for the first weight factor $\upbeta_{\text{nonint},2}=\bar B_{\text{nonint},2}/\bar F_{\text{nonint,sym}}$ and the ratios
$\upbeta_{\text{nonint},2n}/\upbeta_{\text{nonint},2(n+1)}$ for $n=1,2,3$ are displayed in Fig.\ \ref{plots_sec4_0}. Note that the limits $\rho \rightarrow 0$ and $T\rightarrow 0$ do not commute.
The $\rho\rightarrow 0$ limit of the symmetry coefficients at finite temperature is given by
\begin{align} \label{Symcoeffrho0}
\bar B_{\text{nonint},2n}(T\neq 0,\rho \rightarrow 0)&=\frac{T}{2(2n)!}
\frac{\partial^{2n}}{\partial \delta^{2n}}
\Big( (1+\delta)\ln(1+\delta)+(1-\delta)\ln(1-\delta) \Big) \Big|_{\delta=0}=\frac{T}{2n(2n-1)},
\end{align}
which comes entirely from the logarithmic terms, $\sim \ln(-x_{\text{n/p}})=\mu_0^{\text{n/p}}/T$, in the expression for $\bar F_0=(F_0^{\text{n}}+F_0^{\text{p}})/(\rho_{\text{n}}+\rho_{\text{p}})$.
The asymptotic behavior of the weight factors $\upbeta_{\text{nonint},2n}$ for $T\rightarrow \infty$ is determined by the fact that the $\rho\rightarrow 0$ and $T\rightarrow \infty$ limits of $\upbeta_{\text{nonint},2n}=\bar B_{\text{nonint},2n}/\bar F_{\text{nonint,sym}}$ coincide.\footnote{This follows from the fact that $x_\text{n/p} \rightarrow 0$ in both the $\rho \rightarrow 0$ and the $T \rightarrow \infty$ limit.}
As apparent from Fig.\ \ref{plots_sec4_0}, the asymptotic behavior sets in at relatively low values of $T$, causing the convergence rate of the quadratic expansion of $\bar F_{\text{nonint}}$ to decrease significantly with increasing temperature. This behavior is entirely caused by the presence of the logarithmic terms in the nonrelativistic contribution, i.e., by the first term proportional to the sum of the effective one-body chemical potentials, $\sim \mu_0^{\text{n}}+\mu_0^{\text{p}}$. The $\rho\rightarrow 0$ limit of the second term in $\bar F_0$ proportional to $[\text{Li}_{5/2}(x_{\text{n}})+\text{Li}_{5/2}(x_{\text{p}})]/[\text{Li}_{3/2}(x_{\text{n}})+\text{Li}_{3/2}(x_{\text{p}})]$ equals $-T$, which is independent of $\delta$; the convergence rate of the expansion in powers of $\delta^2$ of this term alone increases very strongly with increasing temperature. 
Similarly, the $\rho\rightarrow 0$ limit of $\bar F_{\text{rel}}$ equals $-15 T^{2}/(8M)$, and the quadratic approximation of the relativistic correction term becomes increasingly accurate with increasing temperature. 

\begin{figure}[H] 
\centering
\vspace*{-3.2cm}
\hspace*{-2.2cm}
\includegraphics[width=1.15\textwidth]{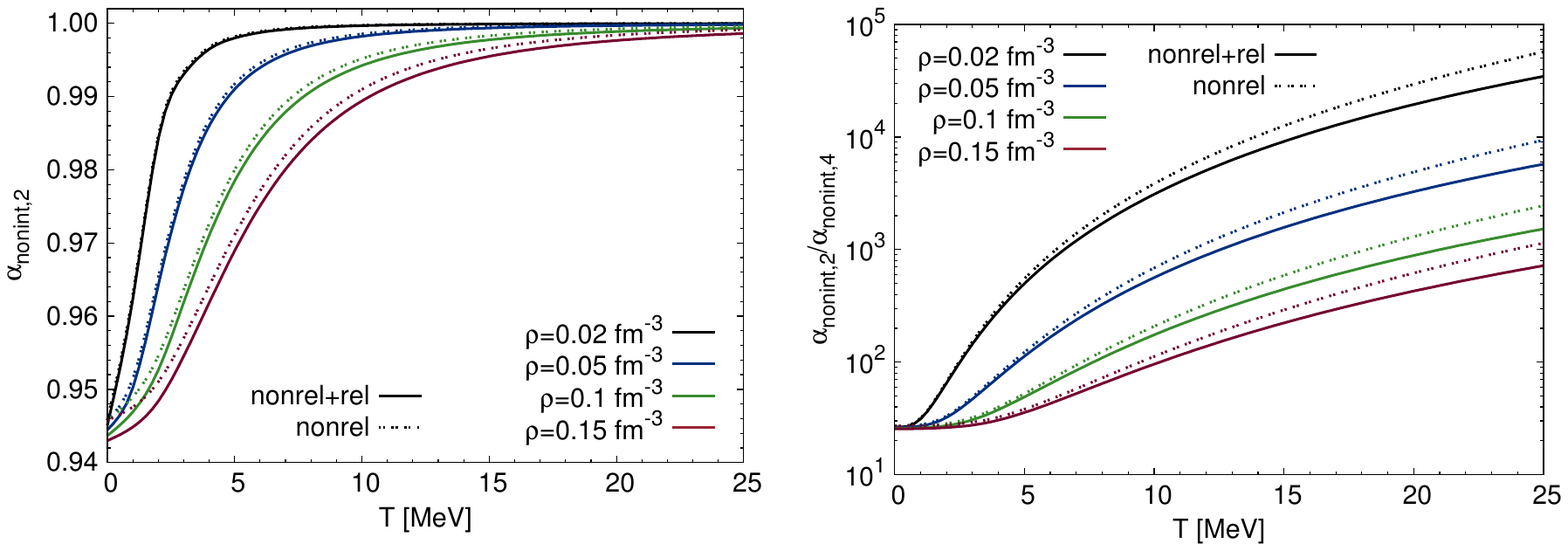} 
\vspace*{-17.6cm}
\caption{(Color online) Temperature dependence of the first weight factor $\upalpha_{\text{nonint},2}$ and the ratio
$\upalpha_{\text{nonint},2}/\upalpha_{\text{nonint},4}$, corresponding to the expansion in powers of $\delta^2$ of the noninteracting contributions to the internal energy per particle, $\bar E_0$ and $\bar E_{\text{rel}}$. Note the logarithmic scale in the second plot.}  
\label{plots_sec4_0b}
\end{figure}

For comparison, in Fig.\ \ref{plots_sec4_0b} we show the results for the weight factors $\upalpha_{\text{nonint},2n}$ in the analogous expansion in powers of $\delta^2$ of the noninteracting contributions to the internal energy per particle, $\bar E_0$ and $\bar E_{\text{rel}}$. The
noninteracting contributions to the neutron and proton internal energy densities are given by
\begin{align} \label{Enpnonint1}
E^{\text{n/p}}_{0}(T,\mu_0^{\text{n/p}})
&=-\frac{3 \alpha T^{5/2}}{2} \text{Li}_{5/2}(x_{\text{n/p}}), \\ \label{Enpnonint2}
E^{\text{n/p}}_{\text{rel}}(T,\mu_0^{\text{n/p}})
&=- \frac{75 \alpha T^{7/2}}{16 M} \text{Li}_{7/2}(x_{\text{n/p}})
+ \frac{45 \alpha T^{7/2}}{16 M } \,
\frac{\text{Li}_{5/2}(x_{\text{n/p}})\;\;\text{Li}_{3/2}(x_{\text{n/p}})}
{\text{Li}_{1/2}(x_{\text{n/p}})}.
\end{align} 
Both $\bar E_0\rightarrow 3T/2$ and $\bar E_{\text{rel}}\rightarrow 15T^2/(8M)$ become independent of $\delta$ as $\rho\rightarrow 0$, and the convergence rate of the expansion of both terms increases strongly with increasing temperature (the increase is significantly more pronounced in the case of $\bar E_0$, which is reflected in the increase of the deviations between relativistically improved and the nonrelativistic results with increasing values of $n$).

\vspace*{1.0cm}
%------------------------------------------------FIGURE sec4-2
%--------------------------------------------END_FIGURE sec4-2

\vspace*{-1.5cm}

%------------------------------------------------FIGURE sec4-2
%--------------------------------------------END_FIGURE sec4-2

\vspace*{-0.5cm}

\subsection{Spinodal and critical temperature}\label{sec41}
The onset of spinodal instability is associated with the violation of a number of equivalent stability criteria \cite{beegle,beegle2}, which are derived from the fundamental principle of maximum entropy.
In the canonical representation the corresponding stability requirement is that
the free energy density $F=\rho \bar F$ at fixed temperature $T$ is a convex function of the component densities $\rho_1$ and $\rho_2$, implying that the Hessian matrix $\mathcal{F}_{ij}$ has no negative eigenvalues. In the unstable region delineated by the spinodal the (analytical) free energy density is concave; in the metastable region between the spinodal and binodal (coexistence boundary) the free energy density is locally convex and the system is protected against phase separation by a nucleation barrier.
The Hessian matrix $\mathcal{F}_{ij}$ is given by 
\begin{align}
\mathcal{F}_{ij}(T,\rho_1,\rho_2)=
\left[
\frac{\partial^2  F(T,\rho_1,\rho_2)}{ \partial \rho_i \partial \rho_j}
\right]
=
\left[
\frac{\partial \mu_i(T,\rho_1,\rho_2)}{ \partial \rho_j}  
\right], \;\;\;\; i,j \in \{1,2\}.
\end{align}
Its eigenvalues are given by
\begin{align}  \label{spinodalEV}
\xi_{\pm}(T,\rho_1,\rho_2)=&
\frac{1}{2} \left[\tr[\mathcal{F}_{ij}]\pm\left({\tr[\mathcal{F}_{ij}]^2-4\det[\mathcal{F}_{ij}]} \right)^{1/2} \right]  \nonumber \\
=&\frac{1}{2} \left[\mathcal{F}_{11}+\mathcal{F}_{22}\pm \big( ({\mathcal{F}_{11}-\mathcal{F}_{22})^{2}+4 \mathcal{F}_{12}^2} \big)^{1/2} \right].
\end{align}
The signs of the eigenvalues are invariant under (linear) basis transformations. From the data $F(T,\rho,\delta)$ they are readily evaluated using as independent density parameters $\rho_\text{1}=\rho_\text{n}+\rho_\text{p}=\rho$ (nucleon density) and $\rho_\text{2}=\rho_\text{n}-\rho_\text{p}=\rho\, \delta$ (isospin asymmetry density).  
In this basis the Hessian matrix becomes diagonal at $\delta=0$ with eigenvalues $\xi_{+}=(\partial^2 F / \partial \rho_2^2)_{T,\rho_1}|_{\rho_2=0}>0$ and $\xi_{-}=(\partial^2 F / \partial \rho_1^2)_{T,\rho_2}|_{\rho_2=0}=\rho^{-1}(\partial P/ \partial \rho)_{T,\delta}|_{\delta=0}$ (this result depends on the neglect of isospin-symmetry breaking effects). Hence, in SNM the region inside the spinodal corresponds to a negative isothermal compressibility $\upkappa_T=\rho^ {-1}(\partial \rho/ \partial P)_{T,\delta}$,
where $\upkappa_T^{-1}>0$ is a stability criterion for a pure substance. 

The exact expressions for the nonrelativistic free Fermi gas contribution to the Hessian matrix components are given by\footnote{Note that $\mathcal{F}_{0,11}=\mathcal{F}_{0,22}$, but  $\mathcal{F}_{\text{int},11}\neq\mathcal{F}_{\text{int},22}$.} 
\begin{align} \label{FermiHess1}
\mathcal{F}_{0,11}(T,\mu_0^{\text{n}},\mu_0^{\text{p}})&
=- \frac{T^{-1/2}}{4 \alpha } 
\left( \frac{1}{\text{Li}_{1/2}(x_{\text{n}})}+ \frac{1}{\text{Li}_{1/2}(x_{\text{p}})}\right)=\mathcal{F}_{0,22}(T,\mu_0^{\text{n}},\mu_0^{\text{p}}),
\\ \label{FermiHess2}
\mathcal{F}_{0,12}(T,\mu_0^{\text{n}},\mu_0^{\text{p}})&
=- \frac{T^{-1/2}}{4 \alpha } 
\left( \frac{1}{\text{Li}_{1/2}(x_{\text{n}})}- \frac{1}{\text{Li}_{1/2}(x_{\text{p}})}\right),
\end{align}
where again $x_{\text{n/p}}=-\exp(\mu_0^{\text{n/p}}/T)$ and $\alpha=2^{-1/2}(M/\pi)^{3/2} $.
In the limit $\delta\rightarrow 1$ the proton density vanishes, $\rho_{\text{p}}\rightarrow 0$, and the proton effective one-body chemical potential diverges (at finite $T$), $\mu_0^{\text{p}}\rightarrow -\infty$, thus $\text{Li}_{1/2}(x_{\text{p}})\rightarrow0$. Hence, the exact calculation of the nonrelativistic free Fermi gas contribution $F_0(T,\rho,\delta)$ leads to divergent behavior of the Hessian components in the limit of vanishing proton concentration. 
The same divergent behavior is obtained in the exact calculation of $\mathcal{F}_{0,ij}$ at zero temperature.
The unstable region then vanishes at a value $\delta<1$ for all values of $T$.
This constraint is lost if the free Fermi gas contribution is approximated by truncating the expansion in powers of $\delta^2$ of $F_0(T,\rho,\delta)$ at a finite order, e.g., at first order as in the usual quadratic isospin approximation. 

\begin{figure}[H] 
\centering
\vspace*{-2.80cm}
\hspace*{0.7cm}
\includegraphics[width=1.3\textwidth]{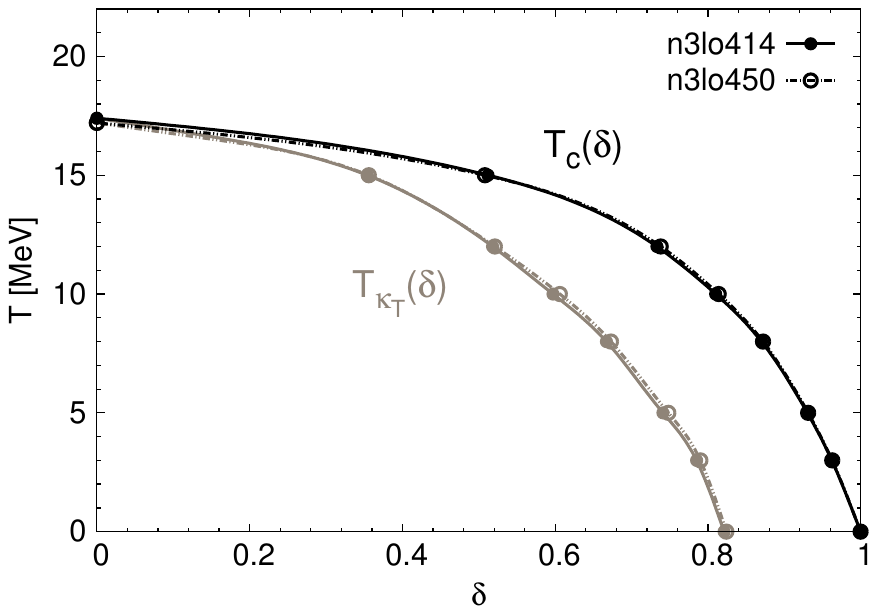} 
\vspace*{-20.5cm}
\caption{Trajectories of the critical temperature $T_c(\delta)$ determined from n3lo450 and n3lo414. The trajectories end at $\delta\simeq 0.9994$. Also shown are the trajectories of the temperatures $T_{\upkappa_T}(\delta)$ where the region with negative isothermal compressibility $\upkappa_T$ vanishes. The calculated data points are shown explicitly.}
\label{plots_CP}
\end{figure}

The evolution of the critical temperature $T_c(\delta)$ where (for a given value of $\delta$) the unstable concave region vanishes is depicted in Fig.\ \ref{plots_CP}. 
The results from n3lo414 and n3lo450 are very similar; in both cases the critical lines end approximately at an isospin asymmetry $\delta_{c}^{\text{end}}\simeq 0.9994$ or a proton concentration $Y_{\text{p},c}^\text{end}\simeq 3\cdot 10^{-4}$. 
The value $\delta_{c}^{\text{end}}\simeq 0.9994$ exceeds the critical line endpoints obtained in Refs.\ \cite{Ducoin:2005aa,Fedoseew:2014cma,PhysRevC.83.044605} using different phenomenological models.
We note that in nuclear matter with $\delta \neq 0$ the coexistence region does not vanish at the critical temperature $T_c(\delta)$ but at a higher temperature $T_{\text{max}}(\delta)$, the so-called maximum temperature \cite{Hempel:2013tfa,Ducoin:2005aa}. 
The existence of a neutron drip point (see Sec. \ref{sec42}) entails that at zero temperature the binodal extends to $\delta=1$ over a finite region of densities or pressures. The trajectory of the maximum temperatures therefore reaches its zero-temperature endpoint at vanishing proton fraction $Y^{\text{end}}_{\text{p,max}}=0$.

For comparison, in Fig.\ \ref{plots_CP} we also show the trajectories of the temperature $T_{\upkappa_T}(\delta)$ where the region with negative isothermal compressibility $\upkappa_T$ vanishes at fixed $\delta$. 
The exact expression for the nonrelativistic free Fermi gas contribution to $\upkappa_T^{-1}$ is given by
\begin{align} \label{Fermikappa}
\upkappa_{T,0}^{-1}&
=- \frac{\alpha T^{5/2}}{4} \big(\text{Li}_{3/2}(x_{\text{n}})+\text{Li}_{3/2}(x_{\text{p}})\big)^{2} 
\left( \frac{(1+\delta)^{2}}{\text{Li}_{1/2}(x_{\text{n}})}+ \frac{(1-\delta)^{2}}{\text{Li}_{1/2}(x_{\text{p}})}\right).
\end{align}
For both n3lo414 and n3lo450 the $T_{\upkappa_T}(\delta)$ trajectories end at approximately $\delta_{\upkappa_T}^{\text{end}}\simeq 0.82$ or $Y_{\text{p},{\upkappa_T}}^{\text{end}}\simeq 0.09$, which exceeds the values $Y_{\text{p},{\upkappa_T}}^{\text{end}}\simeq 0.053$ from Ref.\ \cite{Fiorilla:2011sr} obtained in an in-medium chiral perturbation approach and $Y_{\text{p},{\upkappa_T}}^{\text{end}}\simeq 0.045$ from Ref.\ \cite{PhysRevC.91.035802} obtained by applying the functional renormalization group to a chiral nucleon-meson model.

\vspace*{-0.4cm}
%------------------------------------------------FIGURE sec4-2
%--------------------------------------------END_FIGURE sec4-2

\subsection{Stable self-bound liquid} \label{sec42}
\begin{figure}[H] 
\centering
\vspace*{-3.2cm}
\hspace*{-2.2cm}
\includegraphics[width=1.15\textwidth]{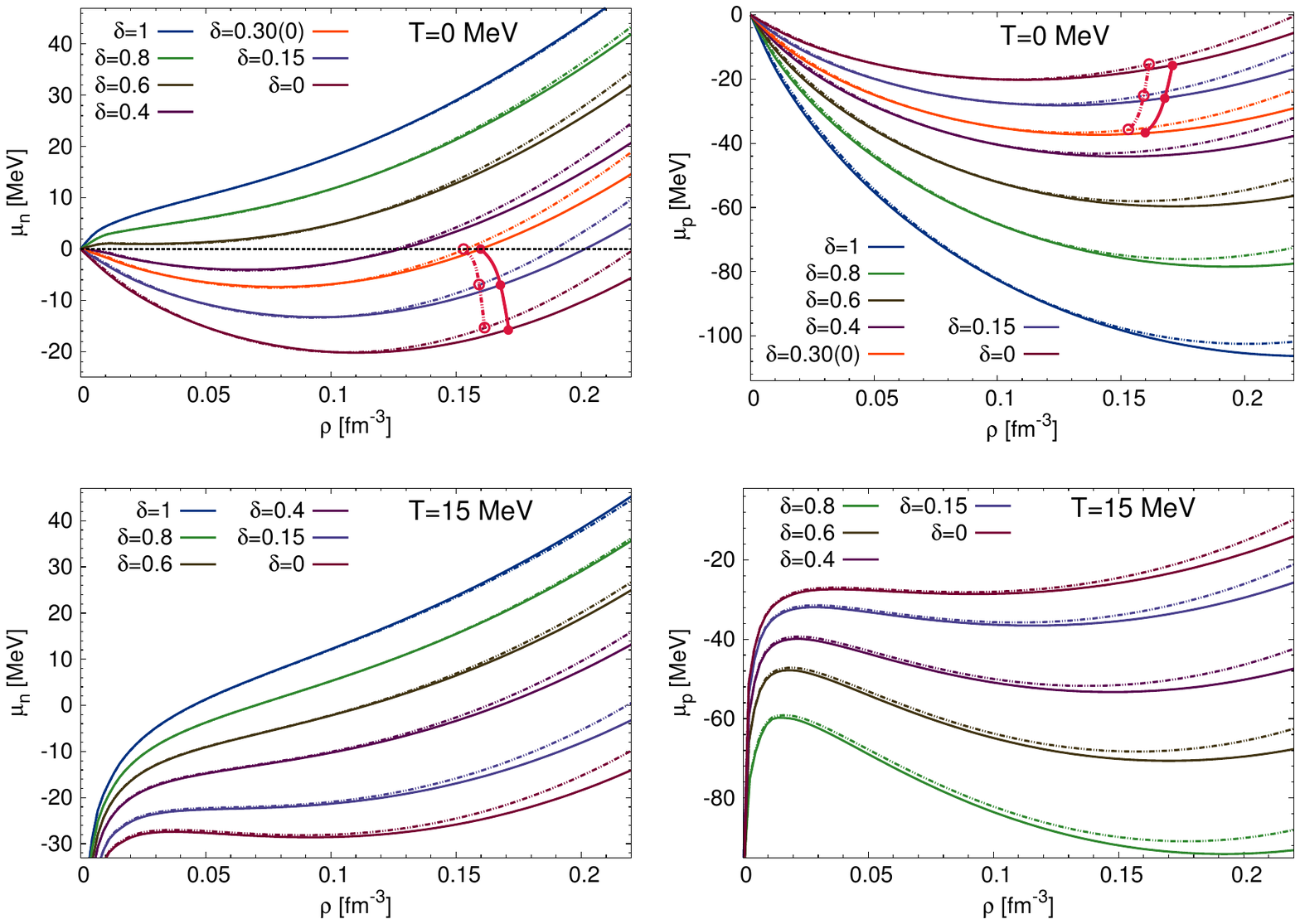} 
\vspace*{-11.6cm}
\caption{(Color online) Neutron and proton chemical potentials, $\mu_{\text{n}}(T,\rho,\delta)$ and $\mu_{\text{p}}(T,\rho,\delta)$, in (homogeneous) isospin-asymmetric nuclear matter at temperatures $T/\text{MeV}=0,15$, calculated using n3lo414 (solid lines) and n3lo450 (dash-dot lines). Stable self-bound states are shown as thick bright red lines with circles (full circles for n3lo414, open circles for n3lo450); the lines end at the neutron drip point $\delta_{\text{ND}}\simeq 0.30$.}
\label{plots_sec4_1}
\end{figure}

From the data $F(T,\rho,\delta)$ the neutron and proton chemical potentials are obtained via
\begin{align} \label{mu_n-mu_p}
\mu_{\text{n/p}}(T,\rho,\delta)&
=
\frac{\partial F(T,\rho,\delta)}{\partial \rho}
\pm
\frac{1 \mp \delta}{\rho}
\frac{\partial F(T,\rho,\delta)}{\partial \delta}.
\end{align}
The results for $\mu_{\text{n}}(T,\rho,\delta)$ and $\mu_{\text{p}}(T,\rho,\delta)$ are displayed in Fig.\ \ref{plots_sec4_1} for temperatures $T/\text{MeV}=0,15$. One sees that $\mu_{\text{n}}(T,\rho,\delta)$ increases and $\mu_{\text{p}}(T,\rho,\delta)$ decreases with $\delta$. The chemical potentials at finite $T$ diverge as $\rho\rightarrow 0$, but at zero temperature $\mu_{\text{n/p}}\rightarrow 0$ for $\rho\rightarrow 0$.
The origin of this feature is the collapse of Fermi-Dirac distribution functions into Heaviside step functions at $T=0$, which eliminates the logarithmic divergence of the free energy per particle at vanishing density \cite{paper1,Ducoin:2005aa}.

The Gibbs conditions for the coexistence of two bulk phases \text{\Rmnum{1}} (liquid) and \text{\Rmnum{2}} (gas) in mutual thermodynamic equilibrium are 
\begin{align} 
T^{\text{\Rmnum{1}}}=T^{\text{\Rmnum{2}}}, \;\;\;\; P^{\text{\Rmnum{1}}}=P^{\text{\Rmnum{2}}}, \;\;\;\;
\mu_{\text{n}}^{\text{\Rmnum{1}}} =\mu_{\text{n}}^{\text{\Rmnum{2}}},  \;\;\;\;
\mu_{\text{p}}^{\text{\Rmnum{1}}} =\mu_{\text{p}}^{\text{\Rmnum{2}}}.
\end{align}
Whereas at finite $T$ liquid-gas equilibrium corresponds to finite values of density $\rho$ and proton fraction $Y$
in both phases, the vanishing 
of $\mu_{\text{n/p}}(T=0,\rho,\delta)$ at vanishing density entails that at $T=0$ the Gibbs conditions for 
the neutron and proton chemical potentials can in most cases not be satisfied, leading to a gas phase 
that is either empty (vacuum) or contains only neutrons \cite{Ducoin:2005aa,Lattimer:1978}. 
The neutron drip point, $\delta_{\text{ND}}$, is given by the value of $\delta$ where the neutron chemical potential at vanishing temperature and pressure becomes positive.
For isospin asymmetries $\delta\leq \delta_{\text{ND}}$ an isolated drop of cold liquid nuclear matter is stable (in equilibrium with the vacuum), defining a stable self-bound state.
As seen from Fig.\ \ref{plots_sec4_1}, in our equation of state neutron drip occurs at an isospin asymmetry $\delta_{\text{ND}}\simeq 0.30$ or a proton concentration $Y_{\text{p,ND}}=(1-\delta_{\text{ND}})/2\simeq 0.35$, which is similar to results obtained with effective Skyrme interactions \cite{Lattimer:1978,Lamb:1981kt}.

\vspace*{0.1cm}
%------------------------------------------------FIGURE sec4-3
%--------------------------------------------END_FIGURE sec4-3

\newpage

\subsection{Metastable self-bound liquid}\label{sec43}

\begin{figure}[H] 
\centering
\vspace*{-3.2cm}
\hspace*{-2.2cm}
\includegraphics[width=1.15\textwidth]{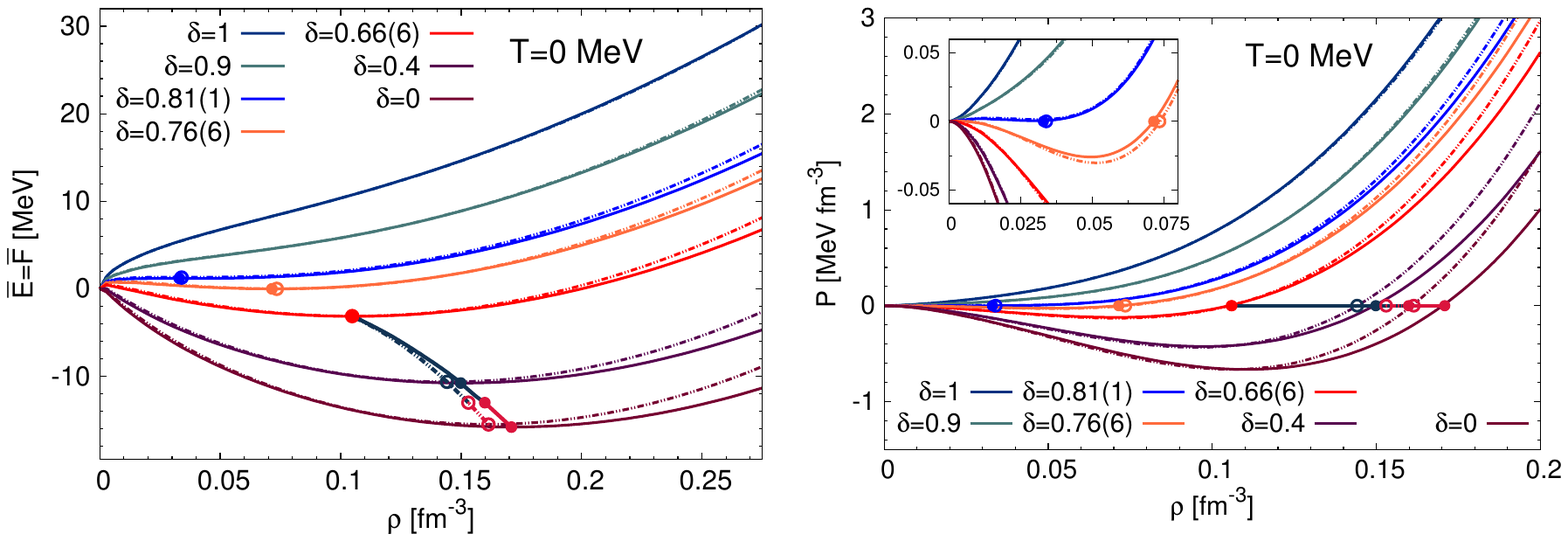} 
\vspace*{-17.5cm}
\caption{(Color online) Energy per particle $\bar E= \bar F$ and pressure $P$ in homogeneous isospin-asymmetric nuclear matter at zero temperature, calculated using n3lo414 (solid lines) and n3lo450 (dash-dot lines). The trajectories of the local energy minima are shown as thick dark gray (bright red below neutron drip) lines with circles (full circles for n3lo414, open circles for n3lo450). The trajectories end at the fragmentation point $\delta_{\text{FP}}\simeq 0.66$. 
The inset magnifies the behavior of the pressure at low densities.}
\label{plots_sec4_3}
\end{figure}

The zero-temperature results for the (ground state) energy per particle $\bar E=\bar F$ and the pressure $P=\rho^2 \partial \bar F/\partial \rho$ are displayed in Fig.\ \ref{plots_sec4_3} as functions of the nucleon density $\rho$ for different values of $\delta$. The trajectory of the points where the energy per particle has a local minimum and the pressure is zero is shown explicitly.
These points correspond to the properties of a drop of cold liquid nuclear matter surrounded by vacuum. For $\delta\leq \delta_{\text{ND}}\simeq 0.30$ the cold drop is stable (the local energy minimum lies on the binodal, cf.\ Sec.\ \ref{sec42}), and for $\delta_{\text{ND}}<\delta<\delta_{\text{FP}}\simeq 0.66$ the local energy minimum lies in the metastable region between the binodal and the spinodal. We refer to the point $\delta_{\text{FP}}\simeq 0.66$ where the trajectory of the local energy minima encounters the spinodal as the fragmentation point (FP).
The energy per particle at neutron drip and at the fragmentation point is $\bar E_{\text{ND}}\simeq -13.0 \,\text{MeV}$ and 
$E_{\text{FP}}\simeq -3.1 \,\text{MeV}$, respectively. 
For comparison we follow the local energy minima also into the unstable spinodal region; i.e., we show also the point where both derivatives of the (analytical) energy per particle vanish (saddle point, SP) at $\delta_{\text{SP}}\simeq 0.81$ as well as the local energy minimum at $\delta\simeq0.76$. For $\delta\gtrsim 0.76$ the energy per particle is positive at all (finite) densities, and for $\delta\geq\delta_{\text{SP}}$ the pressure is a semipositive definite function of density.

\vspace*{-0.4cm}
%------------------------------------------------FIGURE sec4-3
%--------------------------------------------END_FIGURE sec4-3

\begin{figure}[H] 
\centering
\vspace*{-3.1cm}
\hspace*{-2.2cm}
\includegraphics[width=1.15\textwidth]{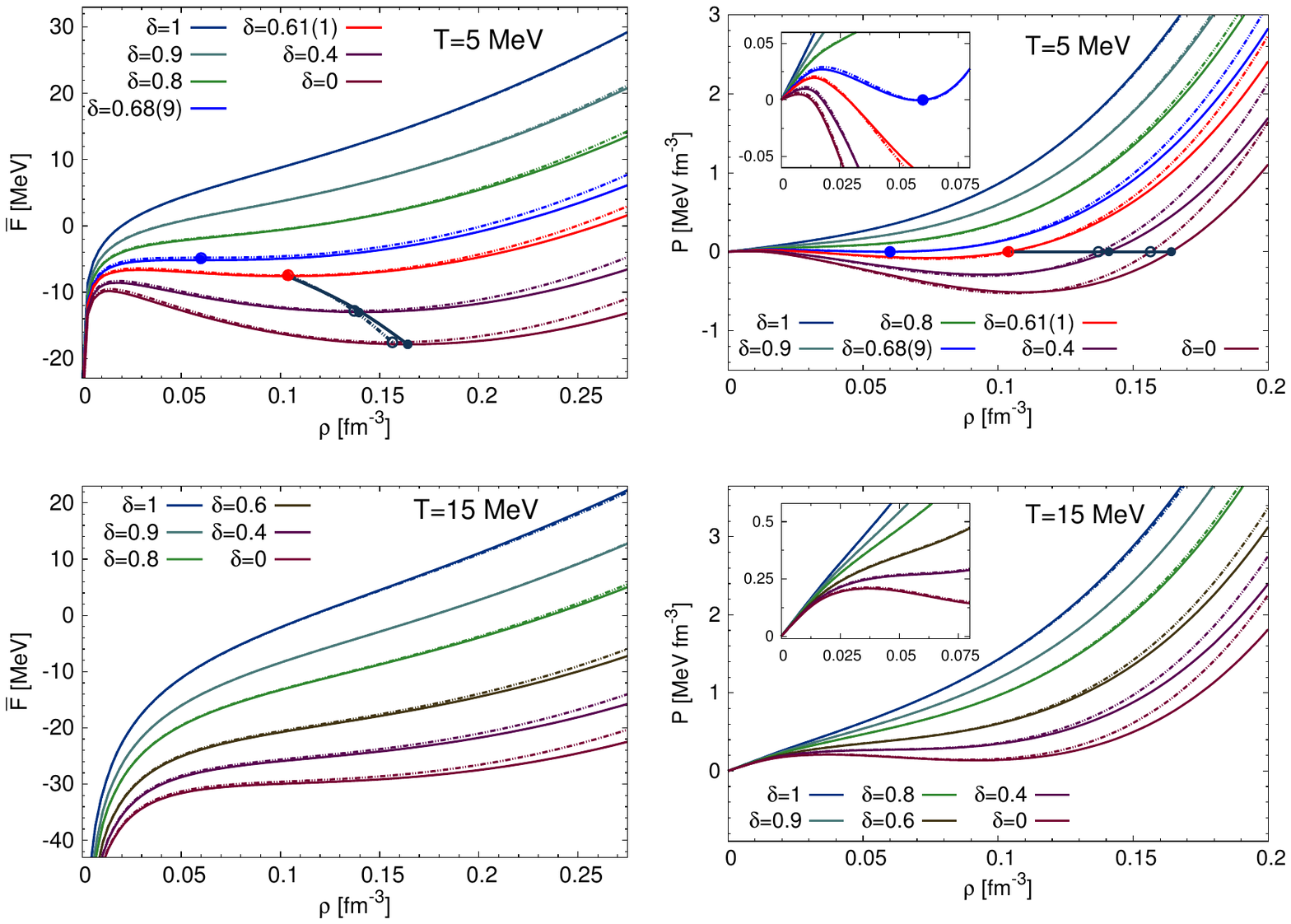} 
\vspace*{-11.7cm}
\caption{(Color online) Free energy per particle $\bar{F}(T,\rho,\delta)$ and pressure $P(T,\rho,\delta)$ at temperatures $T=5,15\,\text{MeV}$. The solid lines show the n3lo414 results, the dash-dot lines the n3lo450 results. The thick dark lines with circles depict the trajectories of the local free energy minima, up to the point where they encounter the spinodal. The insets magnify the behavior of the pressure at low densities.}  
\label{plots_sec4_2}
\end{figure}

The finite-temperature results for the free energy per particle $\bar{F}(T,\rho,\delta)$ and the pressure $P(T,\rho,\delta)$ are shown in Fig.\ \ref{plots_sec4_2} for temperatures $T/\text{MeV}=5,15$. At finite $T$ the trajectories of the local free energy minima lie entirely in the metastable region; an isolated drop of hot liquid nuclear matter has to be stabilized by a surrounding nucleon gas.
At $T=5\,\text{MeV}$ the trajectory ends at $\delta_{\text{FP}}\simeq 0.61$, defining the fragmentation temperature for nuclear matter with proton concentration $Y_{\text{p}}\simeq 0.195$. The $T=5\,\text{MeV}$ saddle point is located at $\delta_{\text{SP}}\simeq 0.68$ for n3lo414 and $\delta_{\text{SP}}\simeq 0.69$ for n3lo450.
For $T= 15\,\text{MeV}$ no local free energy minimum exists; for $T\gtrsim 13.5\,\text{MeV}$ the analytical free energy per particle is a monotonic increasing function of density for all values of $\delta$ (cf.\ Fig.\ \ref{plots_frag}).

\begin{figure}[H] 
\centering
\vspace*{-3.6cm}
\hspace*{-2.2cm}
\includegraphics[width=1.15\textwidth]{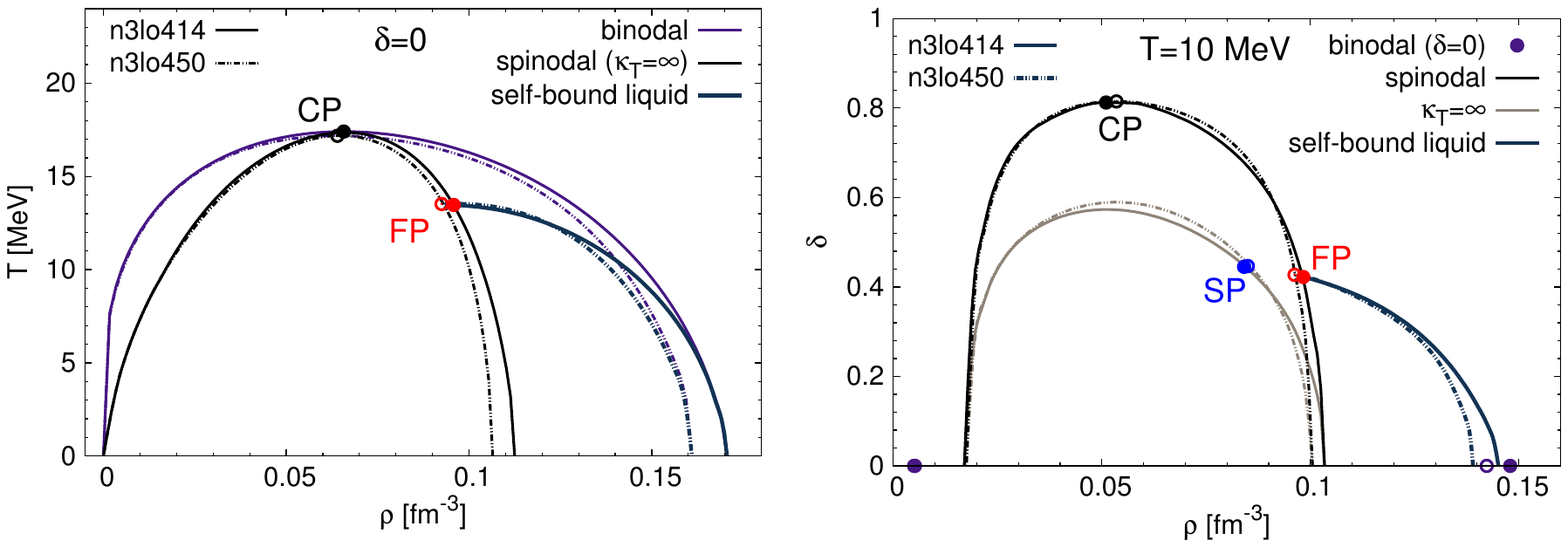} 
\vspace*{-17.5cm}
\caption{(Color online) Left plot: binodal, spinodal, and trajectory of local free energy minima (``self-bound liquid'') in SNM ($\delta=0$). Right plot: $T=10\,\text{MeV}$ cross sections of the spinodal, the $\upkappa_T=\infty$ boundary, and the surface of local free energy minima (``self-bound liquid''); only the $\delta=0$ endpoints of the binodal are shown (we did not construct the binodal for $\delta \neq 0$). The critical points (CP), fragmentation points (FP) and saddle points (SP) are shown explicitly in both plots (in SNM the FP and the SP coincide).
}
\label{plots_sec4_5}
\end{figure}

\begin{figure}[H] 
\centering
\vspace*{-4.5cm}
\hspace*{1.2cm}
\includegraphics[width=1.3\textwidth]{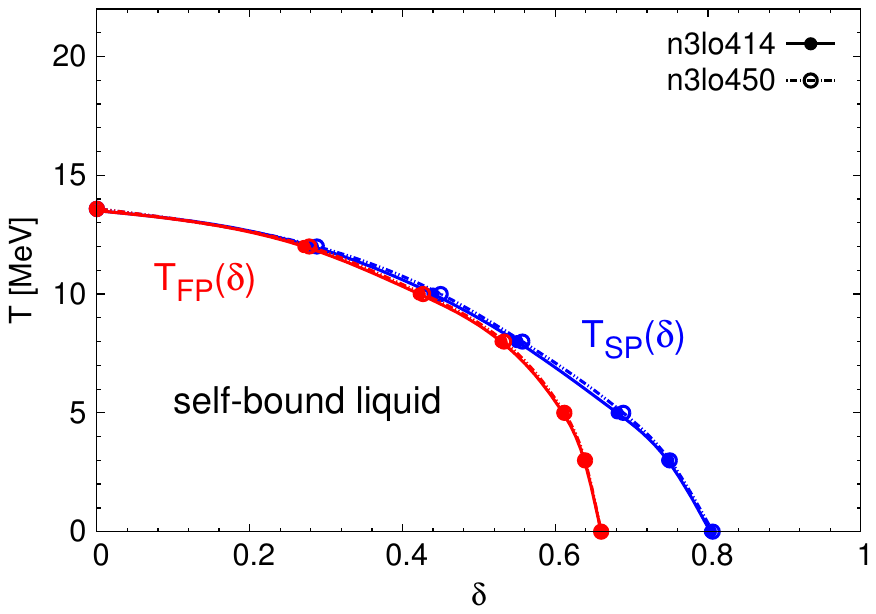} 
\vspace*{-20.5cm}
\caption{(Color online) Trajectory of the fragmentation temperature $T_{\text{FP}}(\delta)$ above which no metastable self-bound state can exist (lower red line). For comparison we also show the trajectory of the points where the (analytical) free energy per particle has a saddle point, $T_{\text{SP}}(\delta)$ (upper blue line). The calculated data points are shown explicitly.}
\label{plots_frag}
\end{figure}

The relation between the spinodal, the binodal, and the trajectories of the local free energy minima is illustrated in Fig.\
\ref{plots_sec4_5}. The two plots in Fig.\ \ref{plots_sec4_5} represent isoplethal ($\delta=\text{const}$) and isothermal cross sections of the respective surfaces (spinodal, binodal, surface of local free energy minima) in $(T,\rho,\delta)$ space (cf.\ also Refs.\ \cite{Muller:1995ji,Ducoin:2005aa}). In the second plot we also show the surface with divergent isothermal compressibility $\upkappa_T=\rho^ {-1}(\partial \rho/ \partial P)_{T,\delta}$, which corresponds to the violation of the stability criterion $\upkappa_T^{-1}>0$ for a one-component system.
Hence, the saddle point (SP) where both derivatives of the free energy per particle with respect to the nucleon density vanish coincides with the fragmentation point (FP) where a liquid drop becomes unstable only for $\delta=0$ where nuclear matter behaves like a pure substance. For $\delta \neq0$ the more restrictive two-component stability criteria are needed \cite{beegle} ($\upkappa_T^{-1}>0$ is not a relevant stability criterion in that case) and the SP is located in the interior of the spinodal. 

Finally, in Fig.\ \ref{plots_frag} the trajectory of the fragmentation temperatures $T_\text{FP}(\delta)$ is shown [for comparison we also show the saddle point temperatures $T_\text{SP}(\delta)$]. This trajectory determines the range of temperatures and isospin asymmetries for which a self-bound liquid state exists.

%------------------------------------------------FIGURE sec4-2
%--------------------------------------------END_FIGURE sec4-2

%-----------------------------------------------------------------------------
%------------------------------------------------FIGURE sec4-5
%------------------------------------------------FIGURE sec4-5

\vspace*{0.5cm}
%--------------------------------------------END_FIGURE sec4-4

\vspace*{0.0cm}
\section{Summary} \label{sec5}
%------------------summary
In this work, we have investigated in detail the temperature and density dependence of the symmetry free energy $\bar F_{\text{sym}}(T,\rho)=\bar F(T,\rho,\delta=1)-\bar F(T,\rho,\delta=0)$ in homogeneous nuclear matter using chiral effective field theory interactions constructed at resolution scales $\Lambda=414,450\,\text{MeV}$. 
The free energy per particle of isospin-symmetric nuclear matter, $\bar F(T,\rho,\delta=0)$, and of pure neutron matter, $\bar F(T,\rho,\delta=1)$, have been calculated in second-order many-body perturbation theory (Kohn-Luttinger-Ward formalism). Constraints from the nuclear saturation 
point, the critical point of the liquid-gas phase transition, 
and the density dependence of $\bar F_{\text{sym}}$ at zero temperature are
reproduced, and our results are in reasonable agreement with the virial expansion of the neutron matter equation of state at low fugacities. 

The four leading coefficients in an expansion of the noninteracting contributions $\bar F_0$ and $\bar F_\text{rel}$ in terms of $\delta^2$ have been examined, and we have found that the convergence rate of the expansion of $\bar F_0$ decreases significantly with temperature. 
Therefore we have used the exact expressions for $\bar F_0$ in computing the free energy per particle $\bar F(T,\rho,\delta)$ in isospin-asymmetric nuclear matter. 
The many-body contributions from nuclear interactions on the other hand have been assumed to have a quadratic dependence on the isospin asymmetry $\delta$. 

From the results for $\bar F(T,\rho,\delta)$ we have computed the pressure $P(T,\rho,\delta)$ and the neutron and proton chemical potentials $\mu_{\text{n/p}}(T,\rho,\delta)$, and we have constructed the trajectory of the critical temperature $T_c(\delta)$. The critical line ends at a proton fraction $Y_{\text{p},c}^\text{end}\simeq 3\cdot 10^{-4}$. The neutron drip point in infinite nuclear matter at zero temperature has been located at $Y_{\text{p,ND}}\simeq 0.35$. Furthermore, we have determined the trajectory of the fragmentation temperature $T_{\text{FP}}(\delta)$ above which no metastable self-bound state exists. 

Future work will be aimed at improving the description of the isospin-asymmetry dependence 
of the interaction contributions to the thermodynamic equation of state, a more detailed treatment of the low-density region including Coulomb and surface effects, and the extrapolation of the equation of state
to higher temperatures and densities required for simulations of core-collapse
supernovae and binary neutron star mergers.

\vspace*{0.5cm}
\acknowledgements
We thank S. Reddy for useful discussions.
This work is supported in part by the DFG and NSFC (CRC 110), and the US DOE Grant No.\ DE-FG02-97ER-41014.

\newpage
\vspace*{0.1cm}
%------------------------------------------------------------
\bibliographystyle{apsrev4-1}		% (uses file "plain.bst")
\bibliography{refs2}		% expects file "myrefs.bib"

\begin{thebibliography}{84}%
\makeatletter
\providecommand \@ifxundefined [1]{%
 \@ifx{#1\undefined}
}%
\providecommand \@ifnum [1]{%
 \ifnum #1\expandafter \@firstoftwo
 \else \expandafter \@secondoftwo
 \fi
}%
\providecommand \@ifx [1]{%
 \ifx #1\expandafter \@firstoftwo
 \else \expandafter \@secondoftwo
 \fi
}%
\providecommand \natexlab [1]{#1}%
\providecommand \enquote  [1]{``#1''}%
\providecommand \bibnamefont  [1]{#1}%
\providecommand \bibfnamefont [1]{#1}%
\providecommand \citenamefont [1]{#1}%
\providecommand \href@noop [0]{\@secondoftwo}%
\providecommand \href [0]{\begingroup \@sanitize@url \@href}%
\providecommand \@href[1]{\@@startlink{#1}\@@href}%
\providecommand \@@href[1]{\endgroup#1\@@endlink}%
\providecommand \@sanitize@url [0]{\catcode `\\12\catcode `\$12\catcode
  `\&12\catcode `\#12\catcode `\^12\catcode `\_12\catcode `\%12\relax}%
\providecommand \@@startlink[1]{}%
\providecommand \@@endlink[0]{}%
\providecommand \url  [0]{\begingroup\@sanitize@url \@url }%
\providecommand \@url [1]{\endgroup\@href {#1}{\urlprefix }}%
\providecommand \urlprefix  [0]{URL }%
\providecommand \Eprint [0]{\href }%
\providecommand \doibase [0]{http://dx.doi.org/}%
\providecommand \selectlanguage [0]{\@gobble}%
\providecommand \bibinfo  [0]{\@secondoftwo}%
\providecommand \bibfield  [0]{\@secondoftwo}%
\providecommand \translation [1]{[#1]}%
\providecommand \BibitemOpen [0]{}%
\providecommand \bibitemStop [0]{}%
\providecommand \bibitemNoStop [0]{.\EOS\space}%
\providecommand \EOS [0]{\spacefactor3000\relax}%
\providecommand \BibitemShut  [1]{\csname bibitem#1\endcsname}%
\let\auto@bib@innerbib\@empty
%</preamble>
\bibitem [{\citenamefont {Steiner}\ \emph {et~al.}(2005)\citenamefont
  {Steiner}, \citenamefont {Prakash}, \citenamefont {Lattimer},\ and\
  \citenamefont {Ellis}}]{Steiner:2004fi}%
  \BibitemOpen
  \bibfield  {author} {\bibinfo {author} {\bibfnamefont {A.~W.}\ \bibnamefont
  {Steiner}}, \bibinfo {author} {\bibfnamefont {M.}~\bibnamefont {Prakash}},
  \bibinfo {author} {\bibfnamefont {J.~M.}\ \bibnamefont {Lattimer}}, \ and\
  \bibinfo {author} {\bibfnamefont {P.~J.}\ \bibnamefont {Ellis}},\ }\href
  {\doibase 10.1016/j.physrep.2005.02.004} {\bibfield  {journal} {\bibinfo
  {journal} {Phys. Rept.}\ }\textbf {\bibinfo {volume} {411}},\ \bibinfo
  {pages} {325} (\bibinfo {year} {2005})}\BibitemShut {NoStop}%
%%CITATION = NUCL-TH/0410066;%%
\bibitem [{\citenamefont {Balantekin}\ \emph {et~al.}(2014)\citenamefont
  {Balantekin}, \citenamefont {Carlson}, \citenamefont {Dean}, \citenamefont
  {Fuller}, \citenamefont {Furnstahl} \emph {et~al.}}]{Balantekin:2014opa}%
  \BibitemOpen
  \bibfield  {author} {\bibinfo {author} {\bibfnamefont {A.~B.}\ \bibnamefont
  {Balantekin}}, \bibinfo {author} {\bibfnamefont {J.}~\bibnamefont {Carlson}},
  \bibinfo {author} {\bibfnamefont {D.}~\bibnamefont {Dean}}, \bibinfo {author}
  {\bibfnamefont {G.}~\bibnamefont {Fuller}}, \bibinfo {author} {\bibfnamefont
  {R.}~\bibnamefont {Furnstahl}},  \emph {et~al.},\ }\href {\doibase
  10.1142/S0217732314300109} {\bibfield  {journal} {\bibinfo  {journal} {Mod.
  Phys. Lett. A}\ }\textbf {\bibinfo {volume} {29}},\ \bibinfo {pages}
  {1430010} (\bibinfo {year} {2014})}\BibitemShut {NoStop}%
%%CITATION = ARXIV:1401.6435;%%
\bibitem [{\citenamefont {Li}\ \emph {et~al.}(2008)\citenamefont {Li},
  \citenamefont {Chen},\ and\ \citenamefont {Ko}}]{Li2008113}%
  \BibitemOpen
  \bibfield  {author} {\bibinfo {author} {\bibfnamefont {B.-A.}\ \bibnamefont
  {Li}}, \bibinfo {author} {\bibfnamefont {L.-W.}\ \bibnamefont {Chen}}, \ and\
  \bibinfo {author} {\bibfnamefont {C.~M.}\ \bibnamefont {Ko}},\ }\href
  {\doibase http://dx.doi.org/10.1016/j.physrep.2008.04.005} {\bibfield
  {journal} {\bibinfo  {journal} {Phys. Rept.}\ }\textbf {\bibinfo {volume}
  {464}},\ \bibinfo {pages} {113 } (\bibinfo {year} {2008})}\BibitemShut
  {NoStop}%
\bibitem [{\citenamefont {Machleidt}\ and\ \citenamefont
  {Entem}(2011)}]{Machleidt:2011zz}%
  \BibitemOpen
  \bibfield  {author} {\bibinfo {author} {\bibfnamefont {R.}~\bibnamefont
  {Machleidt}}\ and\ \bibinfo {author} {\bibfnamefont {D.~R.}\ \bibnamefont
  {Entem}},\ }\href {\doibase 10.1016/j.physrep.2011.02.001} {\bibfield
  {journal} {\bibinfo  {journal} {Phys. Rept.}\ }\textbf {\bibinfo {volume}
  {503}},\ \bibinfo {pages} {1} (\bibinfo {year} {2011})}\BibitemShut {NoStop}%
\bibitem [{\citenamefont {Holt}\ \emph {et~al.}(2013)\citenamefont {Holt},
  \citenamefont {Kaiser},\ and\ \citenamefont {Weise}}]{Holt:2013fwa}%
  \BibitemOpen
  \bibfield  {author} {\bibinfo {author} {\bibfnamefont {J.~W.}\ \bibnamefont
  {Holt}}, \bibinfo {author} {\bibfnamefont {N.}~\bibnamefont {Kaiser}}, \ and\
  \bibinfo {author} {\bibfnamefont {W.}~\bibnamefont {Weise}},\ }\href
  {\doibase 10.1016/j.ppnp.2013.08.001} {\bibfield  {journal} {\bibinfo
  {journal} {Prog. Part. Nucl. Phys.}\ }\textbf {\bibinfo {volume} {73}},\
  \bibinfo {pages} {35} (\bibinfo {year} {2013})}\BibitemShut {NoStop}%
%%CITATION = ARXIV:1304.6350;%%
\bibitem [{\citenamefont {Epelbaum}\ \emph {et~al.}(2009)\citenamefont
  {Epelbaum}, \citenamefont {Hammer},\ and\ \citenamefont
  {Meissner}}]{Epelbaum:2008ga}%
  \BibitemOpen
  \bibfield  {author} {\bibinfo {author} {\bibfnamefont {E.}~\bibnamefont
  {Epelbaum}}, \bibinfo {author} {\bibfnamefont {H.-W.}\ \bibnamefont
  {Hammer}}, \ and\ \bibinfo {author} {\bibfnamefont {U.-G.}\ \bibnamefont
  {Meissner}},\ }\href {\doibase 10.1103/RevModPhys.81.1773} {\bibfield
  {journal} {\bibinfo  {journal} {Rev. Mod. Phys.}\ }\textbf {\bibinfo {volume}
  {81}},\ \bibinfo {pages} {1773} (\bibinfo {year} {2009})}\BibitemShut
  {NoStop}%
\bibitem [{\citenamefont {Bogner}\ \emph {et~al.}(2010)\citenamefont {Bogner},
  \citenamefont {Furnstahl},\ and\ \citenamefont {Schwenk}}]{Bogner:2009bt}%
  \BibitemOpen
  \bibfield  {author} {\bibinfo {author} {\bibfnamefont {S.~K.}\ \bibnamefont
  {Bogner}}, \bibinfo {author} {\bibfnamefont {R.~J.}\ \bibnamefont
  {Furnstahl}}, \ and\ \bibinfo {author} {\bibfnamefont {A.}~\bibnamefont
  {Schwenk}},\ }\href {\doibase 10.1016/j.ppnp.2010.03.001} {\bibfield
  {journal} {\bibinfo  {journal} {Prog. Part. Nucl. Phys.}\ }\textbf {\bibinfo
  {volume} {65}},\ \bibinfo {pages} {94} (\bibinfo {year} {2010})}\BibitemShut
  {NoStop}%
%%CITATION = ARXIV:0912.3688;%%
\bibitem [{\citenamefont {Tews}\ \emph {et~al.}(2013)\citenamefont {Tews},
  \citenamefont {Kr\"{u}ger}, \citenamefont {Hebeler},\ and\ \citenamefont
  {Schwenk}}]{Tews:2012fj}%
  \BibitemOpen
  \bibfield  {author} {\bibinfo {author} {\bibfnamefont {I.}~\bibnamefont
  {Tews}}, \bibinfo {author} {\bibfnamefont {T.}~\bibnamefont {Kr\"{u}ger}},
  \bibinfo {author} {\bibfnamefont {K.}~\bibnamefont {Hebeler}}, \ and\
  \bibinfo {author} {\bibfnamefont {A.}~\bibnamefont {Schwenk}},\ }\href
  {\doibase 10.1103/PhysRevLett.110.032504} {\bibfield  {journal} {\bibinfo
  {journal} {Phys. Rev. Lett.}\ }\textbf {\bibinfo {volume} {110}},\ \bibinfo
  {pages} {032504} (\bibinfo {year} {2013})}\BibitemShut {NoStop}%
\bibitem [{\citenamefont {Furnstahl}\ \emph {et~al.}(2015)\citenamefont
  {Furnstahl}, \citenamefont {Phillips},\ and\ \citenamefont
  {Wesolowski}}]{Furnstahl:2014xsa}%
  \BibitemOpen
  \bibfield  {author} {\bibinfo {author} {\bibfnamefont {R.~J.}\ \bibnamefont
  {Furnstahl}}, \bibinfo {author} {\bibfnamefont {D.~R.}\ \bibnamefont
  {Phillips}}, \ and\ \bibinfo {author} {\bibfnamefont {S.}~\bibnamefont
  {Wesolowski}},\ }\href {\doibase 10.1088/0954-3899/42/3/034028} {\bibfield
  {journal} {\bibinfo  {journal} {J. Phys. G}\ }\textbf {\bibinfo {volume}
  {42}},\ \bibinfo {pages} {034028} (\bibinfo {year} {2015})}\BibitemShut
  {NoStop}%
%%CITATION = ARXIV:1407.0657;%%
\bibitem [{\citenamefont {Sammarruca}\ \emph {et~al.}(2015)\citenamefont
  {Sammarruca}, \citenamefont {Coraggio}, \citenamefont {Holt}, \citenamefont
  {Itaco}, \citenamefont {Machleidt} \emph {et~al.}}]{Sammarruca:2014zia}%
  \BibitemOpen
  \bibfield  {author} {\bibinfo {author} {\bibfnamefont {F.}~\bibnamefont
  {Sammarruca}}, \bibinfo {author} {\bibfnamefont {L.}~\bibnamefont
  {Coraggio}}, \bibinfo {author} {\bibfnamefont {J.~W.}\ \bibnamefont {Holt}},
  \bibinfo {author} {\bibfnamefont {N.}~\bibnamefont {Itaco}}, \bibinfo
  {author} {\bibfnamefont {R.}~\bibnamefont {Machleidt}},  \emph {et~al.},\
  }\href {\doibase 10.1103/PhysRevC.91.054311} {\bibfield  {journal} {\bibinfo
  {journal} {Phys. Rev. C}\ }\textbf {\bibinfo {volume} {91}},\ \bibinfo
  {pages} {054311} (\bibinfo {year} {2015})}\BibitemShut {NoStop}%
%%CITATION = ARXIV:1411.0136;%%
\bibitem [{\citenamefont {Coraggio}\ \emph {et~al.}(2014)\citenamefont
  {Coraggio}, \citenamefont {Holt}, \citenamefont {Itaco}, \citenamefont
  {Machleidt}, \citenamefont {Marcucci},\ and\ \citenamefont
  {Sammarruca}}]{Coraggio:2014nvaa}%
  \BibitemOpen
  \bibfield  {author} {\bibinfo {author} {\bibfnamefont {L.}~\bibnamefont
  {Coraggio}}, \bibinfo {author} {\bibfnamefont {J.~W.}\ \bibnamefont {Holt}},
  \bibinfo {author} {\bibfnamefont {N.}~\bibnamefont {Itaco}}, \bibinfo
  {author} {\bibfnamefont {R.}~\bibnamefont {Machleidt}}, \bibinfo {author}
  {\bibfnamefont {L.~E.}\ \bibnamefont {Marcucci}}, \ and\ \bibinfo {author}
  {\bibfnamefont {F.}~\bibnamefont {Sammarruca}},\ }\href {\doibase
  10.1103/PhysRevC.89.044321} {\bibfield  {journal} {\bibinfo  {journal} {Phys.
  Rev. C}\ }\textbf {\bibinfo {volume} {89}},\ \bibinfo {pages} {044321}
  (\bibinfo {year} {2014})}\BibitemShut {NoStop}%
\bibitem [{\citenamefont {Wellenhofer}\ \emph {et~al.}(2014)\citenamefont
  {Wellenhofer}, \citenamefont {Holt}, \citenamefont {Kaiser},\ and\
  \citenamefont {Weise}}]{paper1}%
  \BibitemOpen
  \bibfield  {author} {\bibinfo {author} {\bibfnamefont {C.}~\bibnamefont
  {Wellenhofer}}, \bibinfo {author} {\bibfnamefont {J.~W.}\ \bibnamefont
  {Holt}}, \bibinfo {author} {\bibfnamefont {N.}~\bibnamefont {Kaiser}}, \ and\
  \bibinfo {author} {\bibfnamefont {W.}~\bibnamefont {Weise}},\ }\href
  {\doibase 10.1103/PhysRevC.89.064009} {\bibfield  {journal} {\bibinfo
  {journal} {Phys. Rev. C}\ }\textbf {\bibinfo {volume} {89}},\ \bibinfo
  {pages} {064009} (\bibinfo {year} {2014})}\BibitemShut {NoStop}%
%%CITATION = ARXIV:1404.2136;%%
\bibitem [{\citenamefont {Bombaci}\ and\ \citenamefont
  {Lombardo}(1991)}]{Bombaci:1991zz}%
  \BibitemOpen
  \bibfield  {author} {\bibinfo {author} {\bibfnamefont {I.}~\bibnamefont
  {Bombaci}}\ and\ \bibinfo {author} {\bibfnamefont {U.}~\bibnamefont
  {Lombardo}},\ }\href {\doibase 10.1103/PhysRevC.44.1892} {\bibfield
  {journal} {\bibinfo  {journal} {Phys. Rev. C}\ }\textbf {\bibinfo {volume}
  {44}},\ \bibinfo {pages} {1892} (\bibinfo {year} {1991})}\BibitemShut
  {NoStop}%
%%CITATION = PHRVA,C44,1892;%%
\bibitem [{\citenamefont {Drischler}\ \emph {et~al.}(2014)\citenamefont
  {Drischler}, \citenamefont {Som\`a},\ and\ \citenamefont
  {Schwenk}}]{PhysRevC.89.025806}%
  \BibitemOpen
  \bibfield  {author} {\bibinfo {author} {\bibfnamefont {C.}~\bibnamefont
  {Drischler}}, \bibinfo {author} {\bibfnamefont {V.}~\bibnamefont {Som\`a}}, \
  and\ \bibinfo {author} {\bibfnamefont {A.}~\bibnamefont {Schwenk}},\ }\href
  {\doibase 10.1103/PhysRevC.89.025806} {\bibfield  {journal} {\bibinfo
  {journal} {Phys. Rev. C}\ }\textbf {\bibinfo {volume} {89}},\ \bibinfo
  {pages} {025806} (\bibinfo {year} {2014})}\BibitemShut {NoStop}%
\bibitem [{\citenamefont {Ducoin}\ \emph {et~al.}(2006)\citenamefont {Ducoin},
  \citenamefont {Chomaz},\ and\ \citenamefont {Gulminelli}}]{Ducoin:2005aa}%
  \BibitemOpen
  \bibfield  {author} {\bibinfo {author} {\bibfnamefont {C.}~\bibnamefont
  {Ducoin}}, \bibinfo {author} {\bibfnamefont {{\relax Ph}.}~\bibnamefont
  {Chomaz}}, \ and\ \bibinfo {author} {\bibfnamefont {F.}~\bibnamefont
  {Gulminelli}},\ }\href {\doibase 10.1016/j.nuclphysa.2006.03.005} {\bibfield
  {journal} {\bibinfo  {journal} {Nucl. Phys. A}\ }\textbf {\bibinfo {volume}
  {771}},\ \bibinfo {pages} {68} (\bibinfo {year} {2006})}\BibitemShut
  {NoStop}%
%%CITATION = NUCL-TH/0512029;%%
\bibitem [{\citenamefont {Hempel}\ \emph {et~al.}(2013)\citenamefont {Hempel},
  \citenamefont {Dexheimer}, \citenamefont {Schramm},\ and\ \citenamefont
  {Iosilevskiy}}]{Hempel:2013tfa}%
  \BibitemOpen
  \bibfield  {author} {\bibinfo {author} {\bibfnamefont {M.}~\bibnamefont
  {Hempel}}, \bibinfo {author} {\bibfnamefont {V.}~\bibnamefont {Dexheimer}},
  \bibinfo {author} {\bibfnamefont {S.}~\bibnamefont {Schramm}}, \ and\
  \bibinfo {author} {\bibfnamefont {I.}~\bibnamefont {Iosilevskiy}},\ }\href
  {\doibase 10.1103/PhysRevC.88.014906} {\bibfield  {journal} {\bibinfo
  {journal} {Phys. Rev. C}\ }\textbf {\bibinfo {volume} {88}},\ \bibinfo
  {pages} {014906} (\bibinfo {year} {2013})}\BibitemShut {NoStop}%
%%CITATION = ARXIV:1302.2835;%%
\bibitem [{\citenamefont {M\"{u}ller}\ and\ \citenamefont
  {Serot}(1995)}]{Muller:1995ji}%
  \BibitemOpen
  \bibfield  {author} {\bibinfo {author} {\bibfnamefont {H.}~\bibnamefont
  {M\"{u}ller}}\ and\ \bibinfo {author} {\bibfnamefont {B.~D.}\ \bibnamefont
  {Serot}},\ }\href {\doibase 10.1103/PhysRevC.52.2072} {\bibfield  {journal}
  {\bibinfo  {journal} {Phys. Rev. C}\ }\textbf {\bibinfo {volume} {52}},\
  \bibinfo {pages} {2072} (\bibinfo {year} {1995})}\BibitemShut {NoStop}%
%%CITATION = NUCL-TH/9505013;%%
\bibitem [{\citenamefont {Chomaz}\ \emph {et~al.}(2004)\citenamefont {Chomaz},
  \citenamefont {Colonna},\ and\ \citenamefont {Randrup}}]{Chomaz:2003dz}%
  \BibitemOpen
  \bibfield  {author} {\bibinfo {author} {\bibfnamefont {{\relax
  Ph}.}~\bibnamefont {Chomaz}}, \bibinfo {author} {\bibfnamefont
  {M.}~\bibnamefont {Colonna}}, \ and\ \bibinfo {author} {\bibfnamefont
  {J.}~\bibnamefont {Randrup}},\ }\href {\doibase
  10.1016/j.physrep.2003.09.006} {\bibfield  {journal} {\bibinfo  {journal}
  {Phys. Rept.}\ }\textbf {\bibinfo {volume} {389}},\ \bibinfo {pages} {263}
  (\bibinfo {year} {2004})}\BibitemShut {NoStop}%
%%CITATION = PRPLC,389,263;%%
\bibitem [{\citenamefont {Margueron}\ and\ \citenamefont
  {Chomaz}(2003)}]{PhysRevC.67.041602}%
  \BibitemOpen
  \bibfield  {author} {\bibinfo {author} {\bibfnamefont {J.}~\bibnamefont
  {Margueron}}\ and\ \bibinfo {author} {\bibfnamefont {{\relax
  Ph}.}~\bibnamefont {Chomaz}},\ }\href {\doibase 10.1103/PhysRevC.67.041602}
  {\bibfield  {journal} {\bibinfo  {journal} {Phys. Rev. C}\ }\textbf {\bibinfo
  {volume} {67}},\ \bibinfo {pages} {041602} (\bibinfo {year}
  {2003})}\BibitemShut {NoStop}%
\bibitem [{\citenamefont {Binder}(1987)}]{0034-4885-50-7-001}%
  \BibitemOpen
  \bibfield  {author} {\bibinfo {author} {\bibfnamefont {K.}~\bibnamefont
  {Binder}},\ }\href {\doibase 10.1088/0034-4885/50/7/001} {\bibfield
  {journal} {\bibinfo  {journal} {Rep. Prog. Phys.}\ }\textbf {\bibinfo
  {volume} {50}},\ \bibinfo {pages} {783} (\bibinfo {year} {1987})}\BibitemShut
  {NoStop}%
\bibitem [{\citenamefont {Abraham}(1979)}]{Abraham197993}%
  \BibitemOpen
  \bibfield  {author} {\bibinfo {author} {\bibfnamefont {F.~F.}\ \bibnamefont
  {Abraham}},\ }\href {\doibase http://dx.doi.org/10.1016/0370-1573(79)90003-6}
  {\bibfield  {journal} {\bibinfo  {journal} {Phys. Rept.}\ }\textbf {\bibinfo
  {volume} {53}},\ \bibinfo {pages} {93 } (\bibinfo {year} {1979})}\BibitemShut
  {NoStop}%
\bibitem [{\citenamefont {Ducoin}\ \emph {et~al.}(2007)\citenamefont {Ducoin},
  \citenamefont {Hasnaoui}, \citenamefont {Napolitani}, \citenamefont
  {Chomaz},\ and\ \citenamefont {Gulminelli}}]{PhysRevC.75.065805}%
  \BibitemOpen
  \bibfield  {author} {\bibinfo {author} {\bibfnamefont {C.}~\bibnamefont
  {Ducoin}}, \bibinfo {author} {\bibfnamefont {K.~H.~O.}\ \bibnamefont
  {Hasnaoui}}, \bibinfo {author} {\bibfnamefont {P.}~\bibnamefont
  {Napolitani}}, \bibinfo {author} {\bibfnamefont {{\relax Ph}.}~\bibnamefont
  {Chomaz}}, \ and\ \bibinfo {author} {\bibfnamefont {F.}~\bibnamefont
  {Gulminelli}},\ }\href {\doibase 10.1103/PhysRevC.75.065805} {\bibfield
  {journal} {\bibinfo  {journal} {Phys. Rev. C}\ }\textbf {\bibinfo {volume}
  {75}},\ \bibinfo {pages} {065805} (\bibinfo {year} {2007})}\BibitemShut
  {NoStop}%
\bibitem [{\citenamefont {Napolitani}\ \emph {et~al.}(2007)\citenamefont
  {Napolitani}, \citenamefont {Chomaz}, \citenamefont {Gulminelli},\ and\
  \citenamefont {Hasnaoui}}]{PhysRevLett.98.131102}%
  \BibitemOpen
  \bibfield  {author} {\bibinfo {author} {\bibfnamefont {P.}~\bibnamefont
  {Napolitani}}, \bibinfo {author} {\bibfnamefont {{\relax Ph}.}~\bibnamefont
  {Chomaz}}, \bibinfo {author} {\bibfnamefont {F.}~\bibnamefont {Gulminelli}},
  \ and\ \bibinfo {author} {\bibfnamefont {K.~H.~O.}\ \bibnamefont
  {Hasnaoui}},\ }\href {\doibase 10.1103/PhysRevLett.98.131102} {\bibfield
  {journal} {\bibinfo  {journal} {Phys. Rev. Lett.}\ }\textbf {\bibinfo
  {volume} {98}},\ \bibinfo {pages} {131102} (\bibinfo {year}
  {2007})}\BibitemShut {NoStop}%
\bibitem [{\citenamefont {Ravenhall}\ \emph {et~al.}(1983)\citenamefont
  {Ravenhall}, \citenamefont {Pethick},\ and\ \citenamefont
  {Wilson}}]{Ravenhall:1983uh}%
  \BibitemOpen
  \bibfield  {author} {\bibinfo {author} {\bibfnamefont {D.~G.}\ \bibnamefont
  {Ravenhall}}, \bibinfo {author} {\bibfnamefont {C.~J.}\ \bibnamefont
  {Pethick}}, \ and\ \bibinfo {author} {\bibfnamefont {J.~R.}\ \bibnamefont
  {Wilson}},\ }\href {\doibase 10.1103/PhysRevLett.50.2066} {\bibfield
  {journal} {\bibinfo  {journal} {Phys. Rev. Lett.}\ }\textbf {\bibinfo
  {volume} {50}},\ \bibinfo {pages} {2066} (\bibinfo {year}
  {1983})}\BibitemShut {NoStop}%
%%CITATION = PRLTA,50,2066;%%
\bibitem [{\citenamefont {Hashimoto}\ \emph {et~al.}(1984)\citenamefont
  {Hashimoto}, \citenamefont {Seki},\ and\ \citenamefont {Yamada}}]{Hashimoto}%
  \BibitemOpen
  \bibfield  {author} {\bibinfo {author} {\bibfnamefont {M.}~\bibnamefont
  {Hashimoto}}, \bibinfo {author} {\bibfnamefont {H.}~\bibnamefont {Seki}}, \
  and\ \bibinfo {author} {\bibfnamefont {M.}~\bibnamefont {Yamada}},\ }\href
  {\doibase 10.1143/PTP.71.320} {\bibfield  {journal} {\bibinfo  {journal}
  {Prog. Theor. Phys.}\ }\textbf {\bibinfo {volume} {71}},\ \bibinfo {pages}
  {320} (\bibinfo {year} {1984})}\BibitemShut {NoStop}%
\bibitem [{\citenamefont {Pethick}\ and\ \citenamefont
  {Potekhin}(1998)}]{Pethick:1998qv}%
  \BibitemOpen
  \bibfield  {author} {\bibinfo {author} {\bibfnamefont {C.~J.}\ \bibnamefont
  {Pethick}}\ and\ \bibinfo {author} {\bibfnamefont {A.~Y.}\ \bibnamefont
  {Potekhin}},\ }\href {\doibase 10.1016/S0370-2693(98)00341-4} {\bibfield
  {journal} {\bibinfo  {journal} {Phys. Lett. B}\ }\textbf {\bibinfo {volume}
  {427}},\ \bibinfo {pages} {7} (\bibinfo {year} {1998})}\BibitemShut {NoStop}%
%%CITATION = ASTRO-PH/9803154;%%
\bibitem [{\citenamefont {Horowitz}\ \emph {et~al.}(2004)\citenamefont
  {Horowitz}, \citenamefont {P\'erez-Garc\'{i}a},\ and\ \citenamefont
  {Piekarewicz}}]{PhysRevC.69.045804}%
  \BibitemOpen
  \bibfield  {author} {\bibinfo {author} {\bibfnamefont {C.~J.}\ \bibnamefont
  {Horowitz}}, \bibinfo {author} {\bibfnamefont {M.~A.}\ \bibnamefont
  {P\'erez-Garc\'{i}a}}, \ and\ \bibinfo {author} {\bibfnamefont
  {J.}~\bibnamefont {Piekarewicz}},\ }\href {\doibase
  10.1103/PhysRevC.69.045804} {\bibfield  {journal} {\bibinfo  {journal} {Phys.
  Rev. C}\ }\textbf {\bibinfo {volume} {69}},\ \bibinfo {pages} {045804}
  (\bibinfo {year} {2004})}\BibitemShut {NoStop}%
\bibitem [{\citenamefont {Maruyama}\ \emph {et~al.}(2005)\citenamefont
  {Maruyama}, \citenamefont {Tatsumi}, \citenamefont {Voskresensky},
  \citenamefont {Tanigawa},\ and\ \citenamefont {Chiba}}]{Maruyama:2005vb}%
  \BibitemOpen
  \bibfield  {author} {\bibinfo {author} {\bibfnamefont {T.}~\bibnamefont
  {Maruyama}}, \bibinfo {author} {\bibfnamefont {T.}~\bibnamefont {Tatsumi}},
  \bibinfo {author} {\bibfnamefont {D.~N.}\ \bibnamefont {Voskresensky}},
  \bibinfo {author} {\bibfnamefont {T.}~\bibnamefont {Tanigawa}}, \ and\
  \bibinfo {author} {\bibfnamefont {S.}~\bibnamefont {Chiba}},\ }\href
  {\doibase 10.1103/PhysRevC.72.015802} {\bibfield  {journal} {\bibinfo
  {journal} {Phys. Rev. C}\ }\textbf {\bibinfo {volume} {72}},\ \bibinfo
  {pages} {015802} (\bibinfo {year} {2005})}\BibitemShut {NoStop}%
\bibitem [{\citenamefont {Watanabe}\ \emph {et~al.}(2005)\citenamefont
  {Watanabe}, \citenamefont {Maruyama}, \citenamefont {Sato}, \citenamefont
  {Yasuoka},\ and\ \citenamefont {Ebisuzaki}}]{Watanabe:2004tr}%
  \BibitemOpen
  \bibfield  {author} {\bibinfo {author} {\bibfnamefont {G.}~\bibnamefont
  {Watanabe}}, \bibinfo {author} {\bibfnamefont {T.}~\bibnamefont {Maruyama}},
  \bibinfo {author} {\bibfnamefont {K.}~\bibnamefont {Sato}}, \bibinfo {author}
  {\bibfnamefont {K.}~\bibnamefont {Yasuoka}}, \ and\ \bibinfo {author}
  {\bibfnamefont {T.}~\bibnamefont {Ebisuzaki}},\ }\href {\doibase
  10.1103/PhysRevLett.94.031101} {\bibfield  {journal} {\bibinfo  {journal}
  {Phys. Rev. Lett.}\ }\textbf {\bibinfo {volume} {94}},\ \bibinfo {pages}
  {031101} (\bibinfo {year} {2005})}\BibitemShut {NoStop}%
%%CITATION = NUCL-TH/0408061;%%
\bibitem [{\citenamefont {Avancini}\ \emph {et~al.}(2012)\citenamefont
  {Avancini}, \citenamefont {Barros}, \citenamefont {Brito}, \citenamefont
  {Chiacchiera}, \citenamefont {Menezes} \emph {et~al.}}]{Avancini:2012bj}%
  \BibitemOpen
  \bibfield  {author} {\bibinfo {author} {\bibfnamefont {S.~S.}\ \bibnamefont
  {Avancini}}, \bibinfo {author} {\bibfnamefont {C.~C.}\ \bibnamefont
  {Barros}}, \bibinfo {author} {\bibfnamefont {L.}~\bibnamefont {Brito}},
  \bibinfo {author} {\bibfnamefont {S.}~\bibnamefont {Chiacchiera}}, \bibinfo
  {author} {\bibfnamefont {D.~P.}\ \bibnamefont {Menezes}},  \emph {et~al.},\
  }\href {\doibase 10.1103/PhysRevC.85.035806} {\bibfield  {journal} {\bibinfo
  {journal} {Phys. Rev. C}\ }\textbf {\bibinfo {volume} {85}},\ \bibinfo
  {pages} {035806} (\bibinfo {year} {2012})}\BibitemShut {NoStop}%
\bibitem [{\citenamefont {Schneider}\ \emph {et~al.}(2013)\citenamefont
  {Schneider}, \citenamefont {Horowitz}, \citenamefont {Hughto},\ and\
  \citenamefont {Berry}}]{Schneider:2013dwa}%
  \BibitemOpen
  \bibfield  {author} {\bibinfo {author} {\bibfnamefont {A.~S.}\ \bibnamefont
  {Schneider}}, \bibinfo {author} {\bibfnamefont {C.~J.}\ \bibnamefont
  {Horowitz}}, \bibinfo {author} {\bibfnamefont {J.}~\bibnamefont {Hughto}}, \
  and\ \bibinfo {author} {\bibfnamefont {D.~K.}\ \bibnamefont {Berry}},\ }\href
  {\doibase 10.1103/PhysRevC.88.065807} {\bibfield  {journal} {\bibinfo
  {journal} {Phys. Rev. C}\ }\textbf {\bibinfo {volume} {88}},\ \bibinfo
  {pages} {065807} (\bibinfo {year} {2013})}\BibitemShut {NoStop}%
%%CITATION = ARXIV:1307.1678;%%
\bibitem [{\citenamefont {Horowitz}\ and\ \citenamefont
  {Schwenk}(2006{\natexlab{a}})}]{Horowitz:2005nd}%
  \BibitemOpen
  \bibfield  {author} {\bibinfo {author} {\bibfnamefont {C.~J.}\ \bibnamefont
  {Horowitz}}\ and\ \bibinfo {author} {\bibfnamefont {A.}~\bibnamefont
  {Schwenk}},\ }\href {\doibase 10.1016/j.nuclphysa.2006.05.009} {\bibfield
  {journal} {\bibinfo  {journal} {Nucl. Phys. A}\ }\textbf {\bibinfo {volume}
  {776}},\ \bibinfo {pages} {55} (\bibinfo {year}
  {2006}{\natexlab{a}})}\BibitemShut {NoStop}%
%%CITATION = NUCL-TH/0507033;%%
\bibitem [{\citenamefont {Shen}\ \emph {et~al.}(2010)\citenamefont {Shen},
  \citenamefont {Horowitz},\ and\ \citenamefont {Teige}}]{PhysRevC.82.045802}%
  \BibitemOpen
  \bibfield  {author} {\bibinfo {author} {\bibfnamefont {G.}~\bibnamefont
  {Shen}}, \bibinfo {author} {\bibfnamefont {C.~J.}\ \bibnamefont {Horowitz}},
  \ and\ \bibinfo {author} {\bibfnamefont {S.}~\bibnamefont {Teige}},\ }\href
  {\doibase 10.1103/PhysRevC.82.045802} {\bibfield  {journal} {\bibinfo
  {journal} {Phys. Rev. C}\ }\textbf {\bibinfo {volume} {82}},\ \bibinfo
  {pages} {045802} (\bibinfo {year} {2010})}\BibitemShut {NoStop}%
\bibitem [{\citenamefont {Typel}\ \emph {et~al.}(2010)\citenamefont {Typel},
  \citenamefont {R\"opke}, \citenamefont {Kl\"ahn}, \citenamefont {Blaschke},\
  and\ \citenamefont {Wolter}}]{PhysRevC.81.015803}%
  \BibitemOpen
  \bibfield  {author} {\bibinfo {author} {\bibfnamefont {S.}~\bibnamefont
  {Typel}}, \bibinfo {author} {\bibfnamefont {G.}~\bibnamefont {R\"opke}},
  \bibinfo {author} {\bibfnamefont {T.}~\bibnamefont {Kl\"ahn}}, \bibinfo
  {author} {\bibfnamefont {D.}~\bibnamefont {Blaschke}}, \ and\ \bibinfo
  {author} {\bibfnamefont {H.~H.}\ \bibnamefont {Wolter}},\ }\href {\doibase
  10.1103/PhysRevC.81.015803} {\bibfield  {journal} {\bibinfo  {journal} {Phys.
  Rev. C}\ }\textbf {\bibinfo {volume} {81}},\ \bibinfo {pages} {015803}
  (\bibinfo {year} {2010})}\BibitemShut {NoStop}%
\bibitem [{\citenamefont {Kohn}\ and\ \citenamefont
  {Luttinger}(1960)}]{Kohn:1960zz}%
  \BibitemOpen
  \bibfield  {author} {\bibinfo {author} {\bibfnamefont {W.}~\bibnamefont
  {Kohn}}\ and\ \bibinfo {author} {\bibfnamefont {J.~M.}\ \bibnamefont
  {Luttinger}},\ }\href {\doibase 10.1103/PhysRev.118.41} {\bibfield  {journal}
  {\bibinfo  {journal} {Phys. Rev.}\ }\textbf {\bibinfo {volume} {118}},\
  \bibinfo {pages} {41} (\bibinfo {year} {1960})}\BibitemShut {NoStop}%
%%CITATION = PHRVA,118,41;%%
\bibitem [{\citenamefont {Luttinger}\ and\ \citenamefont
  {Ward}(1960)}]{Luttinger:1960ua}%
  \BibitemOpen
  \bibfield  {author} {\bibinfo {author} {\bibfnamefont {J.~M.}\ \bibnamefont
  {Luttinger}}\ and\ \bibinfo {author} {\bibfnamefont {J.~C.}\ \bibnamefont
  {Ward}},\ }\href {\doibase 10.1103/PhysRev.118.1417} {\bibfield  {journal}
  {\bibinfo  {journal} {Phys. Rev.}\ }\textbf {\bibinfo {volume} {118}},\
  \bibinfo {pages} {1417} (\bibinfo {year} {1960})}\BibitemShut {NoStop}%
%%CITATION = PHRVA,118,1417;%%
\bibitem [{\citenamefont {Fritsch}\ \emph {et~al.}(2002)\citenamefont
  {Fritsch}, \citenamefont {Kaiser},\ and\ \citenamefont
  {Weise}}]{Fritsch:2002hp}%
  \BibitemOpen
  \bibfield  {author} {\bibinfo {author} {\bibfnamefont {S.}~\bibnamefont
  {Fritsch}}, \bibinfo {author} {\bibfnamefont {N.}~\bibnamefont {Kaiser}}, \
  and\ \bibinfo {author} {\bibfnamefont {W.}~\bibnamefont {Weise}},\ }\href
  {\doibase 10.1016/S0370-2693(02)02559-5} {\bibfield  {journal} {\bibinfo
  {journal} {Phys. Lett. B}\ }\textbf {\bibinfo {volume} {545}},\ \bibinfo
  {pages} {73} (\bibinfo {year} {2002})}\BibitemShut {NoStop}%
%%CITATION = NUCL-TH/0202005;%%
\bibitem [{\citenamefont {Holt}\ \emph {et~al.}(2010)\citenamefont {Holt},
  \citenamefont {Kaiser},\ and\ \citenamefont {Weise}}]{Holt:2009ty}%
  \BibitemOpen
  \bibfield  {author} {\bibinfo {author} {\bibfnamefont {J.~W.}\ \bibnamefont
  {Holt}}, \bibinfo {author} {\bibfnamefont {N.}~\bibnamefont {Kaiser}}, \ and\
  \bibinfo {author} {\bibfnamefont {W.}~\bibnamefont {Weise}},\ }\href
  {\doibase 10.1103/PhysRevC.81.024002} {\bibfield  {journal} {\bibinfo
  {journal} {Phys. Rev. C}\ }\textbf {\bibinfo {volume} {81}},\ \bibinfo
  {pages} {024002} (\bibinfo {year} {2010})}\BibitemShut {NoStop}%
%%CITATION = ARXIV:0910.1249;%%
\bibitem [{\citenamefont {Hebeler}\ and\ \citenamefont
  {Schwenk}(2010)}]{PhysRevC.82.014314}%
  \BibitemOpen
  \bibfield  {author} {\bibinfo {author} {\bibfnamefont {K.}~\bibnamefont
  {Hebeler}}\ and\ \bibinfo {author} {\bibfnamefont {A.}~\bibnamefont
  {Schwenk}},\ }\href {\doibase 10.1103/PhysRevC.82.014314} {\bibfield
  {journal} {\bibinfo  {journal} {Phys. Rev. C}\ }\textbf {\bibinfo {volume}
  {82}},\ \bibinfo {pages} {014314} (\bibinfo {year} {2010})}\BibitemShut
  {NoStop}%
\bibitem [{\citenamefont {Hebeler}\ \emph {et~al.}(2011)\citenamefont
  {Hebeler}, \citenamefont {Bogner}, \citenamefont {Furnstahl}, \citenamefont
  {Nogga},\ and\ \citenamefont {Schwenk}}]{PhysRevC.83.031301}%
  \BibitemOpen
  \bibfield  {author} {\bibinfo {author} {\bibfnamefont {K.}~\bibnamefont
  {Hebeler}}, \bibinfo {author} {\bibfnamefont {S.~K.}\ \bibnamefont {Bogner}},
  \bibinfo {author} {\bibfnamefont {R.~J.}\ \bibnamefont {Furnstahl}}, \bibinfo
  {author} {\bibfnamefont {A.}~\bibnamefont {Nogga}}, \ and\ \bibinfo {author}
  {\bibfnamefont {A.}~\bibnamefont {Schwenk}},\ }\href {\doibase
  10.1103/PhysRevC.83.031301} {\bibfield  {journal} {\bibinfo  {journal} {Phys.
  Rev. C}\ }\textbf {\bibinfo {volume} {83}},\ \bibinfo {pages} {031301}
  (\bibinfo {year} {2011})}\BibitemShut {NoStop}%
\bibitem [{\citenamefont {Holt}\ \emph {et~al.}(2009)\citenamefont {Holt},
  \citenamefont {Kaiser},\ and\ \citenamefont {Weise}}]{PhysRevC.79.054331}%
  \BibitemOpen
  \bibfield  {author} {\bibinfo {author} {\bibfnamefont {J.~W.}\ \bibnamefont
  {Holt}}, \bibinfo {author} {\bibfnamefont {N.}~\bibnamefont {Kaiser}}, \ and\
  \bibinfo {author} {\bibfnamefont {W.}~\bibnamefont {Weise}},\ }\href
  {\doibase 10.1103/PhysRevC.79.054331} {\bibfield  {journal} {\bibinfo
  {journal} {Phys. Rev. C}\ }\textbf {\bibinfo {volume} {79}},\ \bibinfo
  {pages} {054331} (\bibinfo {year} {2009})}\BibitemShut {NoStop}%
\bibitem [{\citenamefont {Coraggio}\ \emph {et~al.}(2007)\citenamefont
  {Coraggio}, \citenamefont {Covello}, \citenamefont {Gargano}, \citenamefont
  {Itaco}, \citenamefont {Entem} \emph {et~al.}}]{Coraggio:2007mc}%
  \BibitemOpen
  \bibfield  {author} {\bibinfo {author} {\bibfnamefont {L.}~\bibnamefont
  {Coraggio}}, \bibinfo {author} {\bibfnamefont {A.}~\bibnamefont {Covello}},
  \bibinfo {author} {\bibfnamefont {A.}~\bibnamefont {Gargano}}, \bibinfo
  {author} {\bibfnamefont {N.}~\bibnamefont {Itaco}}, \bibinfo {author}
  {\bibfnamefont {D.}~\bibnamefont {Entem}},  \emph {et~al.},\ }\href {\doibase
  10.1103/PhysRevC.75.024311} {\bibfield  {journal} {\bibinfo  {journal} {Phys.
  Rev. C}\ }\textbf {\bibinfo {volume} {75}},\ \bibinfo {pages} {024311}
  (\bibinfo {year} {2007})}\BibitemShut {NoStop}%
%%CITATION = NUCL-TH/0701065;%%
\bibitem [{\citenamefont {Coraggio}\ \emph {et~al.}(2013)\citenamefont
  {Coraggio}, \citenamefont {Holt}, \citenamefont {Itaco}, \citenamefont
  {Machleidt},\ and\ \citenamefont {Sammarruca}}]{Coraggio13}%
  \BibitemOpen
  \bibfield  {author} {\bibinfo {author} {\bibfnamefont {L.}~\bibnamefont
  {Coraggio}}, \bibinfo {author} {\bibfnamefont {J.~W.}\ \bibnamefont {Holt}},
  \bibinfo {author} {\bibfnamefont {N.}~\bibnamefont {Itaco}}, \bibinfo
  {author} {\bibfnamefont {R.}~\bibnamefont {Machleidt}}, \ and\ \bibinfo
  {author} {\bibfnamefont {F.}~\bibnamefont {Sammarruca}},\ }\href {\doibase
  10.1103/PhysRevC.87.014322} {\bibfield  {journal} {\bibinfo  {journal} {Phys.
  Rev. C}\ }\textbf {\bibinfo {volume} {87}},\ \bibinfo {pages} {014322}
  (\bibinfo {year} {2013})}\BibitemShut {NoStop}%
\bibitem [{\citenamefont {Brockmann}\ and\ \citenamefont
  {Machleidt}(1990)}]{Brockmann:1990cn}%
  \BibitemOpen
  \bibfield  {author} {\bibinfo {author} {\bibfnamefont {R.}~\bibnamefont
  {Brockmann}}\ and\ \bibinfo {author} {\bibfnamefont {R.}~\bibnamefont
  {Machleidt}},\ }\href {\doibase 10.1103/PhysRevC.42.1965} {\bibfield
  {journal} {\bibinfo  {journal} {Phys. Rev. C}\ }\textbf {\bibinfo {volume}
  {42}},\ \bibinfo {pages} {1965} (\bibinfo {year} {1990})}\BibitemShut
  {NoStop}%
%%CITATION = PHRVA,C42,1965;%%
\bibitem [{\citenamefont {Blaizot}(1980)}]{Blaizot}%
  \BibitemOpen
  \bibfield  {author} {\bibinfo {author} {\bibfnamefont {J.~P.}\ \bibnamefont
  {Blaizot}},\ }\href {\doibase 10.1016/0370-1573(80)90001-0} {\bibfield
  {journal} {\bibinfo  {journal} {Phys. Rept.}\ }\textbf {\bibinfo {volume}
  {64}},\ \bibinfo {pages} {171} (\bibinfo {year} {1980})}\BibitemShut
  {NoStop}%
\bibitem [{\citenamefont {Khoa}\ \emph {et~al.}(1997)\citenamefont {Khoa},
  \citenamefont {Satchler},\ and\ \citenamefont {von Oertzen}}]{Khoa:1997zz}%
  \BibitemOpen
  \bibfield  {author} {\bibinfo {author} {\bibfnamefont {D.~T.}\ \bibnamefont
  {Khoa}}, \bibinfo {author} {\bibfnamefont {G.~R.}\ \bibnamefont {Satchler}},
  \ and\ \bibinfo {author} {\bibfnamefont {W.}~\bibnamefont {von Oertzen}},\
  }\href {\doibase 10.1103/PhysRevC.56.954} {\bibfield  {journal} {\bibinfo
  {journal} {Phys. Rev. C}\ }\textbf {\bibinfo {volume} {56}},\ \bibinfo
  {pages} {954} (\bibinfo {year} {1997})}\BibitemShut {NoStop}%
%%CITATION = PHRVA,C56,954;%%
\bibitem [{\citenamefont {Youngblood}\ \emph {et~al.}(1999)\citenamefont
  {Youngblood}, \citenamefont {Clark},\ and\ \citenamefont
  {Lui}}]{Youngblood:1999zza}%
  \BibitemOpen
  \bibfield  {author} {\bibinfo {author} {\bibfnamefont {D.~H.}\ \bibnamefont
  {Youngblood}}, \bibinfo {author} {\bibfnamefont {H.~L.}\ \bibnamefont
  {Clark}}, \ and\ \bibinfo {author} {\bibfnamefont {Y.-W.}\ \bibnamefont
  {Lui}},\ }\href {\doibase 10.1103/PhysRevLett.82.691} {\bibfield  {journal}
  {\bibinfo  {journal} {Phys. Rev. Lett.}\ }\textbf {\bibinfo {volume} {82}},\
  \bibinfo {pages} {691} (\bibinfo {year} {1999})}\BibitemShut {NoStop}%
%%CITATION = PRLTA,82,691;%%
\bibitem [{\citenamefont {Karnaukhov}\ \emph {et~al.}(2008)\citenamefont
  {Karnaukhov}, \citenamefont {Oeschler}, \citenamefont {Budzanowski},
  \citenamefont {Avdeyev}, \citenamefont {Botvina} \emph
  {et~al.}}]{Karnaukhov:2008be}%
  \BibitemOpen
  \bibfield  {author} {\bibinfo {author} {\bibfnamefont {V.~A.}\ \bibnamefont
  {Karnaukhov}}, \bibinfo {author} {\bibfnamefont {H.}~\bibnamefont
  {Oeschler}}, \bibinfo {author} {\bibfnamefont {A.}~\bibnamefont
  {Budzanowski}}, \bibinfo {author} {\bibfnamefont {S.~P.}\ \bibnamefont
  {Avdeyev}}, \bibinfo {author} {\bibfnamefont {A.~S.}\ \bibnamefont
  {Botvina}},  \emph {et~al.},\ }\href {\doibase 10.1134/S1063778808120077}
  {\bibfield  {journal} {\bibinfo  {journal} {Phys. Atom. Nucl.}\ }\textbf
  {\bibinfo {volume} {71}},\ \bibinfo {pages} {2067} (\bibinfo {year}
  {2008})}\BibitemShut {NoStop}%
\bibitem [{\citenamefont {Natowitz}\ \emph {et~al.}(2002)\citenamefont
  {Natowitz}, \citenamefont {Hagel}, \citenamefont {Ma}, \citenamefont
  {Murray}, \citenamefont {Qin}, \citenamefont {Wada},\ and\ \citenamefont
  {Wang}}]{PhysRevLett.89.212701}%
  \BibitemOpen
  \bibfield  {author} {\bibinfo {author} {\bibfnamefont {J.~B.}\ \bibnamefont
  {Natowitz}}, \bibinfo {author} {\bibfnamefont {K.}~\bibnamefont {Hagel}},
  \bibinfo {author} {\bibfnamefont {Y.}~\bibnamefont {Ma}}, \bibinfo {author}
  {\bibfnamefont {M.}~\bibnamefont {Murray}}, \bibinfo {author} {\bibfnamefont
  {L.}~\bibnamefont {Qin}}, \bibinfo {author} {\bibfnamefont {R.}~\bibnamefont
  {Wada}}, \ and\ \bibinfo {author} {\bibfnamefont {J.}~\bibnamefont {Wang}},\
  }\href {\doibase 10.1103/PhysRevLett.89.212701} {\bibfield  {journal}
  {\bibinfo  {journal} {Phys. Rev. Lett.}\ }\textbf {\bibinfo {volume} {89}},\
  \bibinfo {pages} {212701} (\bibinfo {year} {2002})}\BibitemShut {NoStop}%
\bibitem [{\citenamefont {Ogul}\ and\ \citenamefont
  {Botvina}(2002)}]{Ogul:2002ka}%
  \BibitemOpen
  \bibfield  {author} {\bibinfo {author} {\bibfnamefont {R.}~\bibnamefont
  {Ogul}}\ and\ \bibinfo {author} {\bibfnamefont {A.~S.}\ \bibnamefont
  {Botvina}},\ }\href {\doibase 10.1103/PhysRevC.66.051601} {\bibfield
  {journal} {\bibinfo  {journal} {Phys. Rev. C}\ }\textbf {\bibinfo {volume}
  {66}},\ \bibinfo {pages} {051601} (\bibinfo {year} {2002})}\BibitemShut
  {NoStop}%
%%CITATION = NUCL-TH/0210045;%%
\bibitem [{\citenamefont {Elliott}\ \emph {et~al.}(2013)\citenamefont
  {Elliott}, \citenamefont {Lake}, \citenamefont {Moretto},\ and\ \citenamefont
  {Phair}}]{Elliott:2013pna}%
  \BibitemOpen
  \bibfield  {author} {\bibinfo {author} {\bibfnamefont {J.~B.}\ \bibnamefont
  {Elliott}}, \bibinfo {author} {\bibfnamefont {P.~T.}\ \bibnamefont {Lake}},
  \bibinfo {author} {\bibfnamefont {L.~G.}\ \bibnamefont {Moretto}}, \ and\
  \bibinfo {author} {\bibfnamefont {L.}~\bibnamefont {Phair}},\ }\href
  {\doibase 10.1103/PhysRevC.87.054622} {\bibfield  {journal} {\bibinfo
  {journal} {Phys. Rev. C}\ }\textbf {\bibinfo {volume} {87}},\ \bibinfo
  {pages} {054622} (\bibinfo {year} {2013})}\BibitemShut {NoStop}%
\bibitem [{\citenamefont {de~Oliveira}(2013)}]{bookZc}%
  \BibitemOpen
  \bibfield  {author} {\bibinfo {author} {\bibfnamefont {M.~J.}\ \bibnamefont
  {de~Oliveira}},\ }\href@noop {} {\emph {\bibinfo {title} {Equilibrium
  Thermodynamics}}}\ (\bibinfo  {publisher} {Springer, Heidelberg},\ \bibinfo
  {year} {2013})\BibitemShut {NoStop}%
\bibitem [{\citenamefont {Quintales}\ \emph {et~al.}(1988)\citenamefont
  {Quintales}, \citenamefont {Alvarino},\ and\ \citenamefont
  {Velasco}}]{0143-0807-9-1-010}%
  \BibitemOpen
  \bibfield  {author} {\bibinfo {author} {\bibfnamefont {L.~M.}\ \bibnamefont
  {Quintales}}, \bibinfo {author} {\bibfnamefont {J.~M.}\ \bibnamefont
  {Alvarino}}, \ and\ \bibinfo {author} {\bibfnamefont {S.}~\bibnamefont
  {Velasco}},\ }\href {\doibase 10.1088/0143-0807/9/1/010} {\bibfield
  {journal} {\bibinfo  {journal} {Eur. J. Phys.}\ }\textbf {\bibinfo {volume}
  {9}},\ \bibinfo {pages} {55} (\bibinfo {year} {1988})}\BibitemShut {NoStop}%
\bibitem [{\citenamefont {Debenedetti}(1996)}]{bookML}%
  \BibitemOpen
  \bibfield  {author} {\bibinfo {author} {\bibfnamefont {P.~G.}\ \bibnamefont
  {Debenedetti}},\ }\href@noop {} {\emph {\bibinfo {title} {Metastable Liquids:
  Concepts and Principles}}}\ (\bibinfo  {publisher} {Princeton University
  Press, Princeton},\ \bibinfo {year} {1996})\BibitemShut {NoStop}%
\bibitem [{\citenamefont {Horowitz}\ and\ \citenamefont
  {Schwenk}(2006{\natexlab{b}})}]{Horowitz:2005zv}%
  \BibitemOpen
  \bibfield  {author} {\bibinfo {author} {\bibfnamefont {C.~J.}\ \bibnamefont
  {Horowitz}}\ and\ \bibinfo {author} {\bibfnamefont {A.}~\bibnamefont
  {Schwenk}},\ }\href {\doibase 10.1016/j.physletb.2006.05.055} {\bibfield
  {journal} {\bibinfo  {journal} {Phys. Lett. B}\ }\textbf {\bibinfo {volume}
  {638}},\ \bibinfo {pages} {153} (\bibinfo {year}
  {2006}{\natexlab{b}})}\BibitemShut {NoStop}%
%%CITATION = NUCL-TH/0507064;%%
\bibitem [{\citenamefont {Rrapaj}\ \emph {et~al.}(2015)\citenamefont {Rrapaj},
  \citenamefont {Holt}, \citenamefont {Bartl}, \citenamefont {Reddy},\ and\
  \citenamefont {Schwenk}}]{Rrapaj:2014yba}%
  \BibitemOpen
  \bibfield  {author} {\bibinfo {author} {\bibfnamefont {E.}~\bibnamefont
  {Rrapaj}}, \bibinfo {author} {\bibfnamefont {J.~W.}\ \bibnamefont {Holt}},
  \bibinfo {author} {\bibfnamefont {A.}~\bibnamefont {Bartl}}, \bibinfo
  {author} {\bibfnamefont {S.}~\bibnamefont {Reddy}}, \ and\ \bibinfo {author}
  {\bibfnamefont {A.}~\bibnamefont {Schwenk}},\ }\href {\doibase
  10.1103/PhysRevC.91.035806} {\bibfield  {journal} {\bibinfo  {journal} {Phys.
  Rev. C}\ }\textbf {\bibinfo {volume} {91}},\ \bibinfo {pages} {035806}
  (\bibinfo {year} {2015})}\BibitemShut {NoStop}%
\bibitem [{\citenamefont {Kr\"{u}ger}\ \emph {et~al.}(2013)\citenamefont
  {Kr\"{u}ger}, \citenamefont {Tews}, \citenamefont {Hebeler},\ and\
  \citenamefont {Schwenk}}]{Kruger:2013kua}%
  \BibitemOpen
  \bibfield  {author} {\bibinfo {author} {\bibfnamefont {T.}~\bibnamefont
  {Kr\"{u}ger}}, \bibinfo {author} {\bibfnamefont {I.}~\bibnamefont {Tews}},
  \bibinfo {author} {\bibfnamefont {K.}~\bibnamefont {Hebeler}}, \ and\
  \bibinfo {author} {\bibfnamefont {A.}~\bibnamefont {Schwenk}},\ }\href
  {\doibase 10.1103/PhysRevC.88.025802} {\bibfield  {journal} {\bibinfo
  {journal} {Phys. Rev. C}\ }\textbf {\bibinfo {volume} {88}},\ \bibinfo
  {pages} {025802} (\bibinfo {year} {2013})}\BibitemShut {NoStop}%
\bibitem [{\citenamefont {Wlaz{\l}owski}\ \emph {et~al.}(2014)\citenamefont
  {Wlaz{\l}owski}, \citenamefont {Holt}, \citenamefont {Moroz}, \citenamefont
  {Bulgac},\ and\ \citenamefont {Roche}}]{Wlazlowski2014}%
  \BibitemOpen
  \bibfield  {author} {\bibinfo {author} {\bibfnamefont {G.}~\bibnamefont
  {Wlaz{\l}owski}}, \bibinfo {author} {\bibfnamefont {J.~W.}\ \bibnamefont
  {Holt}}, \bibinfo {author} {\bibfnamefont {S.}~\bibnamefont {Moroz}},
  \bibinfo {author} {\bibfnamefont {A.}~\bibnamefont {Bulgac}}, \ and\ \bibinfo
  {author} {\bibfnamefont {K.~J.}\ \bibnamefont {Roche}},\ }\href {\doibase
  10.1103/PhysRevLett.113.182503} {\bibfield  {journal} {\bibinfo  {journal}
  {Phys. Rev. Lett.}\ }\textbf {\bibinfo {volume} {113}},\ \bibinfo {pages}
  {182503} (\bibinfo {year} {2014})}\BibitemShut {NoStop}%
%%CITATION = ARXIV:1403.3753;%%
\bibitem [{\citenamefont {Hebeler}\ and\ \citenamefont
  {Schwenk}(2014)}]{Hebeler:2014ema}%
  \BibitemOpen
  \bibfield  {author} {\bibinfo {author} {\bibfnamefont {K.}~\bibnamefont
  {Hebeler}}\ and\ \bibinfo {author} {\bibfnamefont {A.}~\bibnamefont
  {Schwenk}},\ }\href {\doibase 10.1140/epja/i2014-14011-4} {\bibfield
  {journal} {\bibinfo  {journal} {Eur. Phys. J. A}\ }\textbf {\bibinfo {volume}
  {50}},\ \bibinfo {pages} {11} (\bibinfo {year} {2014})}\BibitemShut {NoStop}%
%%CITATION = ARXIV:1401.5822;%%
\bibitem [{\citenamefont {Roggero}\ \emph {et~al.}(2014)\citenamefont
  {Roggero}, \citenamefont {Mukherjee},\ and\ \citenamefont
  {Pederiva}}]{PhysRevLett.112.221103}%
  \BibitemOpen
  \bibfield  {author} {\bibinfo {author} {\bibfnamefont {A.}~\bibnamefont
  {Roggero}}, \bibinfo {author} {\bibfnamefont {A.}~\bibnamefont {Mukherjee}},
  \ and\ \bibinfo {author} {\bibfnamefont {F.}~\bibnamefont {Pederiva}},\
  }\href {\doibase 10.1103/PhysRevLett.112.221103} {\bibfield  {journal}
  {\bibinfo  {journal} {Phys. Rev. Lett.}\ }\textbf {\bibinfo {volume} {112}},\
  \bibinfo {pages} {221103} (\bibinfo {year} {2014})}\BibitemShut {NoStop}%
\bibitem [{\citenamefont {Gezerlis}\ \emph {et~al.}(2013)\citenamefont
  {Gezerlis}, \citenamefont {Tews}, \citenamefont {Epelbaum}, \citenamefont
  {Gandolfi}, \citenamefont {Hebeler}, \citenamefont {Nogga},\ and\
  \citenamefont {Schwenk}}]{Gezerlis13}%
  \BibitemOpen
  \bibfield  {author} {\bibinfo {author} {\bibfnamefont {A.}~\bibnamefont
  {Gezerlis}}, \bibinfo {author} {\bibfnamefont {I.}~\bibnamefont {Tews}},
  \bibinfo {author} {\bibfnamefont {E.}~\bibnamefont {Epelbaum}}, \bibinfo
  {author} {\bibfnamefont {S.}~\bibnamefont {Gandolfi}}, \bibinfo {author}
  {\bibfnamefont {K.}~\bibnamefont {Hebeler}}, \bibinfo {author} {\bibfnamefont
  {A.}~\bibnamefont {Nogga}}, \ and\ \bibinfo {author} {\bibfnamefont
  {A.}~\bibnamefont {Schwenk}},\ }\href {\doibase
  10.1103/PhysRevLett.111.032501} {\bibfield  {journal} {\bibinfo  {journal}
  {Phys. Rev. Lett.}\ }\textbf {\bibinfo {volume} {111}},\ \bibinfo {pages}
  {032501} (\bibinfo {year} {2013})}\BibitemShut {NoStop}%
\bibitem [{\citenamefont {Hagen}\ \emph {et~al.}(2014)\citenamefont {Hagen},
  \citenamefont {Papenbrock}, \citenamefont {Ekstr\"{o}m}, \citenamefont
  {Wendt}, \citenamefont {Baardsen}, \citenamefont {Gandolfi}, \citenamefont
  {Hjorth-Jensen},\ and\ \citenamefont {Horowitz}}]{Hagen2014}%
  \BibitemOpen
  \bibfield  {author} {\bibinfo {author} {\bibfnamefont {G.}~\bibnamefont
  {Hagen}}, \bibinfo {author} {\bibfnamefont {T.}~\bibnamefont {Papenbrock}},
  \bibinfo {author} {\bibfnamefont {A.}~\bibnamefont {Ekstr\"{o}m}}, \bibinfo
  {author} {\bibfnamefont {K.~A.}\ \bibnamefont {Wendt}}, \bibinfo {author}
  {\bibfnamefont {G.}~\bibnamefont {Baardsen}}, \bibinfo {author}
  {\bibfnamefont {S.}~\bibnamefont {Gandolfi}}, \bibinfo {author}
  {\bibfnamefont {M.}~\bibnamefont {Hjorth-Jensen}}, \ and\ \bibinfo {author}
  {\bibfnamefont {C.~J.}\ \bibnamefont {Horowitz}},\ }\href {\doibase
  10.1103/PhysRevC.89.014319} {\bibfield  {journal} {\bibinfo  {journal} {Phys.
  Rev. C}\ }\textbf {\bibinfo {volume} {89}},\ \bibinfo {pages} {014319}
  (\bibinfo {year} {2014})}\BibitemShut {NoStop}%
\bibitem [{\citenamefont {Carbone}\ \emph {et~al.}(2014)\citenamefont
  {Carbone}, \citenamefont {Rios},\ and\ \citenamefont
  {Polls}}]{PhysRevC.90.054322}%
  \BibitemOpen
  \bibfield  {author} {\bibinfo {author} {\bibfnamefont {A.}~\bibnamefont
  {Carbone}}, \bibinfo {author} {\bibfnamefont {A.}~\bibnamefont {Rios}}, \
  and\ \bibinfo {author} {\bibfnamefont {A.}~\bibnamefont {Polls}},\ }\href
  {\doibase 10.1103/PhysRevC.90.054322} {\bibfield  {journal} {\bibinfo
  {journal} {Phys. Rev. C}\ }\textbf {\bibinfo {volume} {90}},\ \bibinfo
  {pages} {054322} (\bibinfo {year} {2014})}\BibitemShut {NoStop}%
\bibitem [{\citenamefont {Baardsen}\ \emph {et~al.}(2013)\citenamefont
  {Baardsen}, \citenamefont {Ekstr\"om}, \citenamefont {Hagen},\ and\
  \citenamefont {Hjorth-Jensen}}]{PhysRevC.88.054312}%
  \BibitemOpen
  \bibfield  {author} {\bibinfo {author} {\bibfnamefont {G.}~\bibnamefont
  {Baardsen}}, \bibinfo {author} {\bibfnamefont {A.}~\bibnamefont {Ekstr\"om}},
  \bibinfo {author} {\bibfnamefont {G.}~\bibnamefont {Hagen}}, \ and\ \bibinfo
  {author} {\bibfnamefont {M.}~\bibnamefont {Hjorth-Jensen}},\ }\href {\doibase
  10.1103/PhysRevC.88.054312} {\bibfield  {journal} {\bibinfo  {journal} {Phys.
  Rev. C}\ }\textbf {\bibinfo {volume} {88}},\ \bibinfo {pages} {054312}
  (\bibinfo {year} {2013})}\BibitemShut {NoStop}%
\bibitem [{\citenamefont {Kohno}(2013)}]{PhysRevC.88.064005}%
  \BibitemOpen
  \bibfield  {author} {\bibinfo {author} {\bibfnamefont {M.}~\bibnamefont
  {Kohno}},\ }\href {\doibase 10.1103/PhysRevC.88.064005} {\bibfield  {journal}
  {\bibinfo  {journal} {Phys. Rev. C}\ }\textbf {\bibinfo {volume} {88}},\
  \bibinfo {pages} {064005} (\bibinfo {year} {2013})}\BibitemShut {NoStop}%
\bibitem [{\citenamefont {Gezerlis}\ \emph {et~al.}(2014)\citenamefont
  {Gezerlis}, \citenamefont {Tews}, \citenamefont {Epelbaum}, \citenamefont
  {Freunek}, \citenamefont {Gandolfi} \emph {et~al.}}]{Gezerlis:2014zia}%
  \BibitemOpen
  \bibfield  {author} {\bibinfo {author} {\bibfnamefont {A.}~\bibnamefont
  {Gezerlis}}, \bibinfo {author} {\bibfnamefont {I.}~\bibnamefont {Tews}},
  \bibinfo {author} {\bibfnamefont {E.}~\bibnamefont {Epelbaum}}, \bibinfo
  {author} {\bibfnamefont {M.}~\bibnamefont {Freunek}}, \bibinfo {author}
  {\bibfnamefont {S.}~\bibnamefont {Gandolfi}},  \emph {et~al.},\ }\href
  {\doibase 10.1103/PhysRevC.90.054323} {\bibfield  {journal} {\bibinfo
  {journal} {Phys. Rev. C}\ }\textbf {\bibinfo {volume} {90}},\ \bibinfo
  {pages} {054323} (\bibinfo {year} {2014})}\BibitemShut {NoStop}%
%%CITATION = ARXIV:1406.0454;%%
\bibitem [{\citenamefont {Gezerlis}\ and\ \citenamefont
  {Carlson}(2008)}]{Gezerlis:2007fs}%
  \BibitemOpen
  \bibfield  {author} {\bibinfo {author} {\bibfnamefont {A.}~\bibnamefont
  {Gezerlis}}\ and\ \bibinfo {author} {\bibfnamefont {J.}~\bibnamefont
  {Carlson}},\ }\href {\doibase 10.1103/PhysRevC.77.032801} {\bibfield
  {journal} {\bibinfo  {journal} {Phys. Rev. C}\ }\textbf {\bibinfo {volume}
  {77}},\ \bibinfo {pages} {032801} (\bibinfo {year} {2008})}\BibitemShut
  {NoStop}%
%%CITATION = ARXIV:0711.3006;%%
\bibitem [{\citenamefont {Akmal}\ \emph {et~al.}(1998)\citenamefont {Akmal},
  \citenamefont {Pandharipande},\ and\ \citenamefont
  {Ravenhall}}]{Akmal:1998cf}%
  \BibitemOpen
  \bibfield  {author} {\bibinfo {author} {\bibfnamefont {A.}~\bibnamefont
  {Akmal}}, \bibinfo {author} {\bibfnamefont {V.~R.}\ \bibnamefont
  {Pandharipande}}, \ and\ \bibinfo {author} {\bibfnamefont {D.~G.}\
  \bibnamefont {Ravenhall}},\ }\href {\doibase 10.1103/PhysRevC.58.1804}
  {\bibfield  {journal} {\bibinfo  {journal} {Phys. Rev. C}\ }\textbf {\bibinfo
  {volume} {58}},\ \bibinfo {pages} {1804} (\bibinfo {year}
  {1998})}\BibitemShut {NoStop}%
%%CITATION = NUCL-TH/9804027;%%
\bibitem [{\citenamefont {Danielewicz}\ and\ \citenamefont
  {Lee}(2014)}]{Danielewicz20141}%
  \BibitemOpen
  \bibfield  {author} {\bibinfo {author} {\bibfnamefont {P.}~\bibnamefont
  {Danielewicz}}\ and\ \bibinfo {author} {\bibfnamefont {J.}~\bibnamefont
  {Lee}},\ }\href {\doibase http://dx.doi.org/10.1016/j.nuclphysa.2013.11.005}
  {\bibfield  {journal} {\bibinfo  {journal} {Nucl. Phys. A}\ }\textbf
  {\bibinfo {volume} {922}},\ \bibinfo {pages} {1 } (\bibinfo {year}
  {2014})}\BibitemShut {NoStop}%
\bibitem [{\citenamefont {Chen}\ \emph {et~al.}(2009)\citenamefont {Chen},
  \citenamefont {Cai}, \citenamefont {Ko}, \citenamefont {Li}, \citenamefont
  {Shen},\ and\ \citenamefont {Xu}}]{PhysRevC.80.014322}%
  \BibitemOpen
  \bibfield  {author} {\bibinfo {author} {\bibfnamefont {L.-W.}\ \bibnamefont
  {Chen}}, \bibinfo {author} {\bibfnamefont {B.-J.}\ \bibnamefont {Cai}},
  \bibinfo {author} {\bibfnamefont {C.~M.}\ \bibnamefont {Ko}}, \bibinfo
  {author} {\bibfnamefont {B.-A.}\ \bibnamefont {Li}}, \bibinfo {author}
  {\bibfnamefont {C.}~\bibnamefont {Shen}}, \ and\ \bibinfo {author}
  {\bibfnamefont {J.}~\bibnamefont {Xu}},\ }\href {\doibase
  10.1103/PhysRevC.80.014322} {\bibfield  {journal} {\bibinfo  {journal} {Phys.
  Rev. C}\ }\textbf {\bibinfo {volume} {80}},\ \bibinfo {pages} {014322}
  (\bibinfo {year} {2009})}\BibitemShut {NoStop}%
\bibitem [{\citenamefont {Lattimer}\ and\ \citenamefont
  {Lim}(2013)}]{Lattimer:2012xj}%
  \BibitemOpen
  \bibfield  {author} {\bibinfo {author} {\bibfnamefont {J.~M.}\ \bibnamefont
  {Lattimer}}\ and\ \bibinfo {author} {\bibfnamefont {Y.}~\bibnamefont {Lim}},\
  }\href {\doibase 10.1088/0004-637X/771/1/51} {\bibfield  {journal} {\bibinfo
  {journal} {Astrophys. J.}\ }\textbf {\bibinfo {volume} {771}},\ \bibinfo
  {pages} {51} (\bibinfo {year} {2013})}\BibitemShut {NoStop}%
%%CITATION = ARXIV:1203.4286;%%
\bibitem [{\citenamefont {Newton}\ \emph {et~al.}(2013)\citenamefont {Newton},
  \citenamefont {Gearheart}, \citenamefont {Wen},\ and\ \citenamefont
  {Li}}]{Newton:2012ry}%
  \BibitemOpen
  \bibfield  {author} {\bibinfo {author} {\bibfnamefont {W.~G.}\ \bibnamefont
  {Newton}}, \bibinfo {author} {\bibfnamefont {M.}~\bibnamefont {Gearheart}},
  \bibinfo {author} {\bibfnamefont {D.-H.}\ \bibnamefont {Wen}}, \ and\
  \bibinfo {author} {\bibfnamefont {B.-A.}\ \bibnamefont {Li}},\ }\href
  {\doibase 10.1088/1742-6596/420/1/012145} {\bibfield  {journal} {\bibinfo
  {journal} {J. Phys. Conf. Ser.}\ }\textbf {\bibinfo {volume} {420}},\
  \bibinfo {pages} {012145} (\bibinfo {year} {2013})}\BibitemShut {NoStop}%
%%CITATION = ARXIV:1212.4539;%%
\bibitem [{\citenamefont {Newton}\ \emph {et~al.}(2014)\citenamefont {Newton},
  \citenamefont {Hooker}, \citenamefont {Gearheart}, \citenamefont {Murphy},
  \citenamefont {Wen} \emph {et~al.}}]{Newton:2014tga}%
  \BibitemOpen
  \bibfield  {author} {\bibinfo {author} {\bibfnamefont {W.~G.}\ \bibnamefont
  {Newton}}, \bibinfo {author} {\bibfnamefont {J.}~\bibnamefont {Hooker}},
  \bibinfo {author} {\bibfnamefont {M.}~\bibnamefont {Gearheart}}, \bibinfo
  {author} {\bibfnamefont {K.}~\bibnamefont {Murphy}}, \bibinfo {author}
  {\bibfnamefont {D.-H.}\ \bibnamefont {Wen}},  \emph {et~al.},\ }\href
  {\doibase 10.1140/epja/i2014-14041-x} {\bibfield  {journal} {\bibinfo
  {journal} {Eur. Phys. J. A}\ }\textbf {\bibinfo {volume} {50}},\ \bibinfo
  {pages} {41} (\bibinfo {year} {2014})}\BibitemShut {NoStop}%
%%CITATION = EPHJA,A50,41;%%
\bibitem [{\citenamefont {Lattimer}\ and\ \citenamefont
  {Steiner}(2014)}]{Lattimer:2014sga}%
  \BibitemOpen
  \bibfield  {author} {\bibinfo {author} {\bibfnamefont {J.~M.}\ \bibnamefont
  {Lattimer}}\ and\ \bibinfo {author} {\bibfnamefont {A.~W.}\ \bibnamefont
  {Steiner}},\ }\href {\doibase 10.1140/epja/i2014-14040-y} {\bibfield
  {journal} {\bibinfo  {journal} {Eur. Phys. J. A}\ }\textbf {\bibinfo {volume}
  {50}},\ \bibinfo {pages} {40} (\bibinfo {year} {2014})}\BibitemShut {NoStop}%
%%CITATION = ARXIV:1403.1186;%%
\bibitem [{\citenamefont {Centelles}\ \emph {et~al.}(2009)\citenamefont
  {Centelles}, \citenamefont {Roca-Maza}, \citenamefont {Vi\~nas},\ and\
  \citenamefont {Warda}}]{PhysRevLett.102.122502}%
  \BibitemOpen
  \bibfield  {author} {\bibinfo {author} {\bibfnamefont {M.}~\bibnamefont
  {Centelles}}, \bibinfo {author} {\bibfnamefont {X.}~\bibnamefont
  {Roca-Maza}}, \bibinfo {author} {\bibfnamefont {X.}~\bibnamefont {Vi\~nas}},
  \ and\ \bibinfo {author} {\bibfnamefont {M.}~\bibnamefont {Warda}},\ }\href
  {\doibase 10.1103/PhysRevLett.102.122502} {\bibfield  {journal} {\bibinfo
  {journal} {Phys. Rev. Lett.}\ }\textbf {\bibinfo {volume} {102}},\ \bibinfo
  {pages} {122502} (\bibinfo {year} {2009})}\BibitemShut {NoStop}%
\bibitem [{\citenamefont {Li}\ \emph {et~al.}(2007)\citenamefont {Li},
  \citenamefont {Garg}, \citenamefont {Liu}, \citenamefont {Marks},
  \citenamefont {Nayak}, \citenamefont {Rao}, \citenamefont {Fujiwara} \emph
  {et~al.}}]{PhysRevLett.99.162503}%
  \BibitemOpen
  \bibfield  {author} {\bibinfo {author} {\bibfnamefont {T.}~\bibnamefont
  {Li}}, \bibinfo {author} {\bibfnamefont {U.}~\bibnamefont {Garg}}, \bibinfo
  {author} {\bibfnamefont {Y.}~\bibnamefont {Liu}}, \bibinfo {author}
  {\bibfnamefont {R.}~\bibnamefont {Marks}}, \bibinfo {author} {\bibfnamefont
  {B.~K.}\ \bibnamefont {Nayak}}, \bibinfo {author} {\bibfnamefont {P.~V.~M.}\
  \bibnamefont {Rao}}, \bibinfo {author} {\bibfnamefont {M.}~\bibnamefont
  {Fujiwara}},  \emph {et~al.},\ }\href {\doibase
  10.1103/PhysRevLett.99.162503} {\bibfield  {journal} {\bibinfo  {journal}
  {Phys. Rev. Lett.}\ }\textbf {\bibinfo {volume} {99}},\ \bibinfo {pages}
  {162503} (\bibinfo {year} {2007})}\BibitemShut {NoStop}%
\bibitem [{\citenamefont {Beegle}\ \emph {et~al.}(1974)\citenamefont {Beegle},
  \citenamefont {Modell},\ and\ \citenamefont {Reid}}]{beegle}%
  \BibitemOpen
  \bibfield  {author} {\bibinfo {author} {\bibfnamefont {B.~L.}\ \bibnamefont
  {Beegle}}, \bibinfo {author} {\bibfnamefont {M.}~\bibnamefont {Modell}}, \
  and\ \bibinfo {author} {\bibfnamefont {R.~C.}\ \bibnamefont {Reid}},\ }\href
  {\doibase http://dx.doi.org/10.1002/aic.690200621} {\bibfield  {journal}
  {\bibinfo  {journal} {Amer. Inst. Chem. Eng. J.}\ }\textbf {\bibinfo {volume}
  {20}},\ \bibinfo {pages} {1200} (\bibinfo {year} {1974})}\BibitemShut
  {NoStop}%
\bibitem [{\citenamefont {Beegle}\ \emph {et~al.}(1975)\citenamefont {Beegle},
  \citenamefont {Modell},\ and\ \citenamefont {Reid}}]{beegle2}%
  \BibitemOpen
  \bibfield  {author} {\bibinfo {author} {\bibfnamefont {B.~L.}\ \bibnamefont
  {Beegle}}, \bibinfo {author} {\bibfnamefont {M.}~\bibnamefont {Modell}}, \
  and\ \bibinfo {author} {\bibfnamefont {R.~C.}\ \bibnamefont {Reid}},\ }\href
  {\doibase http://dx.doi.org/10.1002/aic.690210434} {\bibfield  {journal}
  {\bibinfo  {journal} {Amer. Inst. Chem. Eng. J.}\ }\textbf {\bibinfo {volume}
  {21}},\ \bibinfo {pages} {826} (\bibinfo {year} {1975})}\BibitemShut
  {NoStop}%
\bibitem [{\citenamefont {Fedoseew}\ and\ \citenamefont
  {Lenske}(2015)}]{Fedoseew:2014cma}%
  \BibitemOpen
  \bibfield  {author} {\bibinfo {author} {\bibfnamefont {A.}~\bibnamefont
  {Fedoseew}}\ and\ \bibinfo {author} {\bibfnamefont {H.}~\bibnamefont
  {Lenske}},\ }\href {\doibase 10.1103/PhysRevC.91.034307} {\bibfield
  {journal} {\bibinfo  {journal} {Phys. Rev. C}\ }\textbf {\bibinfo {volume}
  {91}},\ \bibinfo {pages} {034307} (\bibinfo {year} {2015})}\BibitemShut
  {NoStop}%
%%CITATION = ARXIV:1407.2643;%%
\bibitem [{\citenamefont {Wu}\ and\ \citenamefont
  {Ren}(2011)}]{PhysRevC.83.044605}%
  \BibitemOpen
  \bibfield  {author} {\bibinfo {author} {\bibfnamefont {C.}~\bibnamefont
  {Wu}}\ and\ \bibinfo {author} {\bibfnamefont {Z.}~\bibnamefont {Ren}},\
  }\href {\doibase 10.1103/PhysRevC.83.044605} {\bibfield  {journal} {\bibinfo
  {journal} {Phys. Rev. C}\ }\textbf {\bibinfo {volume} {83}},\ \bibinfo
  {pages} {044605} (\bibinfo {year} {2011})}\BibitemShut {NoStop}%
\bibitem [{\citenamefont {Fiorilla}\ \emph {et~al.}(2012)\citenamefont
  {Fiorilla}, \citenamefont {Kaiser},\ and\ \citenamefont
  {Weise}}]{Fiorilla:2011sr}%
  \BibitemOpen
  \bibfield  {author} {\bibinfo {author} {\bibfnamefont {S.}~\bibnamefont
  {Fiorilla}}, \bibinfo {author} {\bibfnamefont {N.}~\bibnamefont {Kaiser}}, \
  and\ \bibinfo {author} {\bibfnamefont {W.}~\bibnamefont {Weise}},\ }\href
  {\doibase 10.1016/j.nuclphysa.2012.01.003} {\bibfield  {journal} {\bibinfo
  {journal} {Nucl. Phys. A}\ }\textbf {\bibinfo {volume} {880}},\ \bibinfo
  {pages} {65} (\bibinfo {year} {2012})}\BibitemShut {NoStop}%
%%CITATION = ARXIV:1111.2791;%%
\bibitem [{\citenamefont {Drews}\ and\ \citenamefont
  {Weise}(2015)}]{PhysRevC.91.035802}%
  \BibitemOpen
  \bibfield  {author} {\bibinfo {author} {\bibfnamefont {M.}~\bibnamefont
  {Drews}}\ and\ \bibinfo {author} {\bibfnamefont {W.}~\bibnamefont {Weise}},\
  }\href {\doibase 10.1103/PhysRevC.91.035802} {\bibfield  {journal} {\bibinfo
  {journal} {Phys. Rev. C}\ }\textbf {\bibinfo {volume} {91}},\ \bibinfo
  {pages} {035802} (\bibinfo {year} {2015})}\BibitemShut {NoStop}%
\bibitem [{\citenamefont {Lattimer}\ and\ \citenamefont
  {Ravenhall}(1978)}]{Lattimer:1978}%
  \BibitemOpen
  \bibfield  {author} {\bibinfo {author} {\bibfnamefont {J.~M.}\ \bibnamefont
  {Lattimer}}\ and\ \bibinfo {author} {\bibfnamefont {D.~G.}\ \bibnamefont
  {Ravenhall}},\ }\href {\doibase 10.1086/156265} {\bibfield  {journal}
  {\bibinfo  {journal} {Astrophys. J.}\ }\textbf {\bibinfo {volume} {223}},\
  \bibinfo {pages} {314} (\bibinfo {year} {1978})}\BibitemShut {NoStop}%
%%CITATION = NUPHA,A360,459;%%
\bibitem [{\citenamefont {Lamb}\ \emph {et~al.}(1981)\citenamefont {Lamb},
  \citenamefont {Lattimer}, \citenamefont {Pethick},\ and\ \citenamefont
  {Ravenhall}}]{Lamb:1981kt}%
  \BibitemOpen
  \bibfield  {author} {\bibinfo {author} {\bibfnamefont {D.~Q.}\ \bibnamefont
  {Lamb}}, \bibinfo {author} {\bibfnamefont {J.~M.}\ \bibnamefont {Lattimer}},
  \bibinfo {author} {\bibfnamefont {C.~J.}\ \bibnamefont {Pethick}}, \ and\
  \bibinfo {author} {\bibfnamefont {D.~G.}\ \bibnamefont {Ravenhall}},\ }\href
  {\doibase 10.1016/0375-9474(81)90157-3} {\bibfield  {journal} {\bibinfo
  {journal} {Nucl. Phys. A}\ }\textbf {\bibinfo {volume} {360}},\ \bibinfo
  {pages} {459} (\bibinfo {year} {1981})}\BibitemShut {NoStop}%
%%CITATION = NUPHA,A360,459;%%
\end{thebibliography}%
%-----------------------------------------------------------------

\end {document}